\documentclass{aa}  

\usepackage{graphicx}
\usepackage{txfonts}
\usepackage{hyperref}

\usepackage{subcaption}
\usepackage{natbib}
\usepackage{float}
\usepackage{placeins}
\usepackage{changepage}
\usepackage{comment}
\usepackage{xcolor}
\usepackage[normalem]{ulem}

\begin{document}

   \title{Photometric variability of nitrogen-rich Wolf-Rayet stars in Magellanic Clouds with OGLE}

   \author{J. Marković
          \inst{1} \and
          G. Sáez-Cano \inst{1} \and
          Y. Naz\'e\inst{2}\fnmsep\thanks{F.R.S.-FNRS Senior Research Associate} \and
          M. M. Rubio-Díez\inst{3}  \and
          I. Soszyński\inst{4}
          \and
          A. Udalski\inst{4}
          }

\institute{
  Universidad de Alcalá, Departamento de Física y Matemáticas, Grupo de Física Nuclear y de Partículas, 28805 Alcalá de Henares, Madrid, Spain
  \and 
   Groupe d'Astrophysique des Hautes Energies, STAR, Universit\'e de Li\`ege, Quartier Agora (B5c, Institut d'Astrophysique et de G\'eophysique),
All\'ee du 6 Ao\^ut 19c, B-4000 Sart Tilman, Li\`ege, Belgium
  \and Centro de Astrobiología, CSIC-INTA, Torrejón de Ardoz, Madrid, Spain
  \and Astronomical Observatory, University
of Warsaw, Al. Ujazdowskie 4, 00-478 Warszawa, Poland
}

   \date{Received ; accepted}

 \abstract {} {We present a comprehensive analysis of the photometric
variability of (presumably single) nitrogen-rich Wolf-Rayet (WN) stars
in the Magellanic Clouds, using long-term observations from the OGLE
survey.} {Our sample comprises 47 stars with no nearby Gaia
counterparts. We characterize both overall and short-term variabilities,
examining data dispersion and identifying periodicities. To validate our
findings, we also compare the OGLE light curves with data from the MACHO
and TESS missions.} {Variability is ubiquitous in our WR sample: about
one third of stars display high variability, or four fifths if we include
cases with moderate variations. The observed changes are found to be
periodic in 11 cases, with timescales of 2--56\,d. Such periodic
variations originate in corotating wind structures, binary effects, or
pulsations, thereby increasing the number of systems known to show these
phenomena. Surprisingly, nine targets display (quasi-periodic)
outbursts, making such changes a new type of WR variability. The variability shows a transient character, in about 30$\%$ of the sample,
with changing amplitudes for periodic signals or for outbursts (they even
sometimes completely disappear from view). Finally, we identified six long-period variables, four of which have been confirmed by at least two independent surveys.}{}

   \maketitle

\section{Introduction}

Wolf-Rayet (WR) stars play a fundamental role in the chemical evolution of the Universe due to their strong winds and high mass-loss rates. Furthermore, they represent an important step in the late evolutionary stages of massive stars, as progenitors of supernovae, gamma-ray bursts, or massive black holes. However, despite their importance, there remain significant uncertainties regarding their formation and evolution.

The observed variability in WR stars can provide insights into their physical processes. This variability might be related to intrinsic sources (such as internal pulsations, corotating interaction regions (CIRs), or inhomogeneities in the stellar wind) or extrinsic sources (i.e., a consequence of multiplicity). It can be periodic, with timescales ranging from minutes to decades, or stochastic.
 
Regarding pulsations, two main types that have been defined by their respective restoring force are pressure ($p$) modes and gravity ($g$) modes. Typically, $p$ modes have high frequencies, with pulsation periods on the order of minutes to hours. In contrast, $g$ modes have lower frequencies, with periods ranging from hours to days \citep{Aertsboook2010}. For hot and luminous stars with a high luminosity-to-mass ratio, strange-mode pulsations may also occur in their envelopes. This physical mechanism manifests as cyclic, small-amplitude, and epoch-dependent photometric variability, with periods ranging from minutes to hours \citep{Chene2011}. \citealt{Kar2024} identify strange-mode instabilities as one of the possible origins of this variability, which were initially thought to give rise to rapid pulsations on the order of hundreds of seconds \citep{Glatzel1993}. However, specific cases $-$ such as the detection of a 9.8-hour period in WR 123 \citep{Lefevre2005} $-$ were found to be consistent with models that included strange-mode instability \citep{Dorfi2006}. Such modes may also explain the high-frequency signals detected in other WRs (see \citealt{Rauw1996} on WR 66, \citealt{Toala2022} on WR 7, \citealt{Naze2021b} on WR 7, 66, 79b, 135), although at least WR 134 displays signals at low frequencies (0.44 d$^{-1}$ and harmonics, \citealt{Naze2021b}).
In addition, unstable pulsations have been advocated to explain short-term stochastic low-frequency variability, although it could also result from a combination of processes that includes photospheric perturbations or clumps in the stellar winds \citep{Naze2021b,Lenoir-Craig2022}. 

Rotational modulation may play a role in WR variability, notably through large-scale structures in the stellar wind, such as CIRs, leading to periodic variability as the star rotates \citep{Grassitelli2016}. This variability exhibits epoch-dependent and non-sinusoidal characteristics. Additionally, the presence of harmonics is often observed \citep{St-Louis2020}. The true period of the rotation might be twice the detected one, as CIRs are usually present in pairs \citep{Toala2022}. 

Binary interactions in WR stars with close companions can also induce periodic variability. This may occur through eclipses or tidal distortions, resulting in changes occurring with the orbital period, which can range from days to years. When the extended wind of a WR star partially attenuates the light from its companion, it produces a dip in the light curve. This phenomenon $-$ atmospheric eclipse $-$ typically has a period shorter than 30 days and can be effectively characterized using the analytical model proposed by \citealt{Lamontagne1996}, which provides a robust framework for fitting observed light curves. 
Furthermore, colliding stellar winds in binary systems with two hot, wind-driving components can also produce clear photometric periodic features. This variation is noticeable; for instance, in infrared brightness \citep{Williams1997,White2024}, in the optical range \citep{Luehrs1997}, and in intense X-ray emissions \citep{Pollock2006,Pollock2021}.

Previous light curve analyses have provided initial insights into optical variability in WR stars. However, these studies are significantly affected by observational bias, as the data they rely on usually cover only short timescales. Longer timescales often remain unexplored. In this study, we perform the first photometric analysis of a large sample of WR stars over a long baseline (several years), taking advantage of the time coverage of the OGLE project. In particular, we aim to focus on the nitrogen-rich WR subclass (WN), massive stars that have been observed both in binaries and as single stars, in clusters as well as in isolation (e.g., \citealt{Shenar_2017}, \citealt{Sana_2013}, \citealt{Kalari_2022}, \citealt{Bestenlehner_2011}). 

We present our dataset and introduce the criteria for variability classification and time-series analysis in Sect.~\ref{sec:data_analysis}.
Our analysis is complemented by MACHO and TESS data when available. The cross-match of our target stars with these and other surveys is outlined in Sect.~\ref{sec:X-match}, while the results are presented in Sect.~\ref{sec:results}.

\section{Data and analysis}
\label{sec:data_analysis}
\subsection{OGLE data}

The Optical Gravitational Lensing Experiment $-$ OGLE \citep{Udalski_2015} utilizes a 1.3-m Warsaw telescope located at the Las Campanas Observatory in Chile, operated by the Carnegie Institution for Science. This telescope is unique in providing extensive, long-term observations using {\it I}- and {\it V}-band filters. The OGLE project has been conducted in four phases: OGLE-I (1992-1995), 
OGLE-II (1997-2000), OGLE-III (2001-2009), and OGLE-IV (2010-2025). Each phase marks an improvement in the CCD camera size.
During the COVID-19 pandemic, observations at the Las Campanas observatory were suspended, resulting in a two-and-a-half-year gap in the OGLE-IV data, approximately between HJD 2\,458\,900 and 2\,459\,800.

We selected our sample of WN stars in the Small Magellanic Cloud (SMC) and Large Magellanic Cloud (LMC) from those listed by \citealt{Foellmi_2003a}, \citealt{Foellmi_2003b}, \citealt{Bonanos_2009}, \citealt{Bonanos_2010}, and \citealt{Shenar_2019}. The sample consists of presumably single WN stars and stars whose binary status could not be confirmed by \citealt{Shenar_2019}. The selected stars must appear within the OGLE project's sky coverage and have {\it I}-band magnitudes of 13 or fainter. Our sample ranges from WN3 to WN8 spectral types, and comprises both non-hydrogen and hydrogen-rich stars, allowing us to constrain variability as a function of spectral subtypes. 
The point spread function (PSF) for our sources in OGLE is 1.25$''$-1.35$''$. Therefore, following the approach of \citealt{Kourniotis_2014}, we verified with Gaia DR3 the absence of any bright and nearby companions within a 2$''$ radius around the target stars. Ultimately, our sample consists of 47 stars (Table \ref{tab:tableA}).

More than half of the stars in the sample (27) were observed during the OGLE-III and OGLE-IV phases, 12 stars were observed only during OGLE-IV, two stars were observed only during OGLE-III phase, one star was observed in both OGLE-II and OGLE-III, and five stars were observed in OGLE-II, OGLE-III and OGLE-IV phases of the experiment (see Table \ref{tab:surveys}). None of the stars in our sample was observed in the first phase of the experiment. Moreover, our analysis was restricted to the {\it I} band, owing to its higher data quality and better sampling  compared to the {\it V} band.

There are two sets of data, corresponding to two different reduction methods. The reduction in the first dataset was performed using reference images for each phase separately.
As the OGLE-IV phase extends longer, the reduction of the second dataset, which was provided by coauthor A. Udalski about a year after the first dataset, was done for both OGLE-III and OGLE-IV images together, using OGLE-IV as reference. This second approach minimized any shifts between these OGLE campaigns.
For our analysis, we primarily used the latter dataset, except in cases in which such photometry was unavailable (such as for BAT99-48, BAT99-54, and SMC-AB4) or when the offset between adjacent phases was smaller with the former method (e.g., BAT99-1, BAT99-67, BAT99-82, and BAT99-89). In the case of BAT99-24 and BAT99-26, we combined OGLE-III and OGLE-IV data from the second set with OGLE-II data from the first set, as the second method does not include OGLE-II data. 

Since the majority of the stars (33/47$\sim$70.2$\%$) were observed in more than one OGLE phase, the possibility of  instrumental offsets between them needs to be considered. To determine them, the difference between median {\it I} magnitudes across different OGLE phases was measured and compared to the accuracy of the absolute calibration (0.01-0.02 mag - for OGLE-III and OGLE-IV).  
If the offset between two phases was between 0.01 and 0.03 mag \footnote{propagation of errors $\sqrt{0.02^2 + 0.02^2} \sim 0.03$}, the correction was applied by shifting one of them. 
When the offset was $\lesssim$ 0.01 and/or $\gtrsim$ 0.03 mag, the light curves were not corrected. 

BAT99-31 is the only star that has been observed in two different fields during the third phase of the OGLE experiment. Three other stars (BAT99-41, BAT99-94, and BAT99-124) fell onto two different OGLE-IV fields.  
The choice of which field to use was made based on the offset value and number of data points in each field. For BAT99-41 and BAT99-124, we chose the sub-frames with a greater number of data points, despite their higher offsets. 
The offset-corrected light curves were then $\sigma$-clipped using a 95th-percentile cut over the median to remove possible outliers. 
Appendix~\ref{app:LCs} shows two example figures of the final light curves without outliers. The remaining light curve figures are provided on Zenodo.

\subsection{Variability criteria} \label{subsec:variabilityCriteria}

The selected light curves exhibit a range of variability, differing not only in amplitude but also in timescale. Based on the scatter (i.e., its standard deviation, $\sigma = \sqrt {\sum_{i=1}^{N} (x_i-\bar{x})^2/(N-1)}$) in the non-detrended and detrended light curves (see Sect.~\ref{subsec:TSanalysis}), we categorized our sample into three groups. Stars are classified as having a low level of variability if their scatter values are less than twice the mean error. Those with scatter greater than four times the mean error are classified as highly variable, while those falling between these two thresholds are considered moderately variable. The scatter values are listed in Table~\ref{tab:tableA}, along with the corresponding status of each source. 

\subsection{Time-series analysis}
\label{subsec:TSanalysis}
The photometric data we use are particularly well suited for studying changes in stellar brightness over long periods of time. Some stars in our sample exhibit both short- and long-term variability, often superimposed on one another. The presence of short- and long-term signals in the frequency spectrum renders the analysis more complex. Therefore, it is beneficial to separate them prior to analysis. To achieve this, we first examined the offset-corrected and outlier-removed light curves to identify long-term signals. We then removed the underlying long-term trend from the light curves in order to analyze short-term variability. This detrending was performed using a one-year sliding window.

We conducted the period analysis using the modified generalized Lomb-Scargle (GLS) algorithm \citep{Heck1985,zechmeister-kurster2009} in the version of \citealt{wbhatti_astrobase}. The GLS is particularly well suited for our dataset, as it is designed for unevenly spaced time series. 
The generalized method fits the mean during the calculation, rather than fixing it from the start as was done in the initial Lomb-Scargle method.

The GLS analysis was conducted over a period range from 1.5 days (the average cadence of OGLE light curves for Magellanic clouds (MCs) is 1-3 days) to the total observational time span for long-term variability and from 1.5 days to 100 days for short-term variability. We calculated the spectral window function for all stars to ensure that the detected signals were not artifacts of the sampling.

Signal-to-noise ratios (S/N)  were calculated using the Python \texttt{lightkurve}\footnote{\href{https://lightkurve.github.io/lightkurve/}{https://lightkurve.github.io/lightkurve/}} package to identify significant peaks. We used the \texttt{periodogram.flatten} function, which normalizes the power spectrum by dividing it by a background estimate obtained via a moving filter applied in logarithmic space.
This function is not sensitive to low frequencies, so we developed a custom function to calculate it. The signal is considered significant if it exceeds the threshold value of 4 \citep{Breger_1992}.
Once a period is identified and deemed significant, the first step is to fold the light curve using the first OGLE observation as $T_{0}$. Next, a spline is fit to the folded data, and the time of minimum of the folded light curve is adopted as the refined $T_{0}$ used for the final phased light curve plot. 
The amplitudes of variability were determined from the phase-folded data. The periods corresponding to the highest S/N values, along with the amplitudes, the adopted $T_{0}$ values of the signal, and their uncertainties, are presented in Appendix~\ref{app:ephemeris}. We also binned the phase-folded light curves, which facilitates the visualization of the periodic signal by reducing noise and highlighting the overall trend (e.g., as is shown in the bottom panels of the figures in Appendix~\ref{app:detrend_panels}). 

The GLS frequency spectra and binned phase-folded light curves of stars with a significant long-period signals are discussed and presented in the Sect.~\ref{subsec:OGLE}, while the panels containing two examples of the detrended light curves, GLS frequency spectra, and phase-folded and binned data are shown in Appendix~\ref{app:detrend_panels}. The panels for the rest of the stars with confirmed periodicities (Sect.~\ref{sec:results}) are available on Zenodo.

To test whether the variability across the different OGLE campaigns remains consistent for each star exhibiting significant periodicity, a $\chi^2$ analysis was performed. For this purpose, each light curve was folded using the ephemeris specific to the target under consideration and divided into a fixed number of phase bins for each campaign.
A a $\chi^2$ test was then applied pairwise between campaigns for each star, adopting a significance level (p-value) of 0.05. Results are discussed in Appendix~\ref{app:comments} for individual stars.

Finally, we generated time-frequency diagrams (Appendix~\ref{app:TFD}) using a sliding window technique to evaluate if any existing periodic signal is temporary or coherent throughout the whole light curve. For each star, the step size was fixed to 30 days and the window size was set to 1 year.
Our light curves vary in cadence and duration, and contain numerous gaps in the observations (due to i.e., bad weather, instrumental upgrades, or the ``COVID gap"). This required the implementation of a threshold for 30 data points in each window: if the data points in a window fell below this threshold, the GLS was not computed for that window, and the power was set to zero.

\section{Cross-match with the literature and catalogs} \label{sec:X-match}
   
\subsection{WDS catalog}

The ICRS coordinates of the stars in our sample were cross-matched with the ones from the Washington Double Star Catalog (WDS), which is intended to contain all known visual double stars for which at least one differential measure has been published. We found no matches.

\subsection{Gaia DR3}
\label{sec:Gaia}
\noindent Astrometric perturbations in single-source pipeline fits are quantified in various ways within the Gaia pipeline. 
We checked the values of the astrometric excess noise parameter, $\epsilon_i$, and its significance, $D$, which measures the disagreement between observations and the best-fitting standard astrometric model (using five astrometric parameters). If the value of $\epsilon_i$ is close to zero, the residuals of the fit statistically agree with the assumed observational noise \citep{Lindegren2012}. Sources that deviate from the standard five-parameter astrometric model (e.g., unresolved binaries, or exoplanet systems) may have positive excess noise, if their significance is $D\geq$ 2. 
Along with $\epsilon_i$, Gaia data releases also provide a statistical parameter known as RUWE (renormalized unit weight error), whose values larger than unity are a signpost for intrinsic source complexity. Thus, RUWE and $\epsilon_i$ are complementary to each other. 
There are three stars with RUWE$>$1.4, D$\geq$2, and $\epsilon_{i}>$0.01 \citep{Gandhi2021}: BAT99-40, SMC-AB4, and SMC-AB9. These objects thus probably have undetected companions. 

\subsection{Photometric surveys} \label{sec:Photometric_surveys}

To corroborate our findings, we searched for photometric data from publicly available surveys. Different surveys cover different wavelength ranges, implying that the contribution of the continuum versus strong lines of the WR stars will be different for each photometric dataset that was examined. Furthermore, different lines may be produced over different volumes, which may again lead to differences between datasets. The amplitude (and in some cases even the shape) may thus vary between bands -- particularly for variability processes influenced by stellar winds. However, the presence of the variability and, in particular, its periodicity should remain unaffected.

First, we examined data from the All Sky Automated Survey (ASAS) and the All Sky Automated Survey for SuperNovae (ASAS-SN), two ground-based photometric sky surveys. We focused on several stars that appear bright and display large-amplitude photometric variations in OGLE (BAT99-3, BAT99-26, BAT99-48, and BAT99-56).
After folding the ASAS and ASAS-SN light curves with the OGLE ephemeris, we found that these data were too noisy to reveal the subtle variations detected by OGLE, even in these ``best" cases. Consequently, the ASAS and ASAS-SN data are not suitable for our purposes. 

Second, we could retrieve data for 33 of our target stars (see Table \ref{tab:surveys}) from the Massive Compact Halo Objects (MACHO) Project public archive\footnote{\href{https://macho.nci.org.au/}{https://macho.nci.org.au/}}. After removing corrupted data and outliers, we performed a GLS analysis on both blue-band and red-band MACHO light curves, considering both non-detrended data and data detrended using a one-year sliding window.

Finally, the Transiting Exoplanet Survey Satellite (TESS) provides high-cadence photometry for selected regions of the sky, with its pointing changing approximately every 27 days (i.e., one sector). Some of the fields of view overlap between sectors, and this is the case for the MC observations, for which many sectors are available (34 for LMC and 5 for SMC). However, TESS pixels are 21$^{\prime\prime}$ in size and the stellar signal is usually distributed across several pixels. The low spatial resolution of TESS for extragalactic objects makes it challenging to fully resolve individual sources, leading to possible contamination from nearby stars. Therefore, this poses a significant challenge for stars in the MCs in general, and for our targets in particular, since they are faint and located in crowded areas. Indeed, BAT99-7 and BAT99-35 are the only targets without close ($<1'$) and bright ($\Delta(G)<2$mag) companions in Gaia DR3 catalog, while other stars are often not the brightest objects within a one-arcmin distance. Nevertheless, we examined the TESS light curves of some of our targets (see Table \ref{tab:surveys}). Preliminary attempts using the usual light curve extraction showed mixed results, most probably because of crowding issues. Therefore, we next used the TGLC \citep{Han2023} package. This tool uses the Gaia DR3 catalog to obtain a full list of sources in a chosen region, determines the PSF shape, and then cleans the image cutouts for all sources (assumed to be constant) except the chosen target. Aperture photometry is then performed on the cleaned images in a 3 $\times$ 3 pixel area. This was carried out for stars that OGLE detected as having a short period and that did not have a companion brighter than themselves within 1$'$ (BAT99-3, BAT99-26, BAT99-31, BAT99-47, BAT99-56, BAT99-65, and SMC-AB4). These objects rank among the most highly variable stars of our sample.
Note that we only considered the aperture TGLC light curves, not the so-called calibrated ones (as those are detrended for daily changes, and we are interested in variations on timescales longer than 1 day, not shorter ones) or the PSF-corrected light curves (which are more sensitive to imperfect decontamination). The only two exceptions are BAT99-26 and BAT99-31, for which the PSF method produced better results. This represents the best effort to get uncontaminated light curves but, because of the high crowding, a perfect elimination of contamination cannot be guaranteed, even for those best cases. Section \ref{subsec:OGLE&TESS} further discusses the results obtained with those data.

\subsection{Spectroscopic surveys}
\label{sec:spec_surveys}
To further validate our findings, we examined the ESO archives, but found no spectroscopic data that were not discussed before. Therefore, a bibliographic review summarizes the general spectroscopic variability results.

\citealt{Foellmi_2003b} investigated 61 nitrogen-rich WNE stars in the LMC, using the CTIO telescope over time spans ranging from 4 to 50 days. Their aim was to identify single versus binary WR stars and determine their spectral types. They performed a similar analysis in SMC \citep{Foellmi_2003b}, with time coverages of between 40-49 days. Seventeen of the stars from their study overlap with our LMC targets, along with six from the SMC. Of this sample, three LMC stars and two SMC stars are classified as short-period binary candidates in our study (see Table \ref{tab:tableA} and Sect. \ref{sec:results}).
\citealt{Schnurr2008} analyzed 41 late-type, nitrogen-rich WR stars in the LMC observable with ground-based telescopes. Five of these stars are included in our sample, but none exceed the RV variability threshold.
\citealt{Hainich_2014} built upon the datasets of \citealt{Foellmi_2003b} and \citealt{Schnurr2008} to derive fundamental physical parameters for WN stars in the LMC, classify their spectra, and identify potential binary candidates based on published X-ray information. BAT99-31 is classified as a binary candidate, based on the findings of \citealt{Foellmi_2003b}. BAT99-47 and BAT99-82 are considered to be questionably single due to their unusually high X-ray emission.
\citealt{Shenar_2019} studied a sample of 44 WN stars, all either confirmed binaries or suspected binary candidates, using data from \citealt{Foellmi_2003b} and \citealt{Schnurr2008}. Of these, only three stars -- BAT99-31, BAT99-47, and BAT99-82 -- are included in our dataset, all of which are classified as binary candidates.
Most recently, \citealt{Schootemeijer2024} analyzed high-resolution UVES spectra from 2006, which include all stars in our SMC sample. Their Monte Carlo simulations exclude the presence of companions more massive than 5 $M_{\odot}$ with orbital periods shorter than one year. Appendix \ref{app:comments} provides details on these spectroscopic results for each target.

\section{Assessing variability} \label{sec:results}
\subsection{OGLE} \label{subsec:OGLE}

Using the scatter in the light curves (Sect.~\ref{subsec:variabilityCriteria}), we classified the targets into three categories for both overall variability (using the non-detrended light curves) and short-term variability (using the detrended curves). Seventeen stars exhibit high overall variability, while 16 show high short-term variability. Moderate variability is observed in 21 targets for overall and 20 cases for short timescales. The remaining stars fall into the low variability category, with 9 for overall timescales and 11 for short timescales (Table~\ref{tab:tableA}).

\begin{figure*}[ht!]
\centering

\begin{subfigure}[t]{0.33\linewidth}
    \centering
    \includegraphics[width=\linewidth]{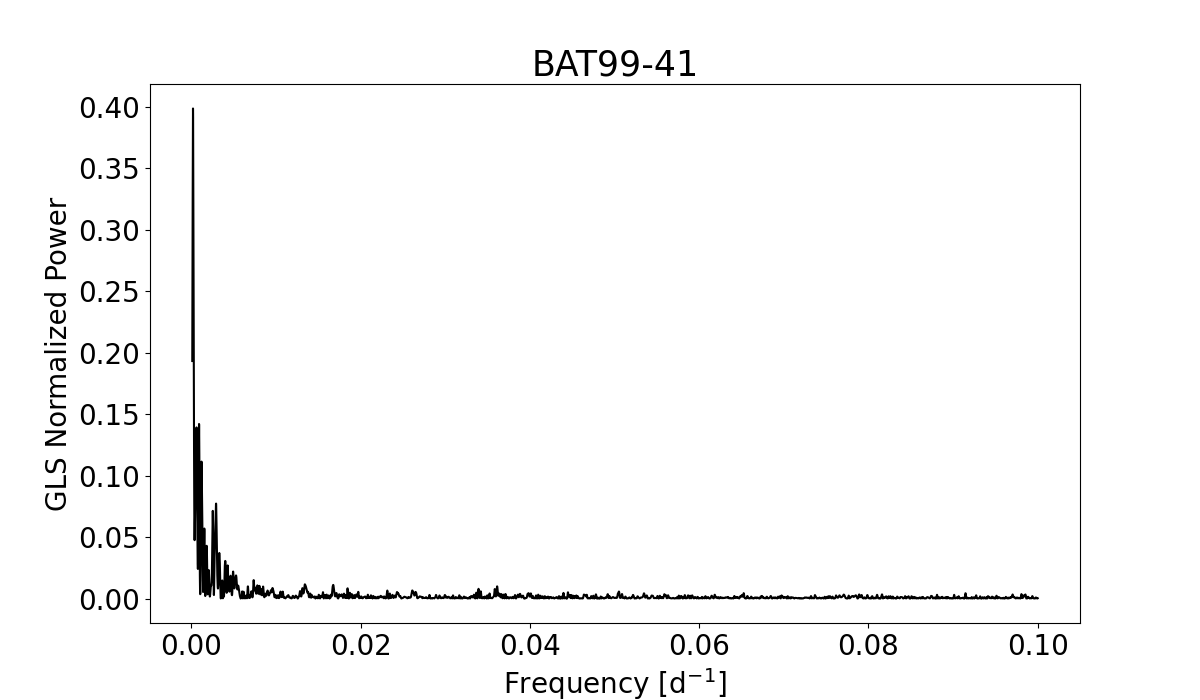}
\end{subfigure}
\begin{subfigure}[t]{0.33\linewidth}
    \centering
    \includegraphics[width=\linewidth]{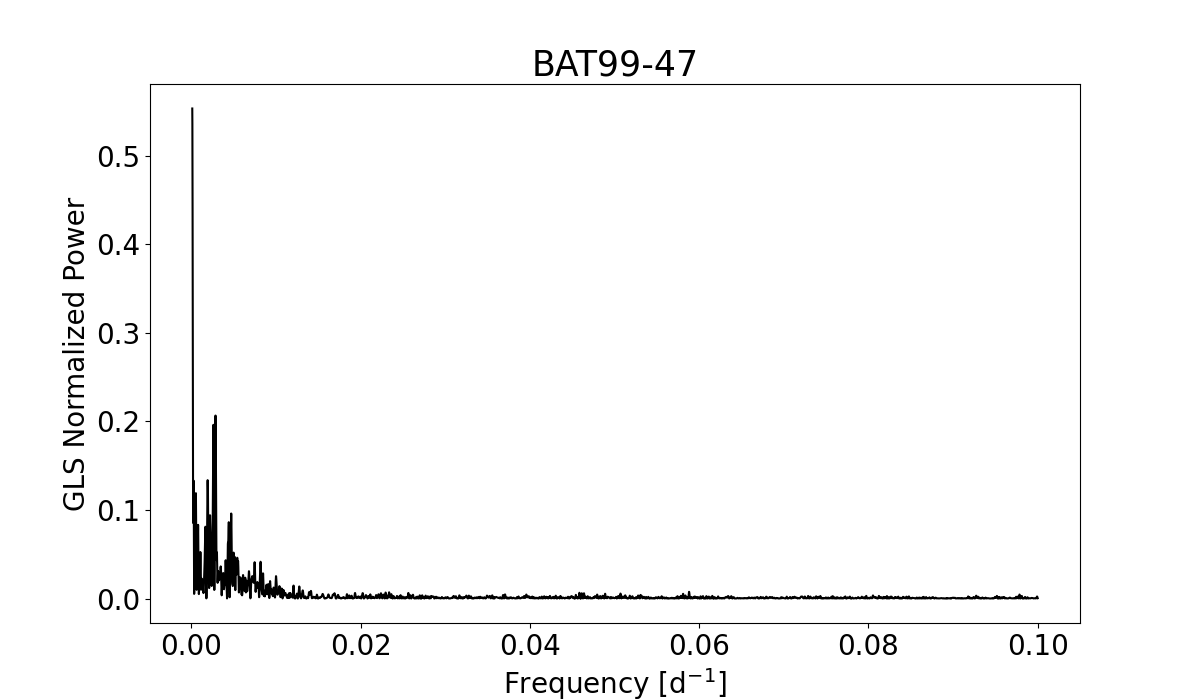}
\end{subfigure}
\begin{subfigure}[t]{0.33\linewidth}
    \centering
    \includegraphics[width=\linewidth]{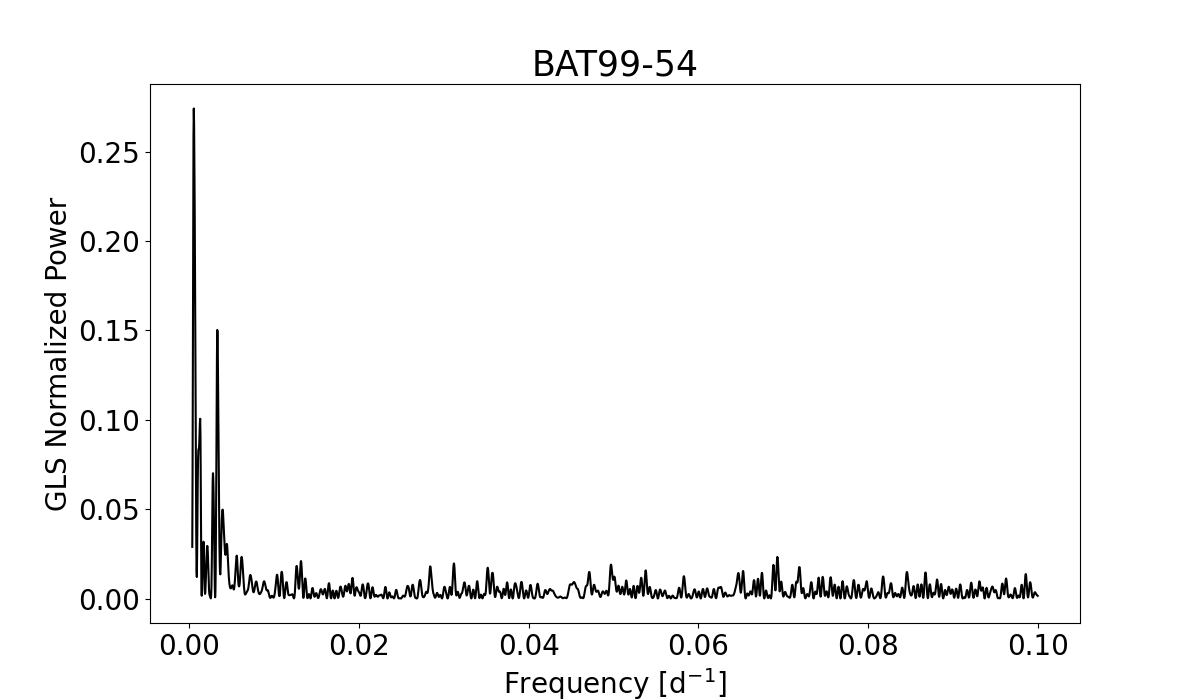}
\end{subfigure}

\vspace{2mm}

\begin{subfigure}[t]{0.33\linewidth}
    \centering
    \includegraphics[width=\linewidth]{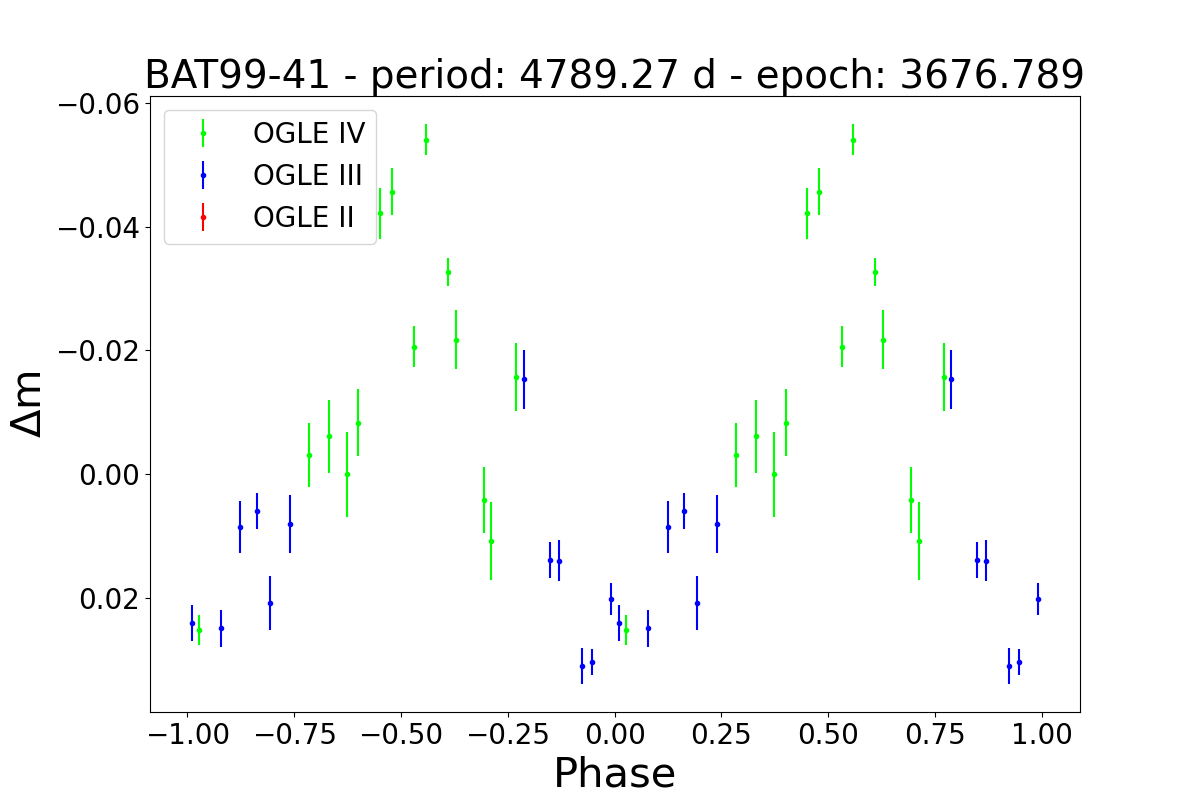}
\end{subfigure}
\begin{subfigure}[t]{0.33\linewidth}
    \centering
    \includegraphics[width=\linewidth]{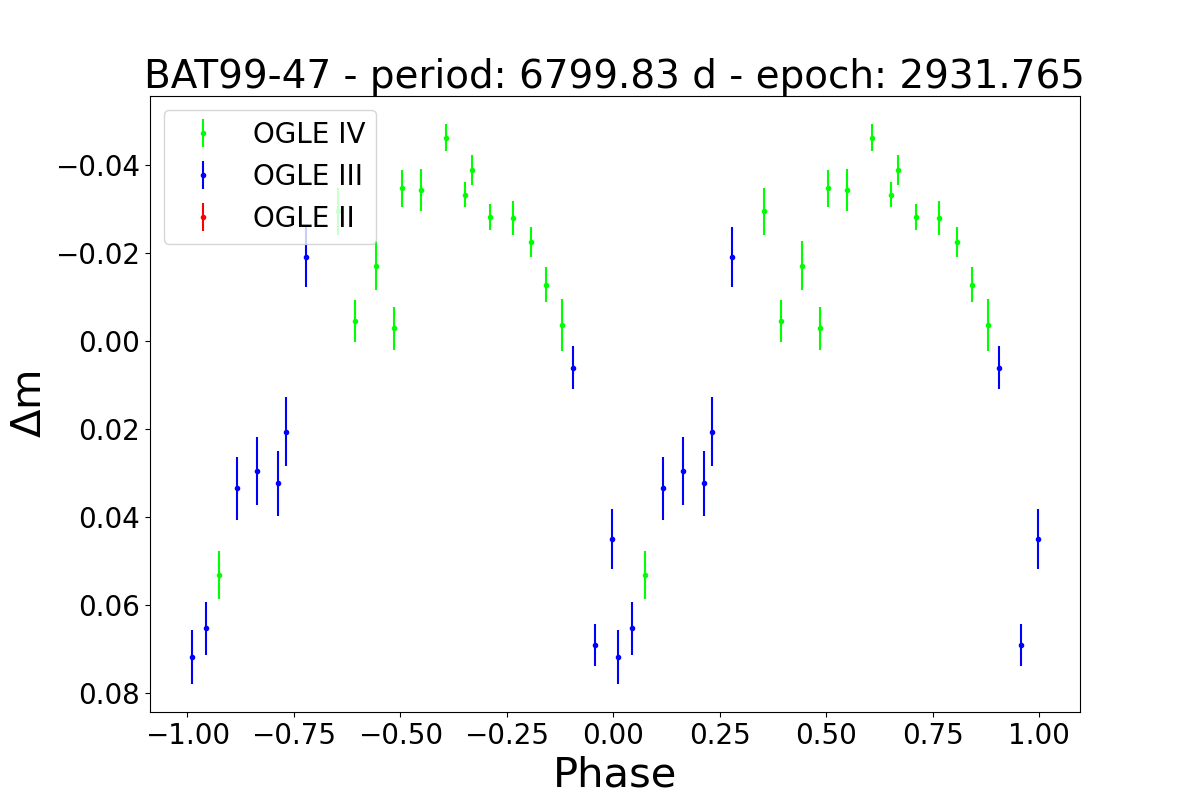}
\end{subfigure}
\begin{subfigure}[t]{0.33\linewidth}
    \centering
    \includegraphics[width=\linewidth]{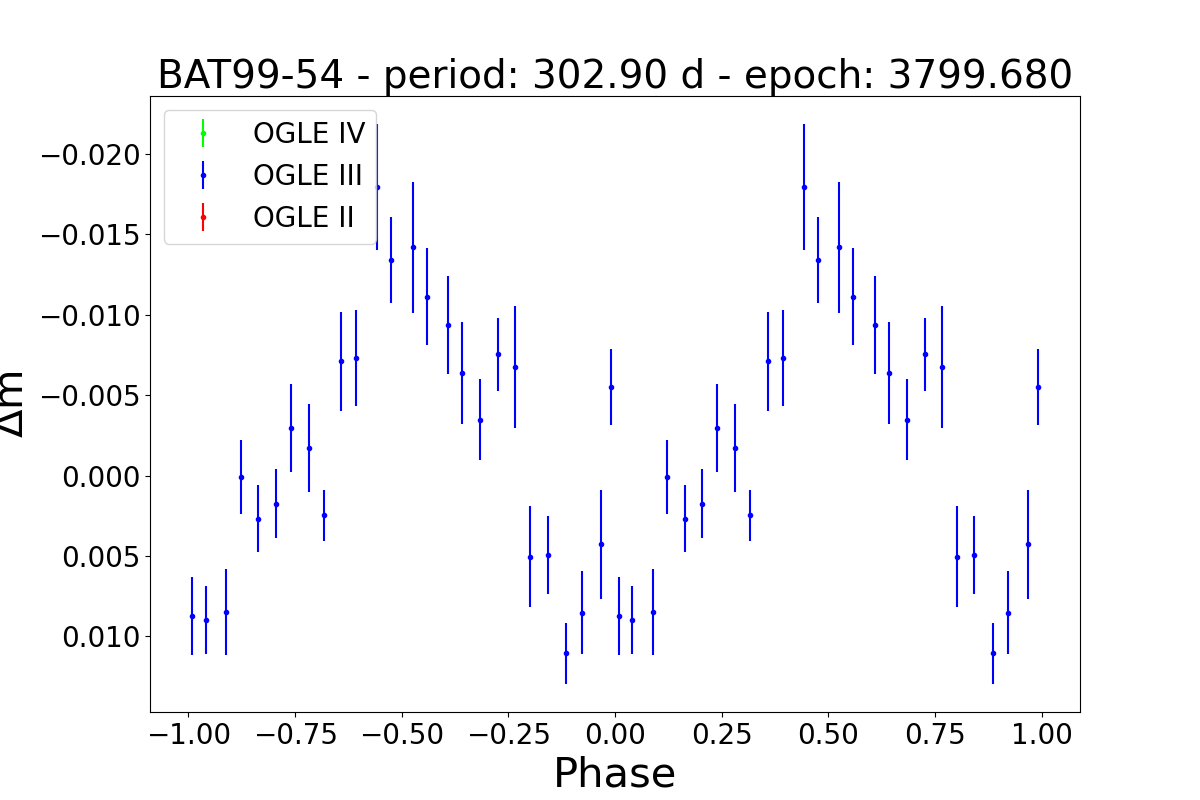}
\end{subfigure}

\begin{subfigure}[t]{0.33\linewidth}
    \centering
    \includegraphics[width=\linewidth]{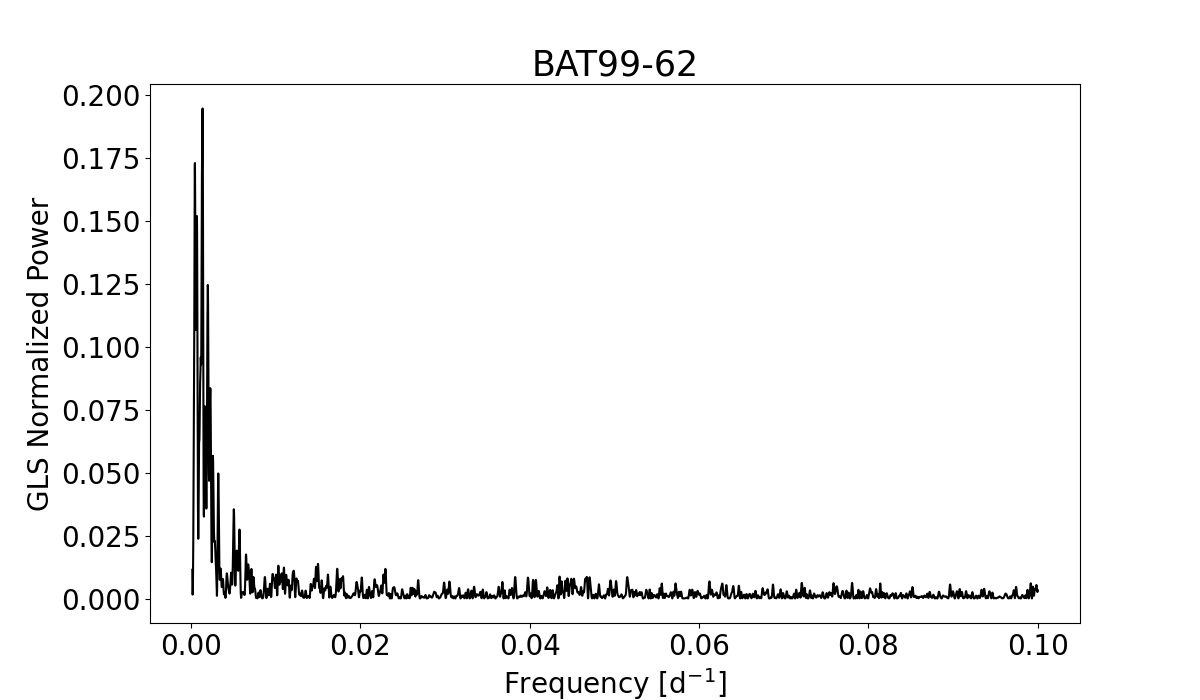}
\end{subfigure}
\begin{subfigure}[t]{0.33\linewidth}
    \centering
    \includegraphics[width=\linewidth]{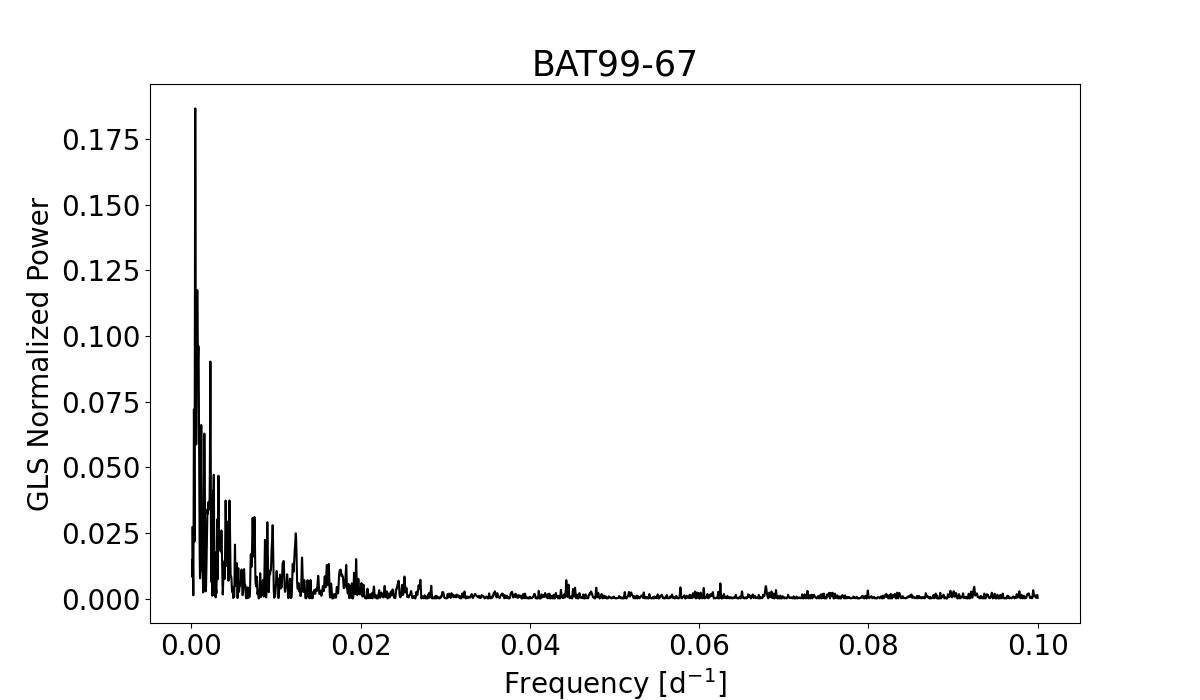}
\end{subfigure}
\begin{subfigure}[t]{0.33\linewidth}
    \centering
    \includegraphics[width=\linewidth]{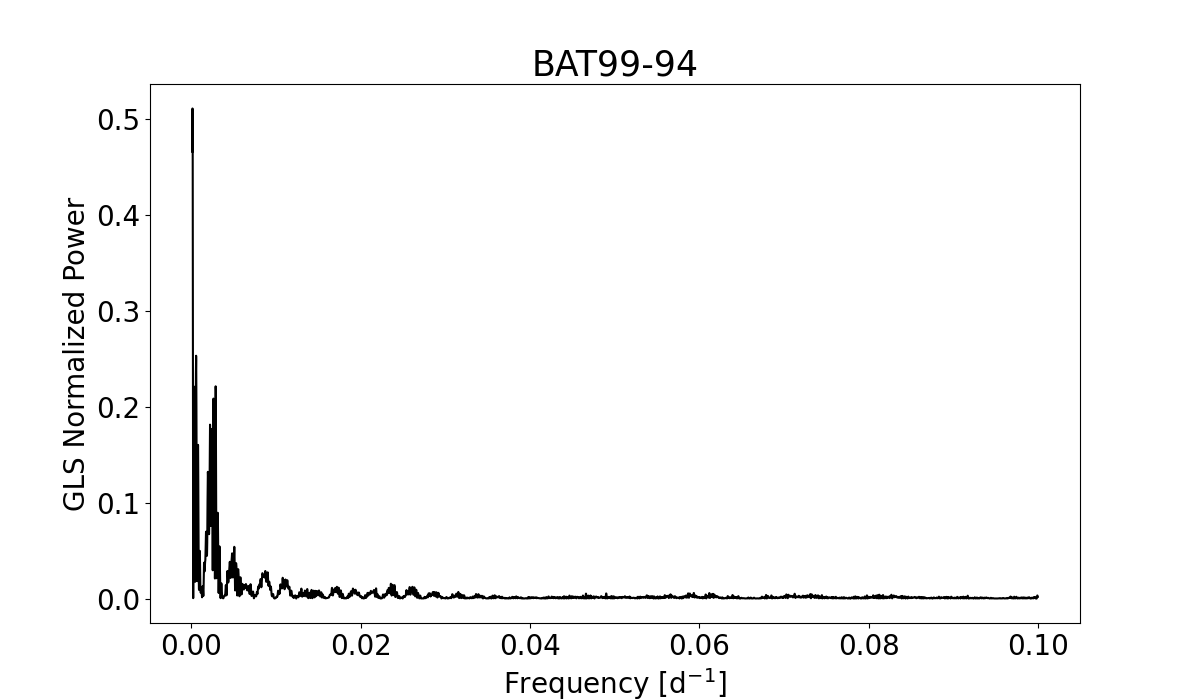}
\end{subfigure}

\vspace{2mm}

\begin{subfigure}[t]{0.33\textwidth}
    \centering
    \includegraphics[width=\textwidth]{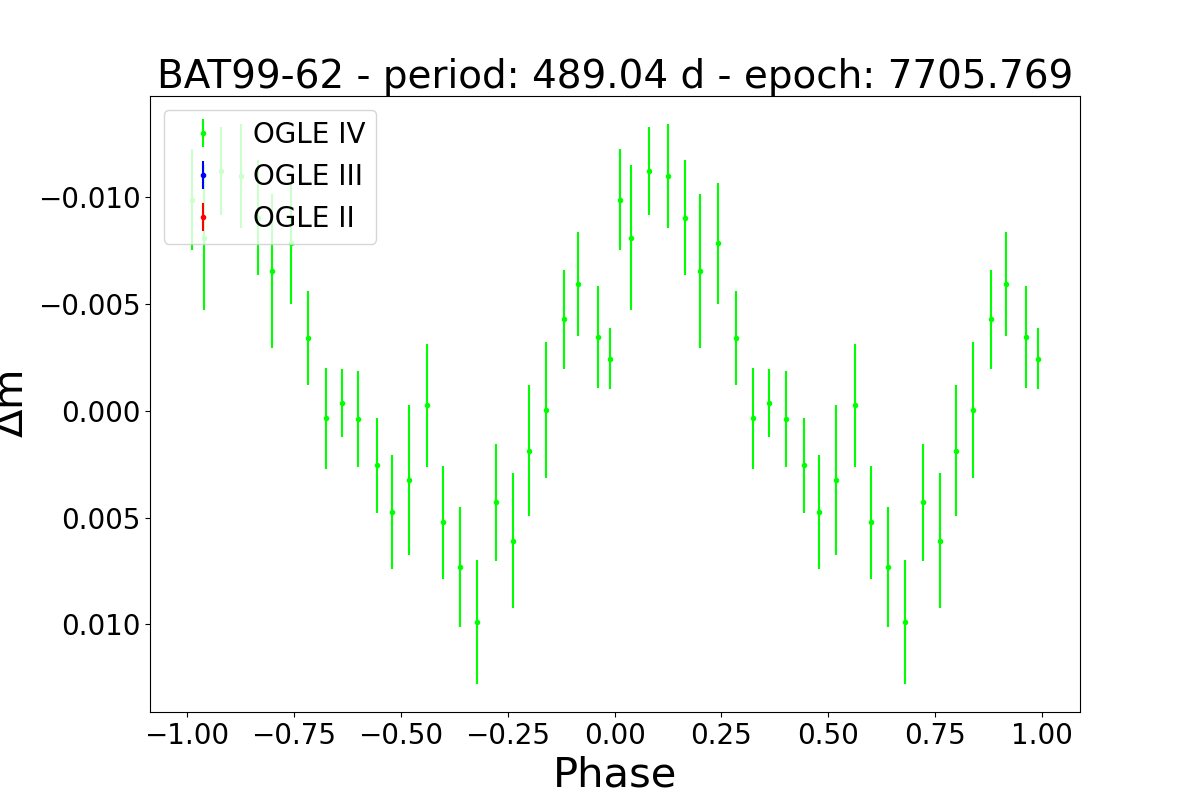}
\end{subfigure}
\hfill
\begin{subfigure}[t]{0.33\linewidth}
    \centering
    \includegraphics[width=\linewidth]{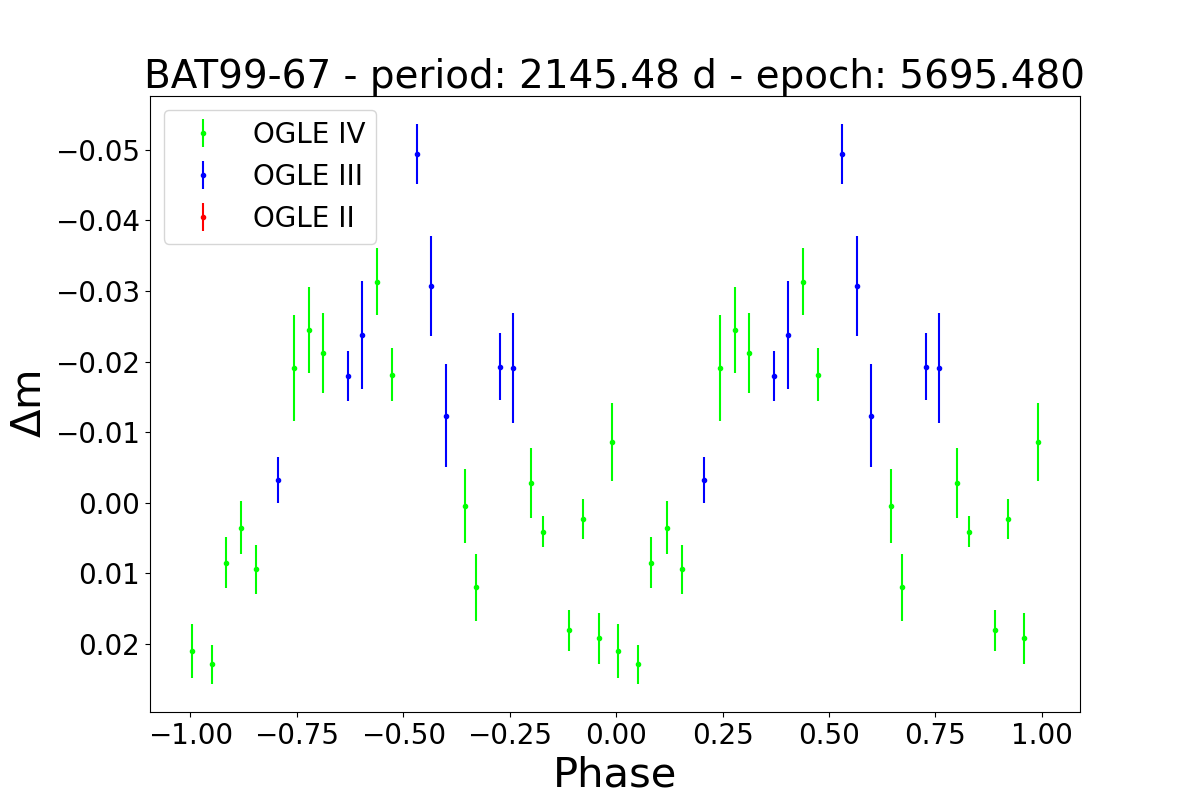}
\end{subfigure}
\begin{subfigure}[t]{0.33\linewidth}
    \centering
    \includegraphics[width=\linewidth]{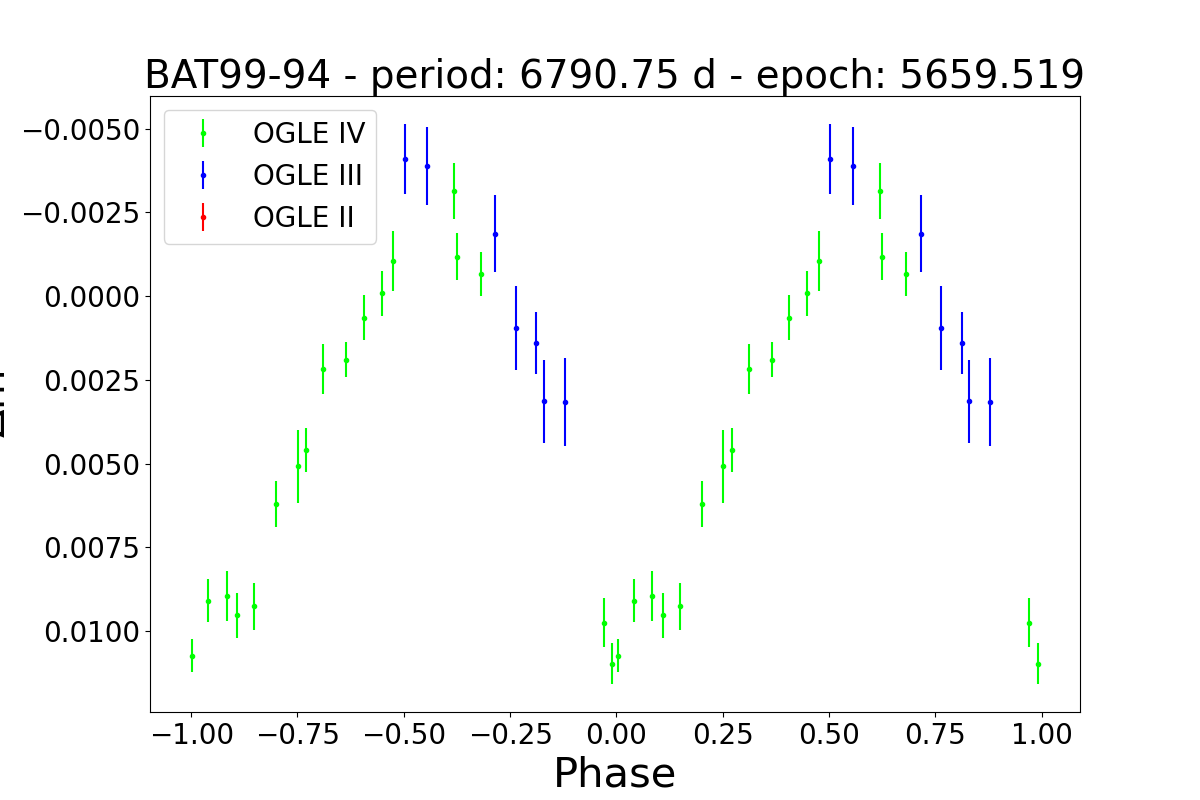}
\end{subfigure}

\caption{First and third rows: GLS periodograms zoomed-in to the frequencies lower than 0.1 d$^{-1}$. Second and fourth rows: Binned, phase-folded light curves for the stars BAT99-41, BAT99-47, BAT99-54, BAT99-62, BAT99-67,\, and BAT99-94. The period and reference epoch corresponding to the most prominent peak are noted in the title of each bin-folded light curve (see also Table~\ref{tab:ephemeris}).}
\label{fig:longterm}
\end{figure*}

The GLS analysis on the non-detrended OGLE light curves revealed six stars (BAT99-41, BAT99-47, BAT99-54, BAT99-62, BAT99-67, and BAT99-94) exhibiting significant peaks at long periods (the ephemeris are presented in Table~\ref{tab:ephemeris_long}). All are considered to be convincing detections, as their bin-folded light curves fold well; i.e., they follow a clear and smooth periodic pattern (Fig.~\ref{fig:longterm}).

The analysis of the detrended light curves revealed 16 stars with significant short-term periodicities (one of them, BAT99-47, shows marginal significance, with a S/N slightly below yet very close to 4). The corresponding ephemerides are listed in Table~\ref{tab:ephemeris}. Two stars -- BAT99-48 and BAT99-56 -- were observed in only one OGLE campaign. In both cases, their phase-binned light curves exhibit a clear periodic pattern. For those stars observed in more than one OGLE campaign, we performed a $\chi ^2$ test with a confidence level of $95\%$ comparing their phase-binned light curves. Based on the $\chi ^2$ test result, seven stars exhibit the same variability across OGLE campaigns: BAT99-2, BAT99-5, BAT99-31, BAT99-47, BAT99-51, BAT99-67, and SMC-AB4. For two stars, two campaigns show the same variability while the other campaign is significantly different from them: BAT99-24 (OGLE-III and OGLE-IV show the same variability, although they differ compared to OGLE-II) and BAT99-26 (OGLE-II and OGLE-III show the same variability, although they differ compared to OGLE-IV). Finally, significant differences between all available OGLE campaigns are found for five stars, implying a changing variability character: BAT99-1, BAT99-3, BAT99-65, BAT99-124, and SMC-AB9.

Based on the number of significant peaks shown by the periodograms, these 16 stars can be classified as follows (Table~\ref{tab:tableA}):
\begin{itemize}
    \item Single period: BAT99-51 (visible mainly in short time spans), BAT99-67, SMC-AB4 (throughout the entire observation period), and SMC-AB9 (visible mainly after the COVID gap).
    \item A period and its (sub-)harmonic/s: BAT99-3, BAT99-24, BAT99-48, BAT99-56, BAT99-65, and BAT99-124. For all of these, both signals exhibit slight evolution in the time-frequency diagram (see Appendix \ref{app:TFD}).
    \item Two independent periods: BAT99-31 and BAT99-47. These peaks also exhibit slight evolution in the time-frequency diagram (see Appendix \ref{app:TFD}).
    \item Multiple periods (forest of peaks): BAT99-1, BAT99-2, BAT99-5, and BAT99-26, with their most significant periods ranging from around 40 to 80 days.
\end{itemize}

\subsection{OGLE and MACHO}

\begin{figure*}[h!]
    \centering
    \subfloat{
        \includegraphics[width=0.45\textwidth]{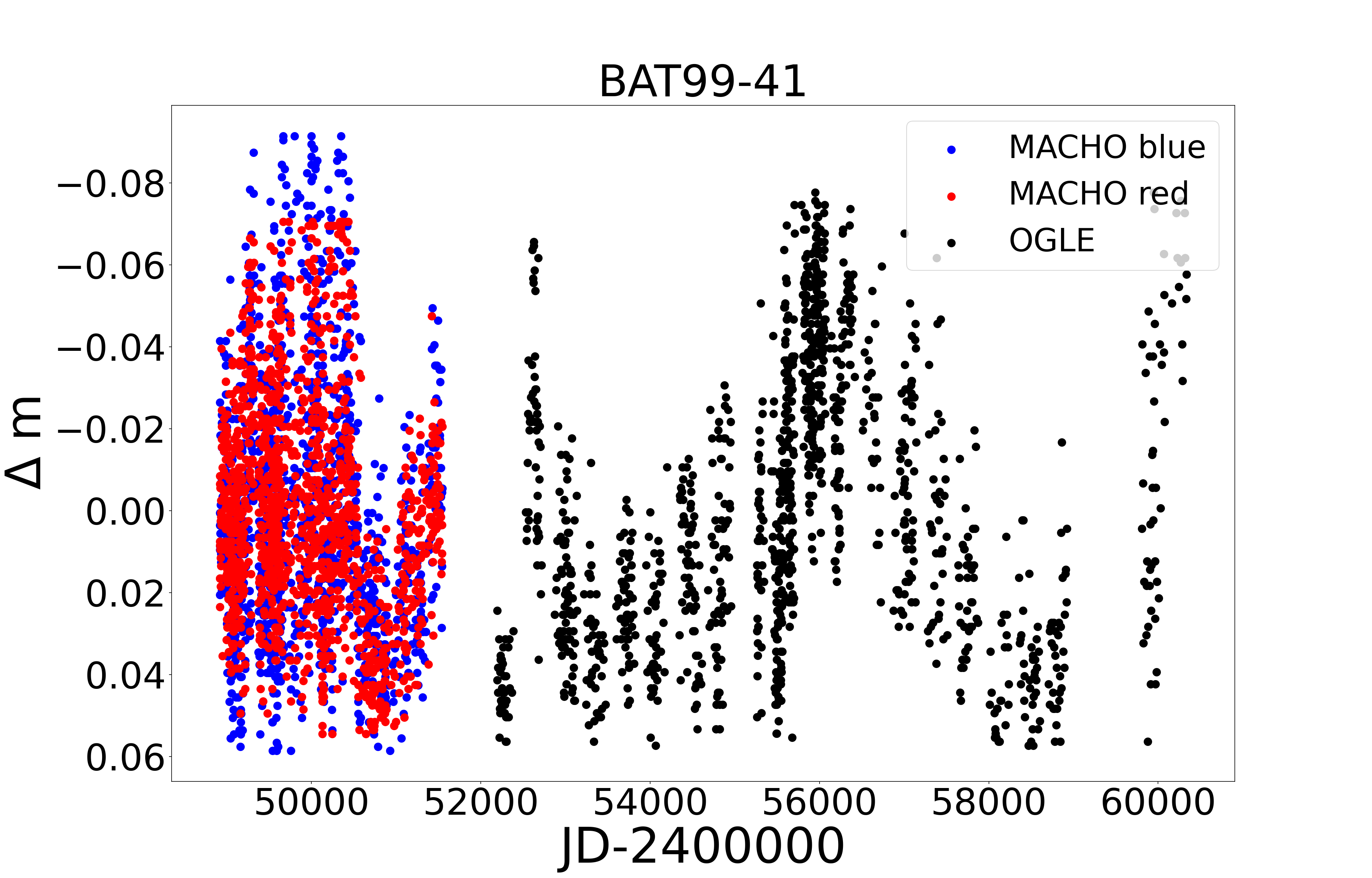}
    }
    \subfloat{
        \includegraphics[width=0.4\textwidth]{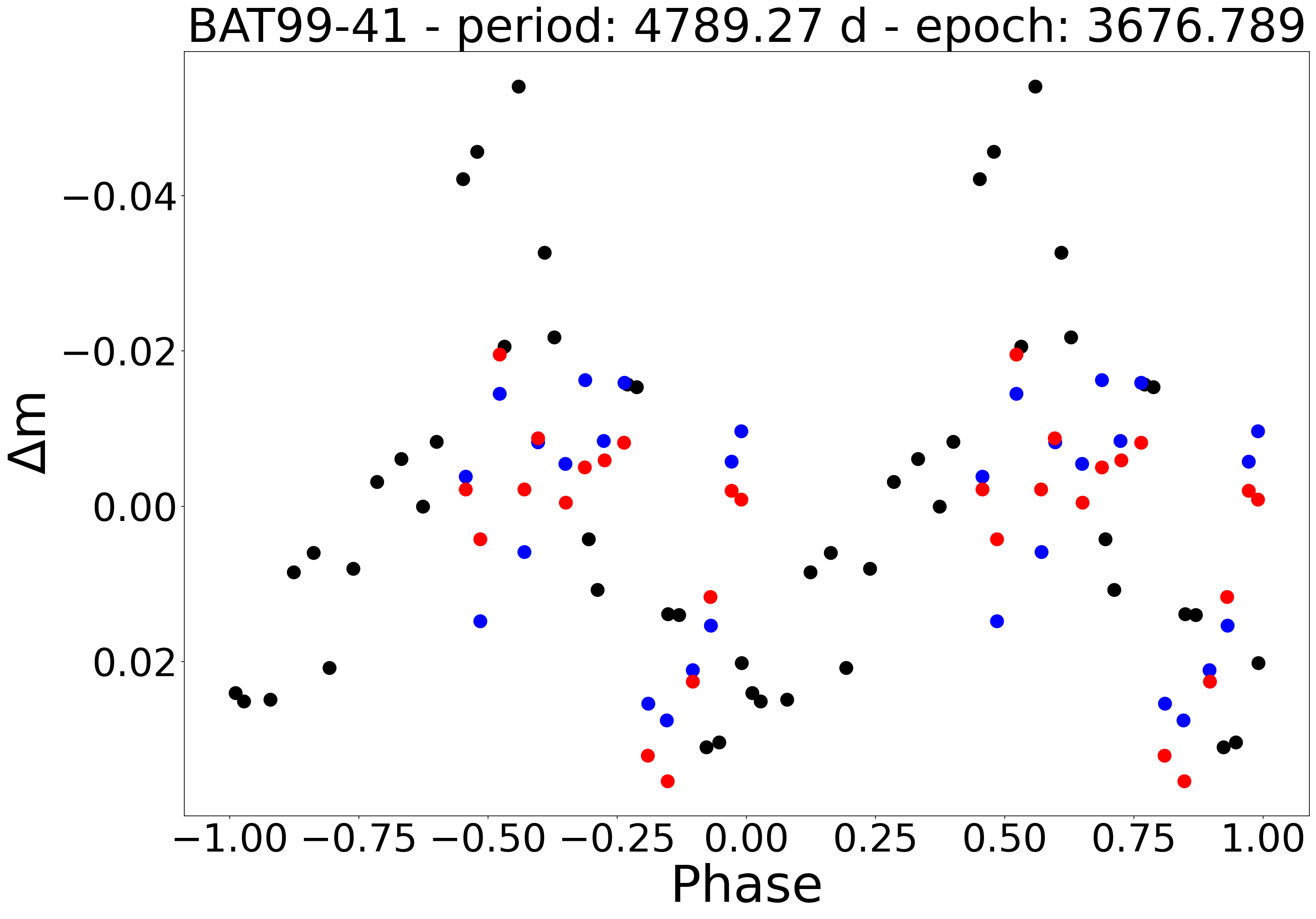}
    }\\
    \subfloat{
        \includegraphics[width=0.45\textwidth]{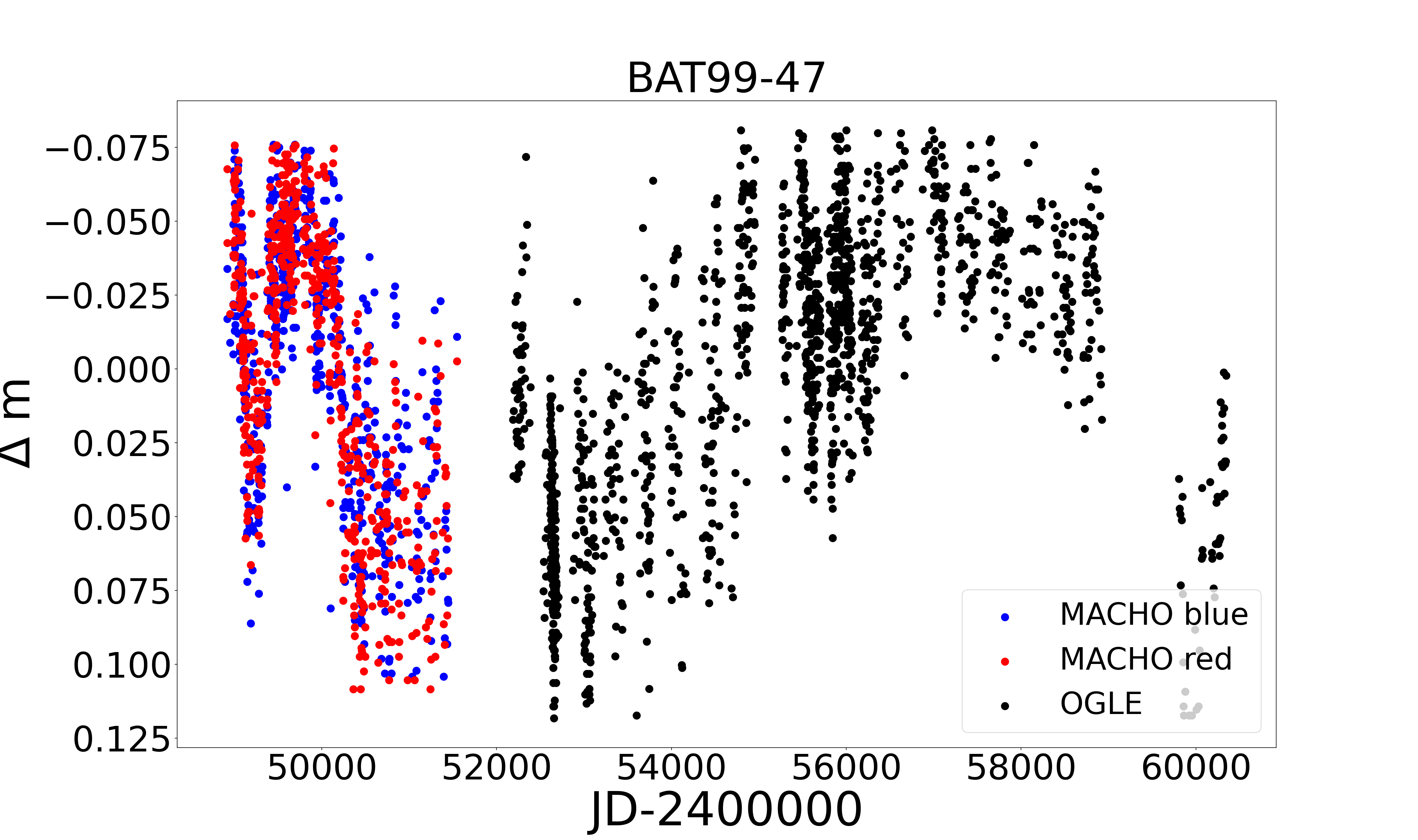}
    }
    \subfloat{
        \includegraphics[width=0.4\textwidth]{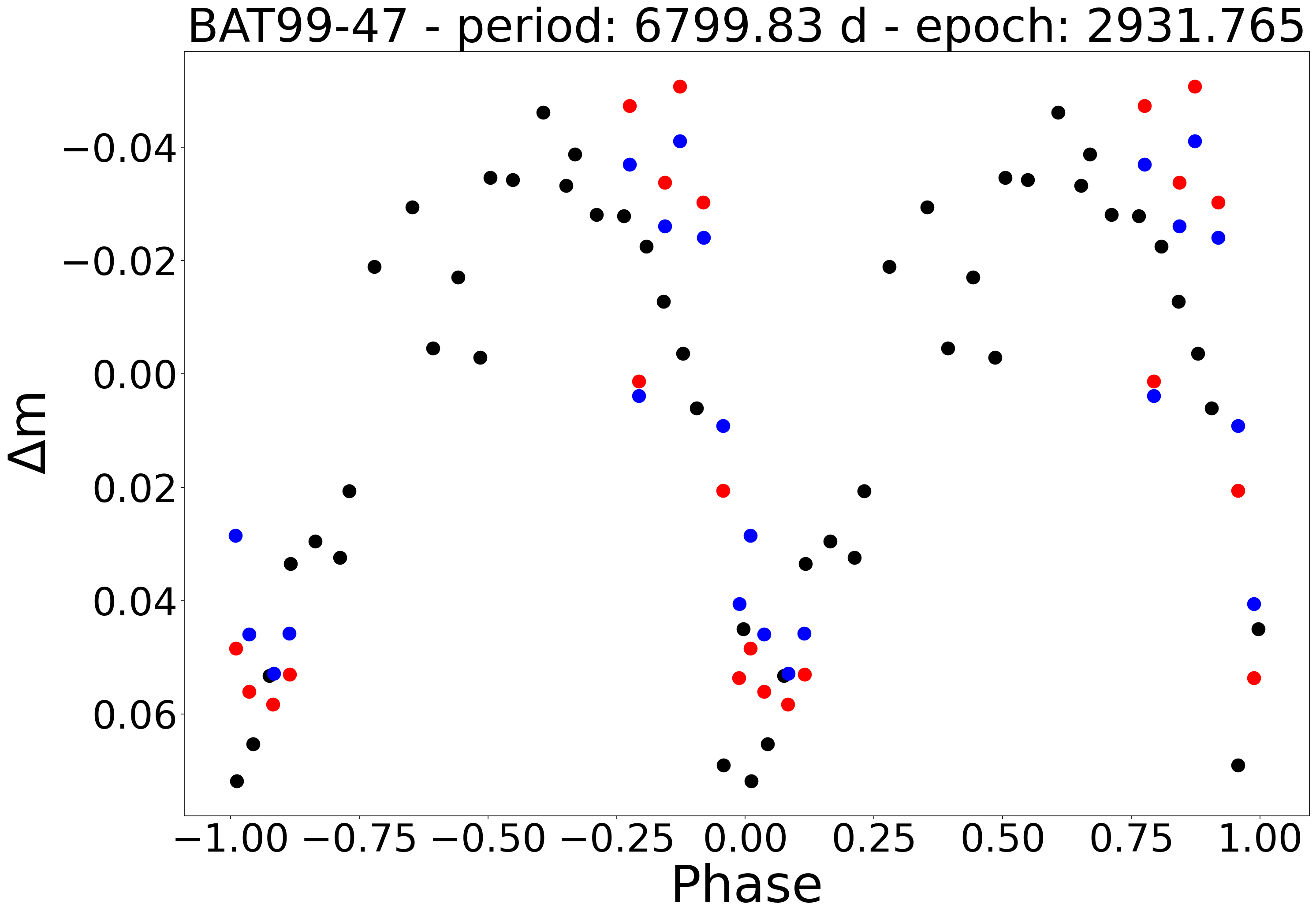}
    }\\
    \subfloat{
        \includegraphics[width=0.45\textwidth]{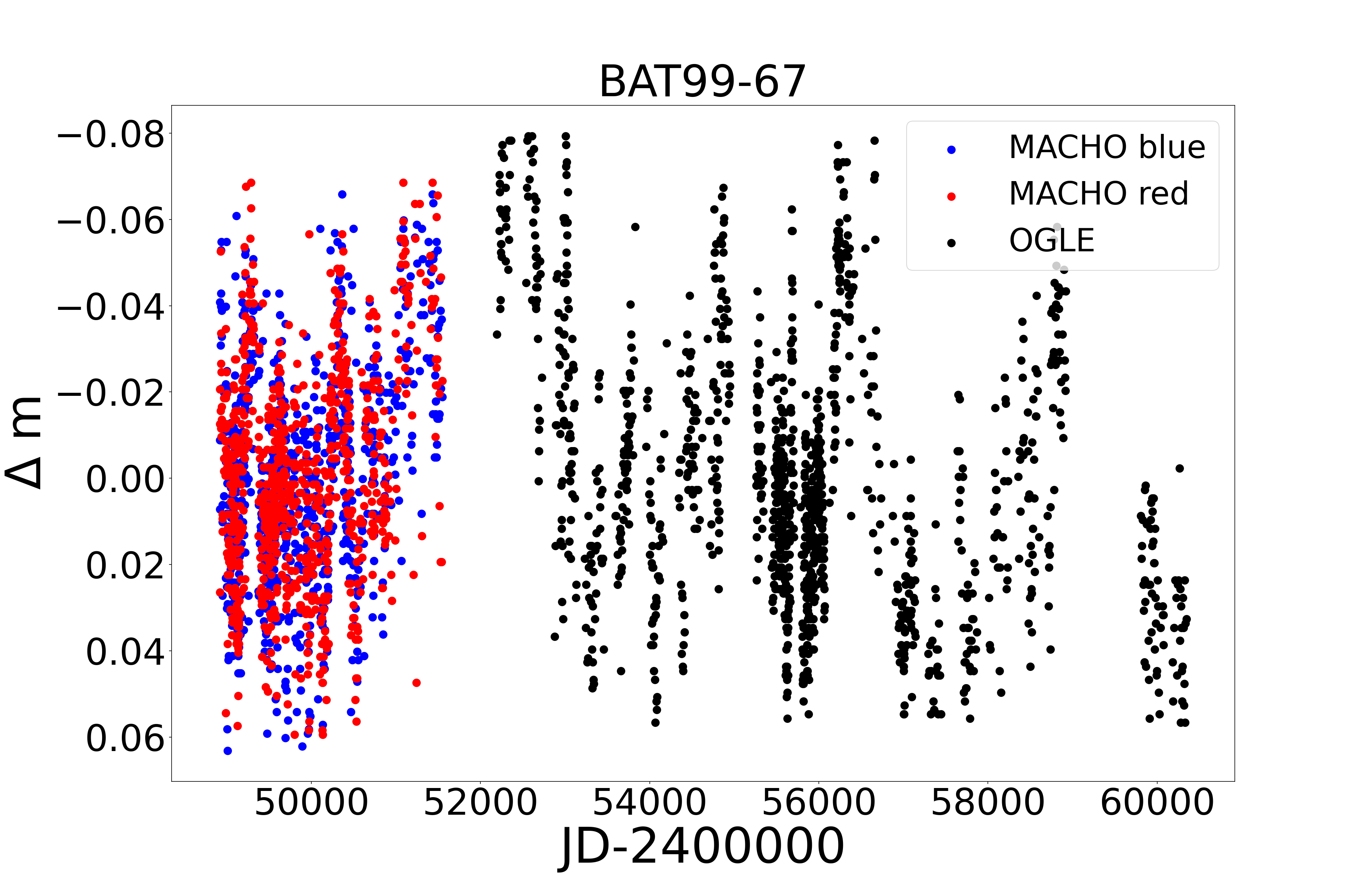}
    }
    \subfloat{
        \includegraphics[width=0.4\textwidth]{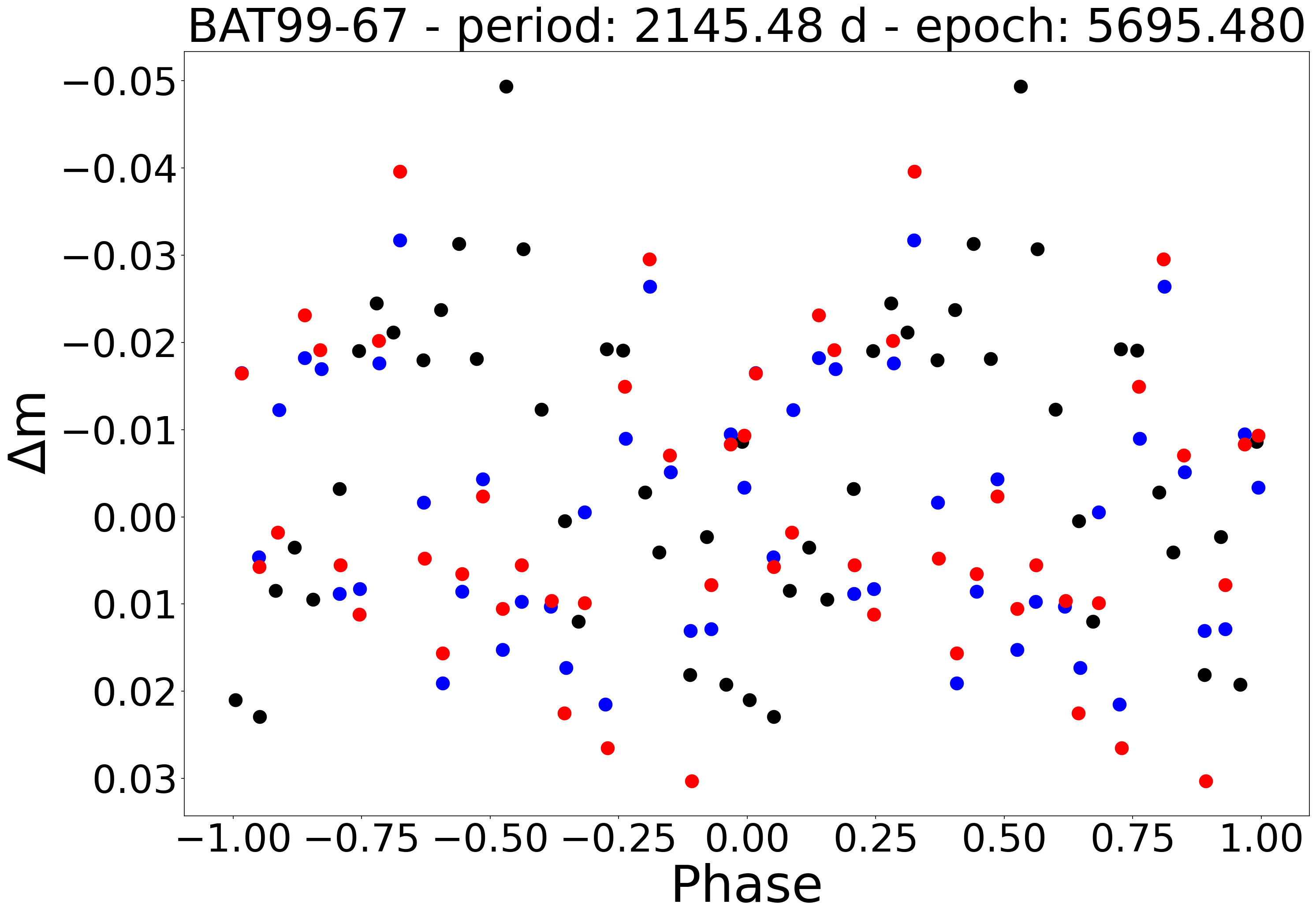}
    }\\
    \subfloat{
        \includegraphics[width=0.45\textwidth]{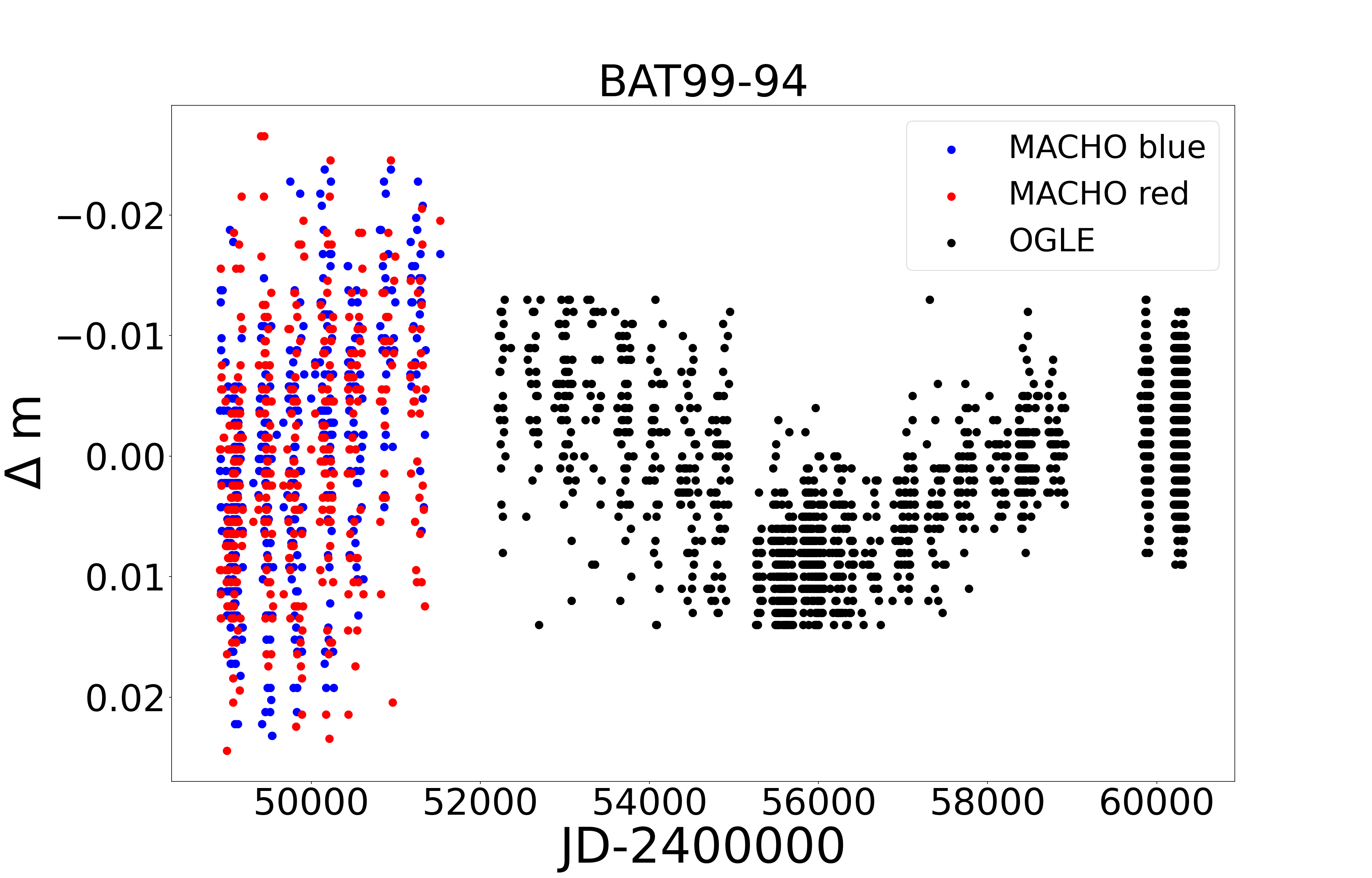}
    }
    \subfloat{
        \includegraphics[width=0.4\textwidth]{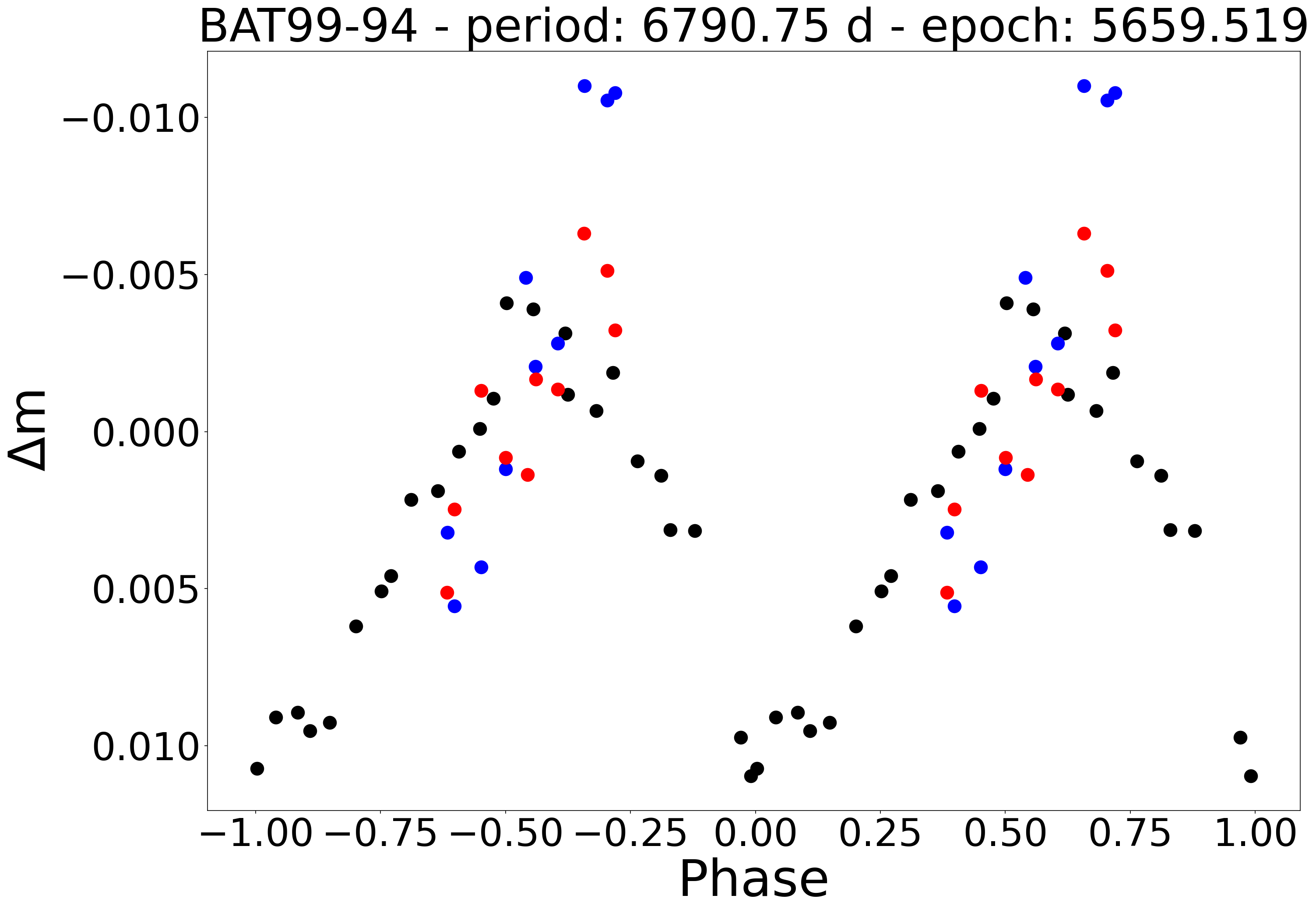}
    }
    \caption{Left column: OGLE (black circles) and MACHO (red and blue circles) light curves showing long-term variability trends.
    Right column: Binned light curves folded using the OGLE ephemerides from Table~\ref{tab:ephemeris}.
    The OGLE light curve is represented in the JD-2400000 instead of HJD-2450000.}
    \label{fig:MACHOlongterm}
\end{figure*}

Thirty-three of the forty-seven OGLE targets have also been observed by the MACHO survey. While MACHO provides multiyear coverage, it is sometimes insufficient to fully sample the extended periodicities identified in OGLE (i.e., BAT99-94 has a period of 6790 days in OGLE, while MACHO coverage spans only approximately 2500 days). 
MACHO detected long-term periodicities in four stars: BAT99-41, BAT99-47, BAT99-67, and BAT99-94 (Table~\ref{tab:surveys}).
To compare the OGLE and MACHO results, both datasets were folded together with the OGLE-detected ephemerides (Fig.~\ref{fig:MACHOlongterm}). The light curves of BAT99-41, BAT99-47, and BAT99-94 show consistent variability patterns and scatter. On the other hand, MACHO data of BAT99-67 does not fold well with the OGLE ephemeris, as can be seen in Fig.~\ref{fig:MACHOlongterm}. 
Moreover, the peak detected in the OGLE data is not found with the MACHO dataset.

\begin{figure*}[!h]
\centering

\subfloat{\includegraphics[width=0.35\linewidth]{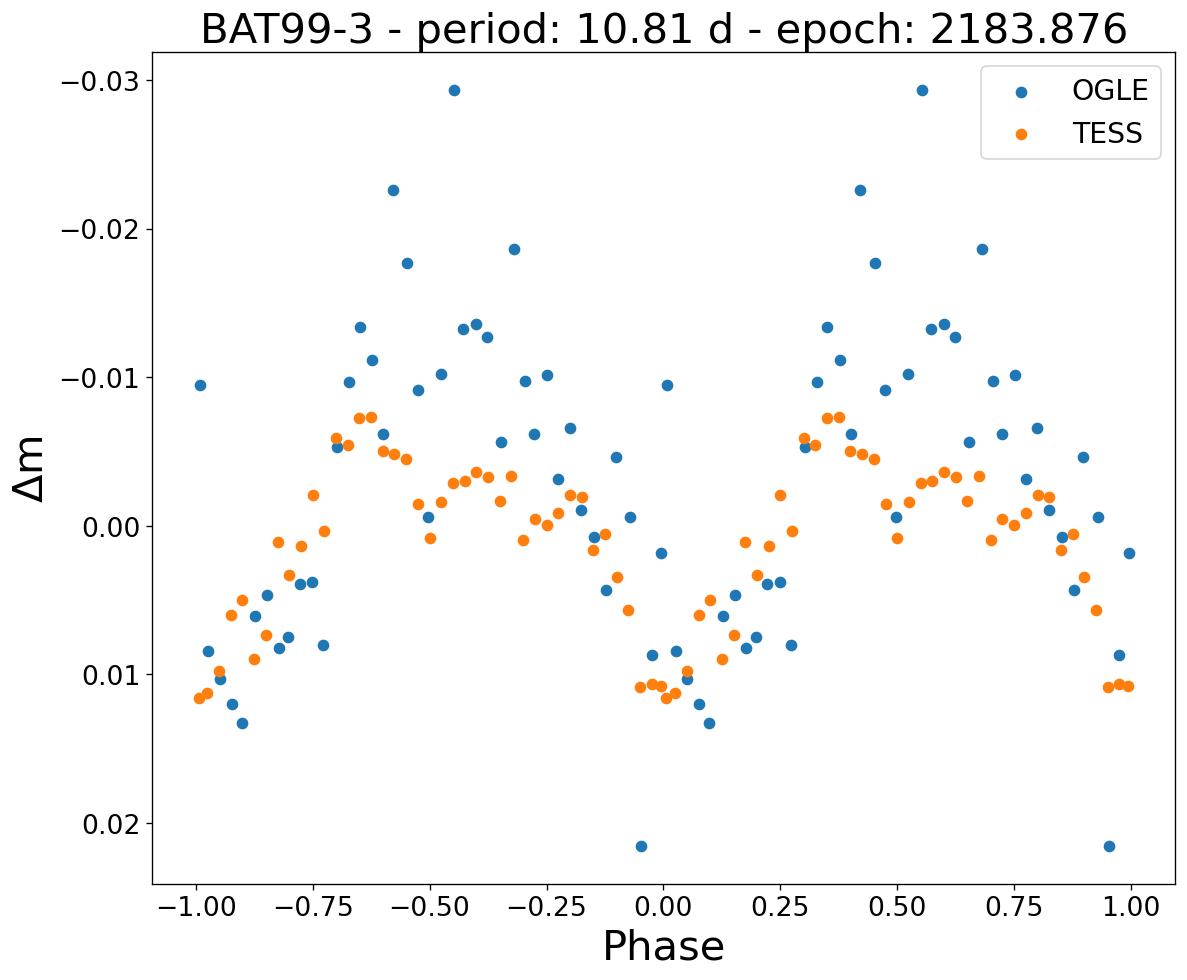}}%
\subfloat{\includegraphics[width=0.35\linewidth]{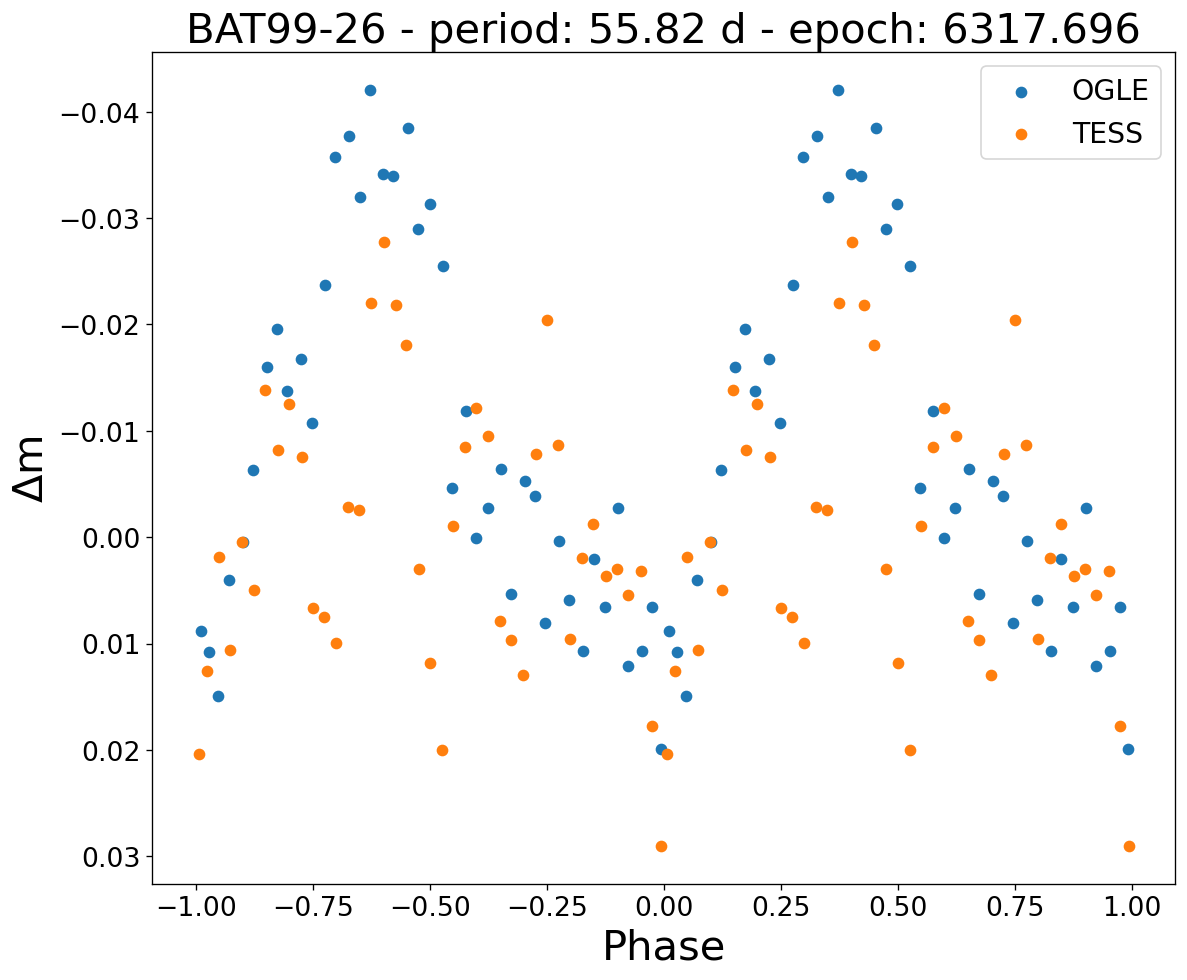}}%
\subfloat{\includegraphics[width=0.35\linewidth]{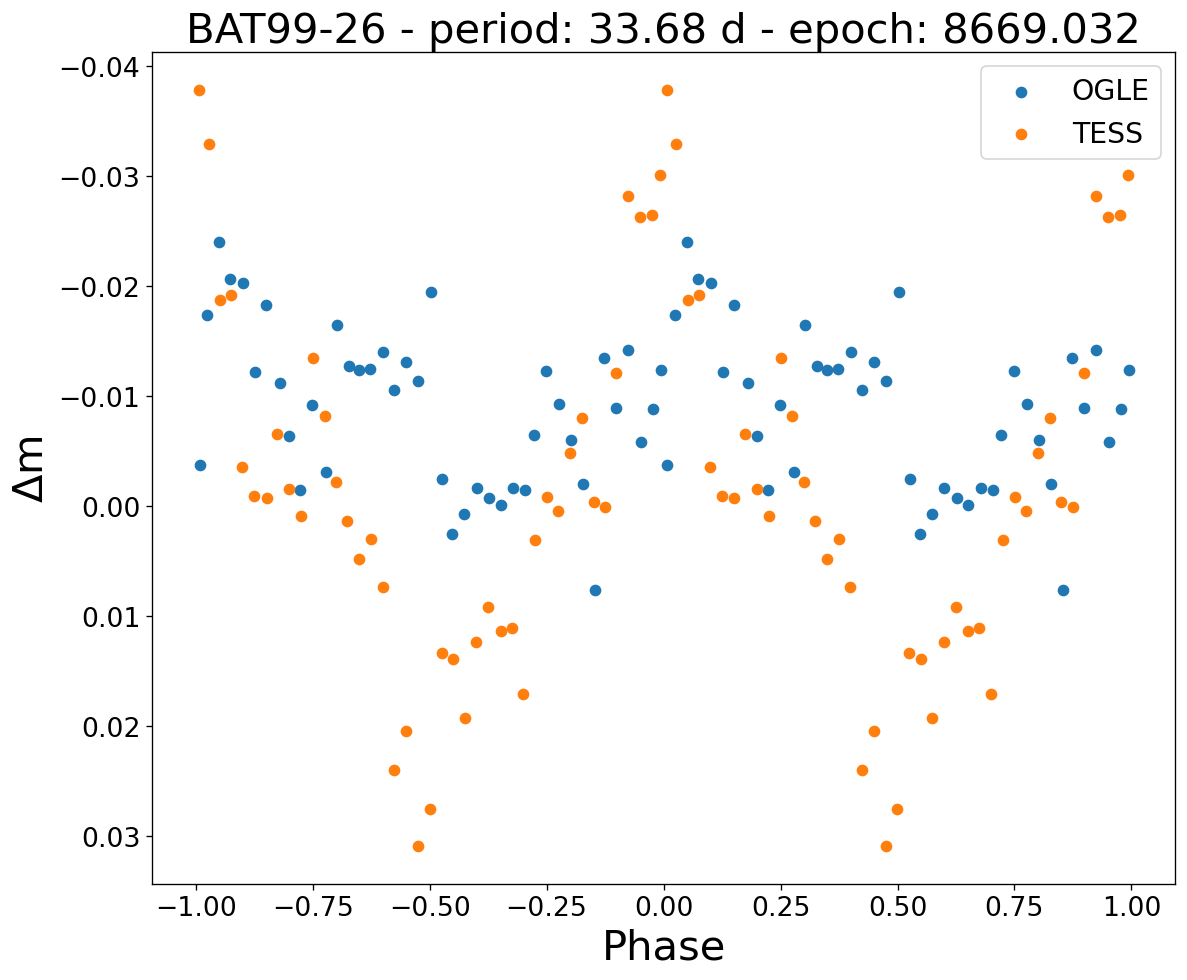}}\hspace{0.2mm}%
\subfloat{\includegraphics[width=0.35\linewidth]{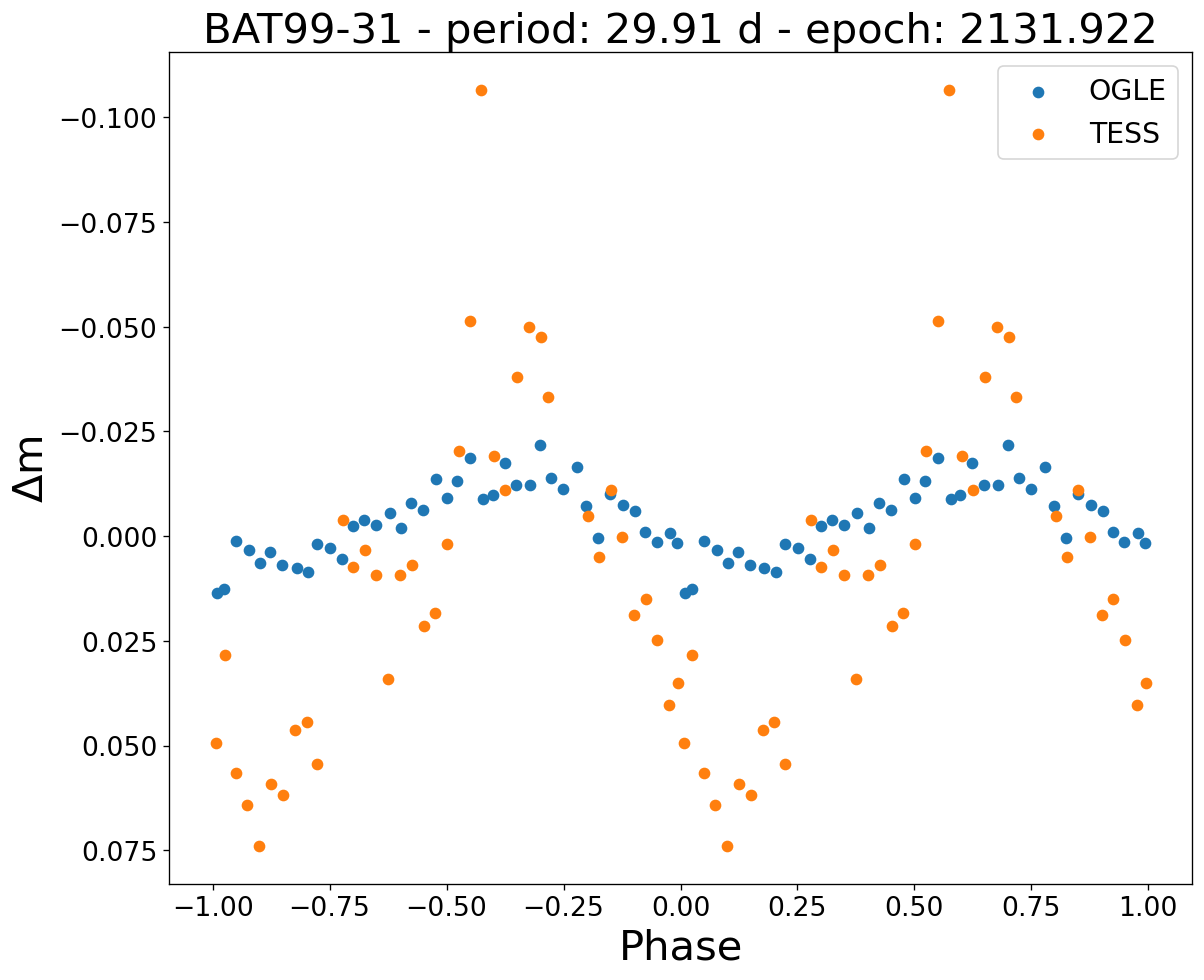}}%
\subfloat{\includegraphics[width=0.35\linewidth]{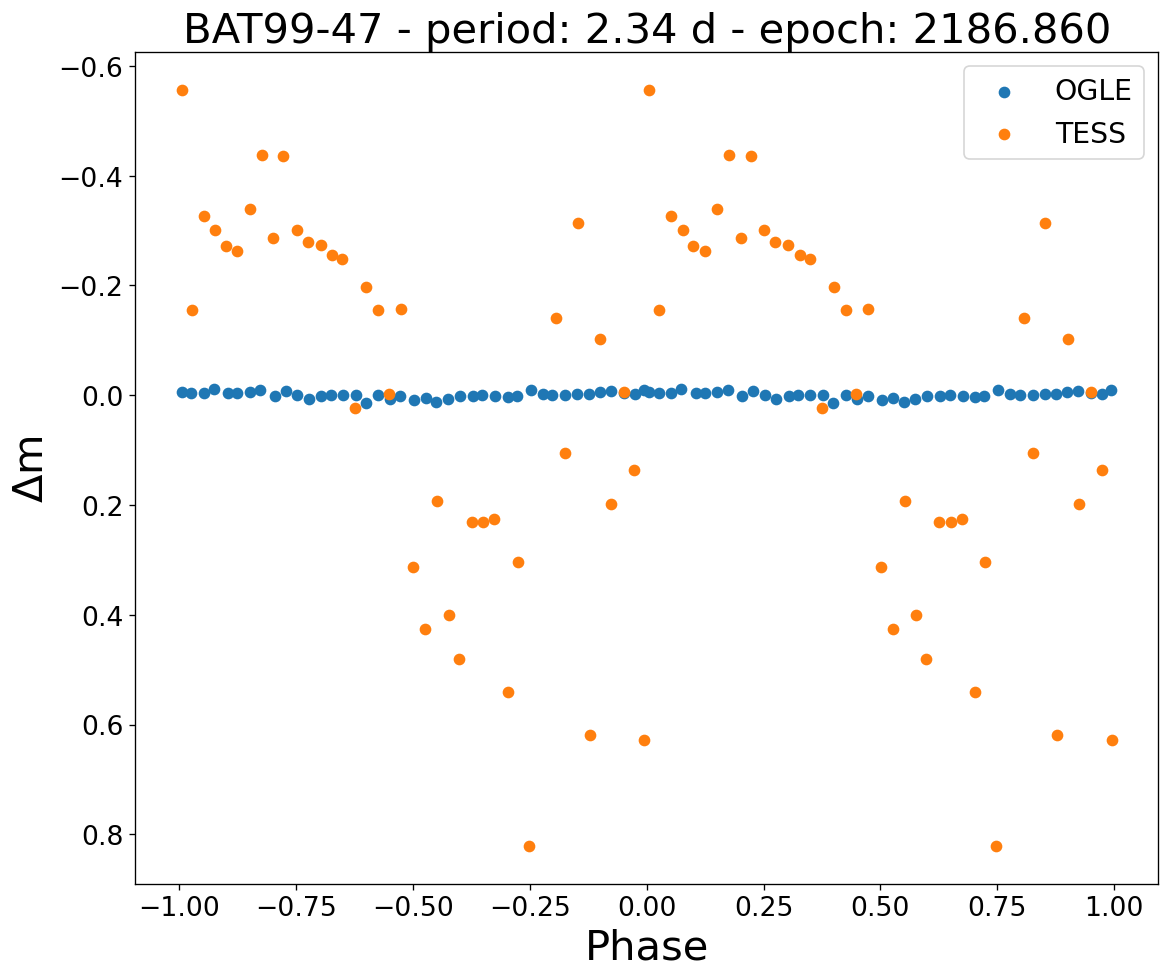}}
\subfloat{\includegraphics[width=0.35\linewidth]{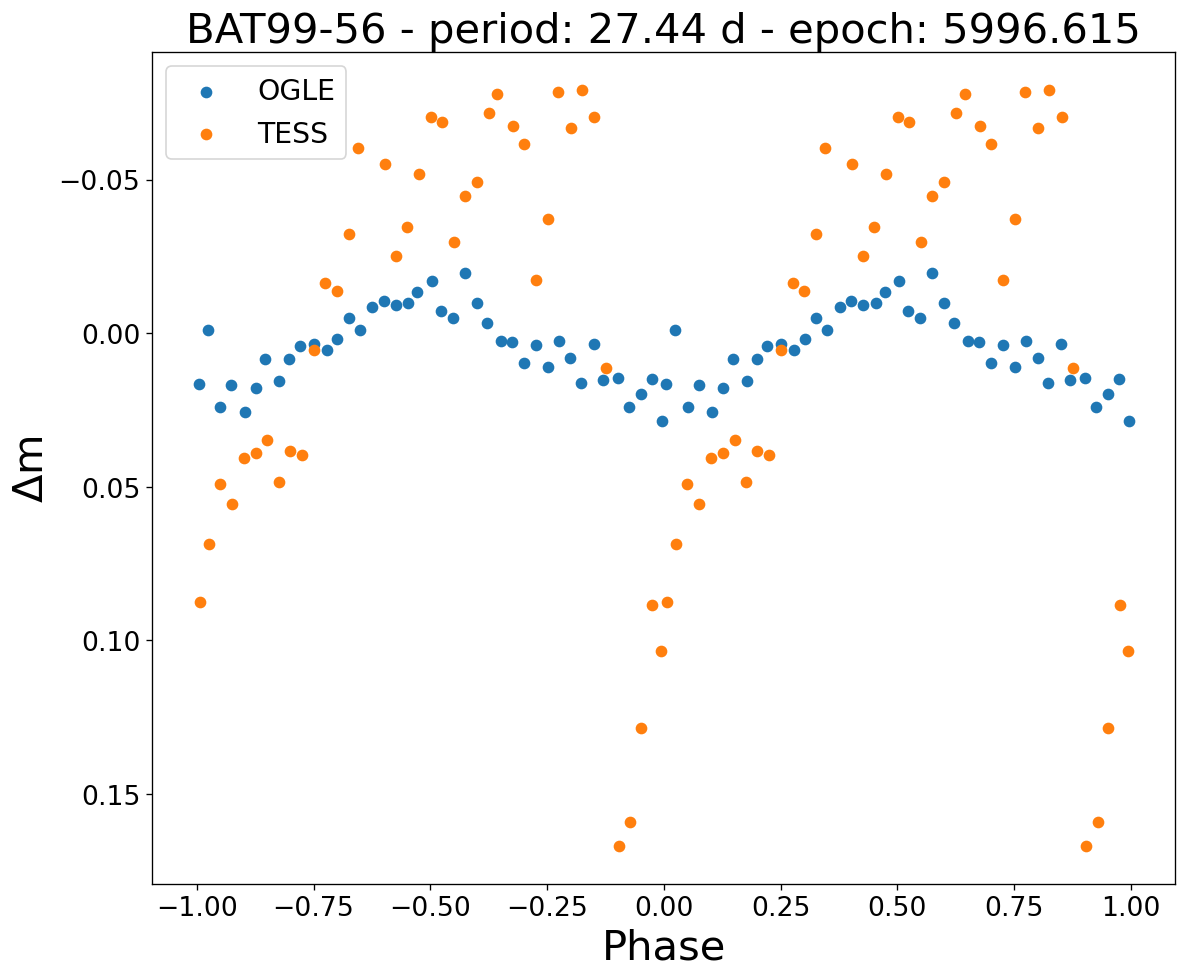}}\hspace{1mm}%
\subfloat{\includegraphics[width=0.35\linewidth]{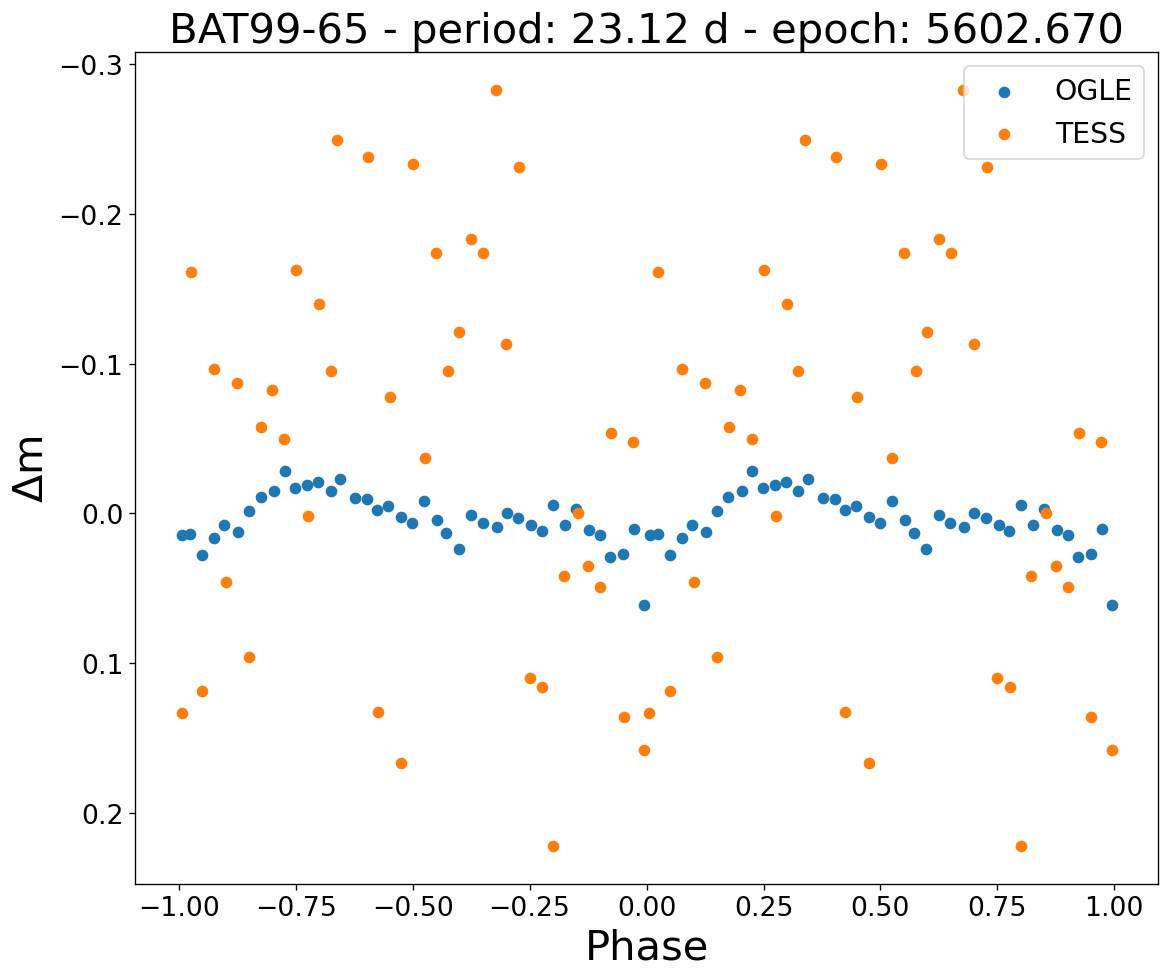}}
\subfloat{\includegraphics[width=0.35\linewidth]{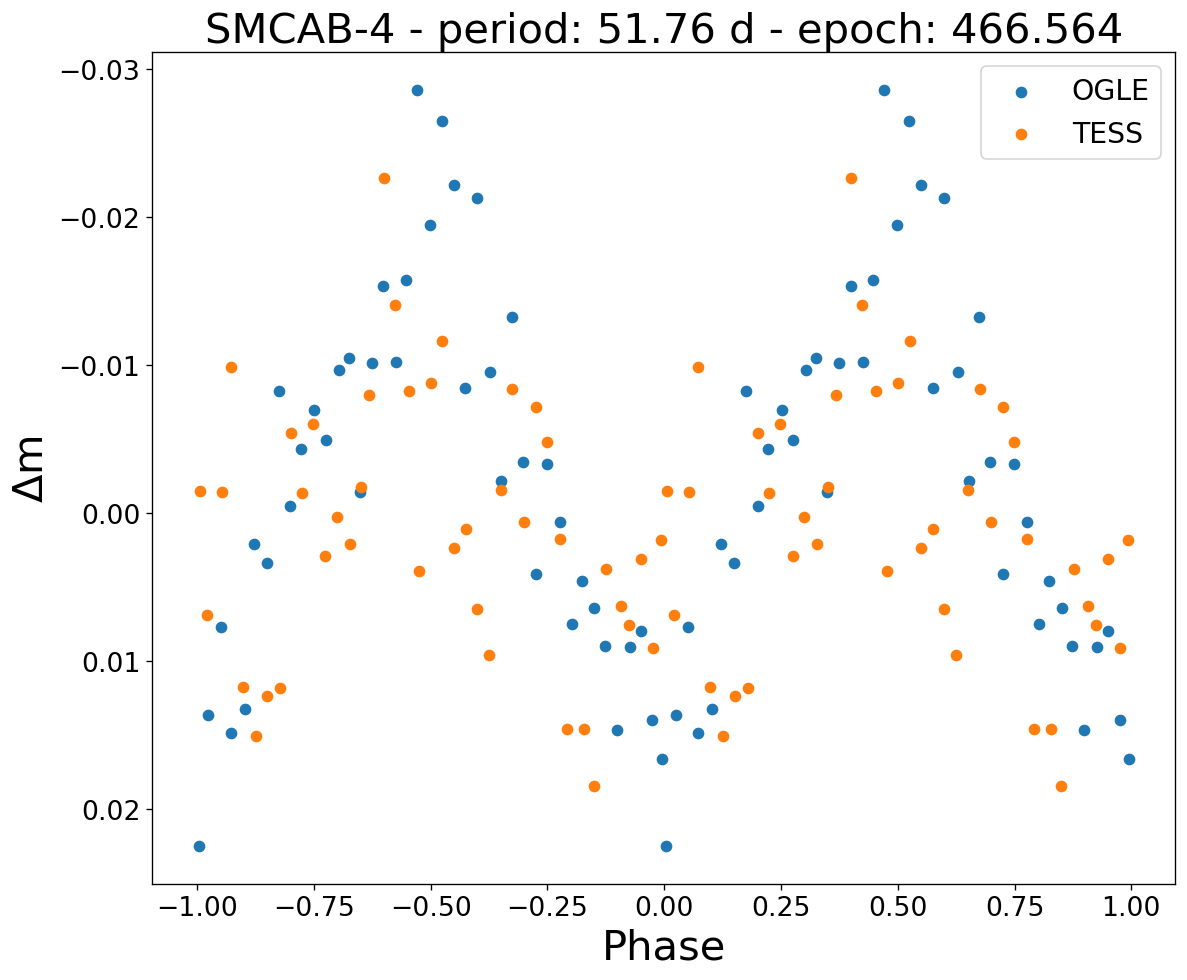}}
\caption{Binned OGLE and TESS light curves, folded using ephemeris from OGLE (as well as the ones from TESS for BAT99-26).
The number of TESS sectors used for BAT99-3 is 22; for BAT99-26 is 13;
1 sector for BAT99-31, BAT99-47, BAT99-56, and BAT99-65; and for SMC-AB4 the number of used sectors is 5.}
\label{fig:TESSbinned}
\end{figure*}

Periodogram analysis of the detrended MACHO curves of these 33 stars reveals significant short-term periodicities in 8 cases: BAT99-5, BAT99-24, BAT99-41, BAT99-48, BAT99-51, BAT99-65, BAT99-124, and SMC-AB4 (Table~\ref{tab:surveys}). 
Among these, seven sources also exhibit signals with \text{S/N} $\geq 4$ in OGLE. Five of them -- BAT99-24, BAT99-48, BAT99-51, BAT99-65, and BAT99-124 -- show similar period values in their periodogram and variability patterns in both OGLE and MACHO. Regarding BAT99-24, MACHO and OGLE light curves fold well to the $\sim$6-day period, if the epoch $T_{0}$ (Table~\ref{tab:ephemeris}) is slightly modified (+3.0913 days) for MACHO data. For BAT99-48, BAT99-51, and BAT99-124, some difference in amplitudes is most likely due to imperfect crowding correction and the use of different instrumental bands.
As is shown in the time-frequency diagrams of OGLE data for BAT99-48 and BAT99-51 (Appendix~\ref{app:TFD}), their periods exhibit slight temporal variations.
The OGLE periodogram of BAT99-65 displays a strong peak at (23.118 $\pm$ 0.007) days, along with its harmonic at half this period value. However, the light curve from MACHO does not fold well using the OGLE period, and vice versa. According to the time-frequency diagram, both signals might be evolving over time (Fig.\ref{fig:BAT9965-TF}).
On the other hand, BAT99-5 and SMC-AB4 exhibit different variability behavior between the two surveys. 
With OGLE we registered a 42-day period in BAT99-5, while the period detected with MACHO is $\sim$88 days.
In regard to SMC-AB4, OGLE identified a 51-day period and MACHO detected a 6-day period. The MACHO data were collected during the same epoch as OGLE-II, when the cadence was lower compared to the other two OGLE campaigns. Therefore, it is possible that the 6-day period was not detected due to the insufficient sampling.

Furthermore, one star (BAT99-41) exhibits a short period with a $\text{S/N}\geq$ 4 in MACHO but OGLE fails to detect significant peaks at these values, and the OGLE light curves do not fold coherently with the MACHO-derived periods. A closer look at the time-frequency diagrams reveal that the signal detected by MACHO has not been observed in OGLE data in any epoch. Since MACHO and OGLE data were not taken at the same epoch, one possible explanation is that the variability detected by MACHO may have evolved over time. Indeed, this variability is not consistently visible throughout the entire MACHO light curve either.

Due to the limited number of data points for BAT99-1 and BAT99-7, a reliable analysis could not be performed. Finally, the remaining 20 stars observed by MACHO did not show significant peaks in their periodograms.  

\subsection{OGLE and TESS} \label{subsec:OGLE&TESS}

We focused on TESS data to reinforce our findings. We aimed to carry out a comprehensive analysis; however, faint, high-contamination targets and weak signals strongly limited the applicability of TESS to seven stars that have OGLE-detected periods shorter than 50–100 days and no nearby object brighter than the target (Sect.~\ref{sec:Photometric_surveys}). As a first step, the TESS light curves of one sector were folded using the OGLE ephemeris from Table ~\ref{tab:ephemeris}. Next, they were binned and overlaid onto the folded OGLE light curves to assess whether they exhibited the same variability pattern and comparable amplitudes (Table~\ref{tab:surveys}). For BAT99-3, BAT99-31, BAT99-47, BAT99-56, and SMC-AB4, the TESS and OGLE data show agreement in the variability pattern. However, the TESS amplitudes appear higher for all stars except BAT99-3 and SMC-AB4, for which they are comparable (Fig.~\ref{fig:TESSbinned}). The discrepancy between amplitudes is most likely due to the imperfect crowding correction, although different instrumental bands may also play a role. Indeed, the TGLC correction assumes neighbors to remain stable, which might not be the case, leading to potential biases.
The OGLE periodogram for the star BAT99-47 reveals two peaks at approximately 1.8 and 2.3 days but the TESS one only detects the latter. Finally, the TESS data helped refine the ephemeris of the five stars mentioned earlier (reported in Table~\ref{tab:ephemeris}).
The OGLE period of BAT99-26 exceeds the duration of a single TESS sector. Therefore, more sectors were needed to assess its variability. Using 13 sectors, the variability in BAT99-26 appears to differ from that observed in the OGLE dataset, with a different significant period. The time-frequency diagram from OGLE shows that the signal is not constant over time. Additionally, no signal is detected in OGLE data between approximately 
 HJD 2\,457\,500  and 2\,459\,000, a “quiet” period during which TESS data were taken. The corresponding light curves fold with their respective periods. However, folding the entire OGLE dataset using the TESS period yields a light curve with reasonably good phase coherence, thereby supporting the identification in the OGLE data of the variability detected by TESS.
The OGLE variability of BAT99-65 is not confirmed with TESS data, as the light curves folded to OGLE ephemeris do not show the same variability pattern.

\begin{table}[h!]
\centering
{
\caption{Survey observations for each source.}
\label{tab:surveys}
\resizebox{0.5\textwidth}{!}{
\begin{tabular}{llll}
\hline
Name & OGLE phases & MACHO & TESS sectors \\
\hline
BAT99-1  & OIII \& OIV & * &  \\
BAT99-2  & OIII \& OIV &  &  \\
BAT99-3  & OIII \& OIV &  & 1, 27-38, 61-68, 87 ($=P_s$)\\
BAT99-5  & OIII \& OIV &  $\neq P_s$ &  \\
BAT99-7  & OIII \& OIV &  *  &  \\
BAT99-18 & OIII \& OIV &  no $P$ &  \\
BAT99-23 & OIII \& OIV &  &  \\
BAT99-24 & OII, OIII \& OIV & $=P_s$ &  \\
BAT99-25 & OIV  & no $P$ &  \\
BAT99-26 & OII, OIII \& OIV   & no $P$ & 1-13 ($\neq P_s$)\\
BAT99-30 & OIV  &  &  \\
BAT99-31 & OIII \& OIV  &  & 1 ($=P_s$) \\
BAT99-35 & OIV  & no $P$ &  \\
BAT99-37 & OIV  & no $P$ &  \\
BAT99-40 & OIII \& OIV  & no $P$ &  \\
BAT99-41 & OIII \& OIV  & new $P_s$, $=P_l$ &  \\
BAT99-44 & OIII \& OIV  &  &  \\
BAT99-46 & OIII \& OIV  &  &  \\
BAT99-47 & OIII \& OIV  & $=P_l$ & 1, 2 ($=P_s$)\\
BAT99-48 & OIII  & $=P_s$ &  \\
BAT99-50 & OIII \& OIV  & no $P$ &  \\
BAT99-51 & OIII \& OIV  & $=P_s$ &  \\
BAT99-54 & OIII  &  &  \\
BAT99-56 & OIV  &  & 1 ($=P_s$) \\
BAT99-57 & OIV  & no $P$ &  \\
BAT99-58 & OIII \& OIV  &  &  \\
BAT99-60 & OIII \& OIV  & no $P$ &  \\
BAT99 62 & OIV  &  &  \\
BAT99-63 & OIV  & no $P$ &  \\
BAT99-65 & OIII \& OIV  & $=P_s$ & 1 (no $P$) \\
BAT99-66 & OIV  & no $P$ &  \\
BAT99-67 & OIII \& OIV  &  $\neq P_l$ &  \\
BAT99-73 & OIII \& OIV  & no $P$ &  \\
BAT99-74 & OIII \& OIV  & no $P$ &  \\
BAT99-75 & OIV  & no $P$ &  \\
BAT99-81 & OIV  & no $P$ &  \\
BAT99-82 & OIII \& OIV  & no $P$ &  \\
BAT99-89 & OIII \& OIV  &  &  \\
BAT99-94 & OIII \& OIV  & $=P_l$ &  \\
BAT99-124 & OIII \& OIV  & $=P_s$ &  \\
BAT99-128 & OIII \& OIV  &  &  \\
BAT99-131 & OIV  & no $P$ &  \\
SMC-AB1  & OII, OIII \& OIV  & no $P$ &  \\
SMC-AB2  & OII, OIII \& OIV  & no $P$ &  \\
SMC-AB4  & OII \& OIII  & $\neq P_s$ & 1, 27, 28, 67, 68 ($=P_s$)\\
SMC-AB9  & OII, OIII \& OIV  &  &  \\
SMC-AB10 & OIII \& OIV  & no $P$ &  \\
\hline
\end{tabular}}
\tablefoot{Summary of the OGLE observing campaigns for each star. The MACHO column denotes whether the star was observed by MACHO and, if so, whether the same period ($=P$), a different period ($\neq P$), a newly identified period (new $P$), no significant period (no $P$), or insufficient data (*) were recorded. The TESS column lists the sectors employed in the period search using the TESS dataset, with symbols defined as for MACHO. Subscripts “s” and “l” indicate short and long variability timescales, respectively.}
}
\end{table}

\section{Discussion}

A considerable fraction of WR stars exhibit high photometric variability: 36\% overall, and  34\% on short timescales. When both high and moderate variability are considered, these fractions increase to 81\% and 77\%, respectively. There is no similar analysis of Galactic WN stars to compare our results with. 

\begin{figure*}[ht!]
    \centering

    \begin{subfigure}[t]{0.32\textwidth}
        \centering
        \includegraphics[width=\textwidth]{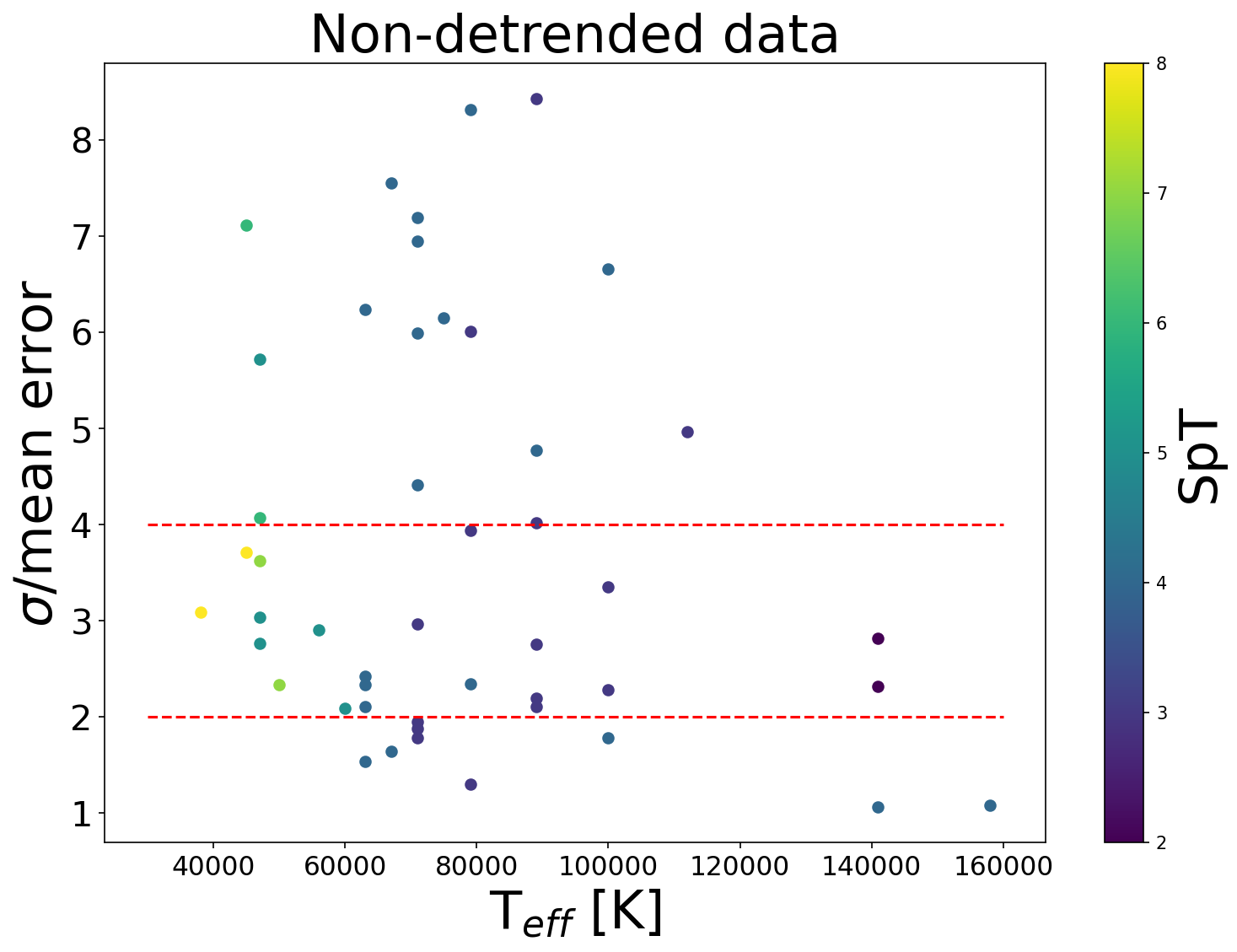}
    \end{subfigure}
    \hfill
    \begin{subfigure}[t]{0.32\textwidth}
        \centering
        \includegraphics[width=\textwidth]{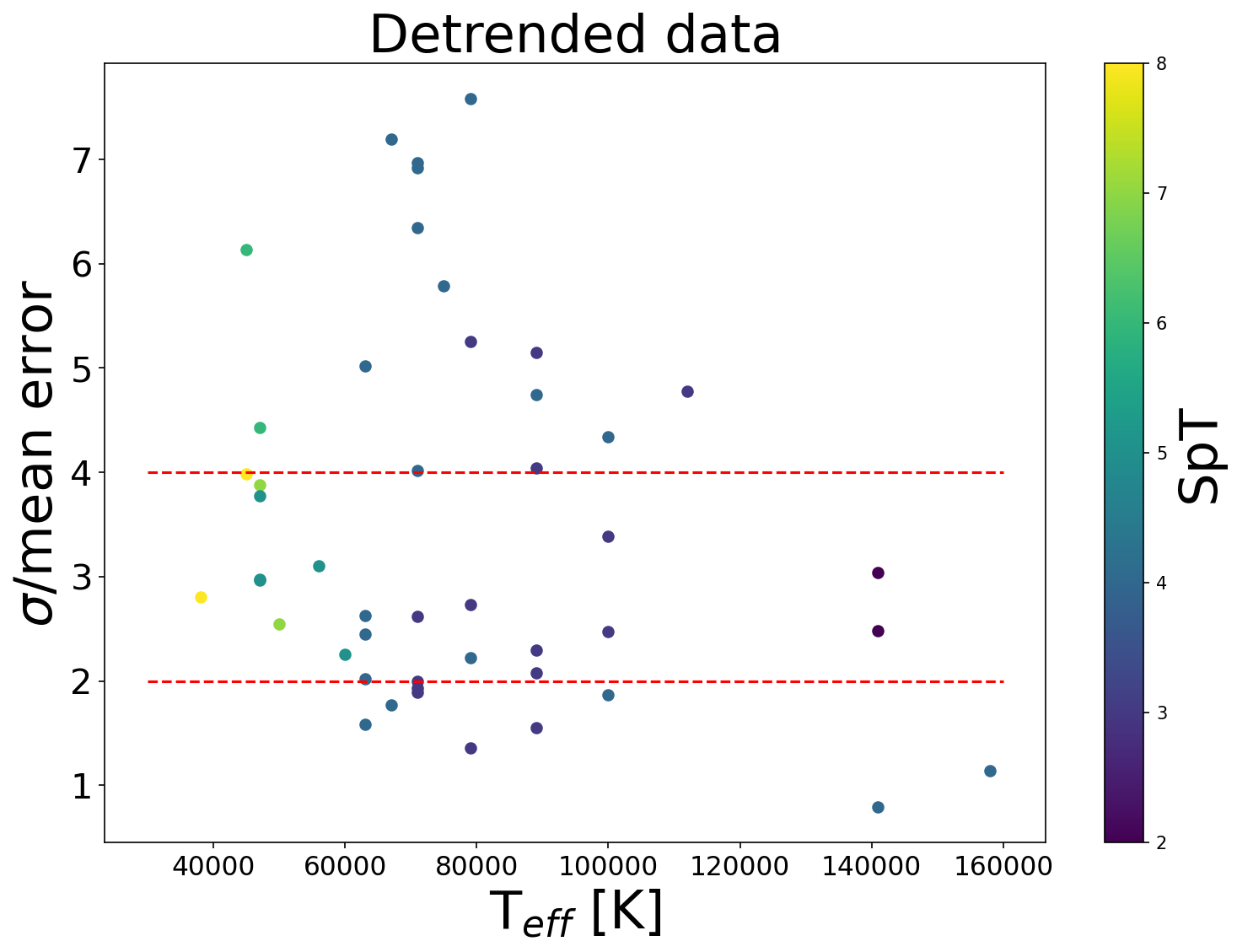}
    \end{subfigure}
    \hfill
    \begin{subfigure}[t]{0.30\textwidth}
        \centering
        \includegraphics[width=\textwidth]{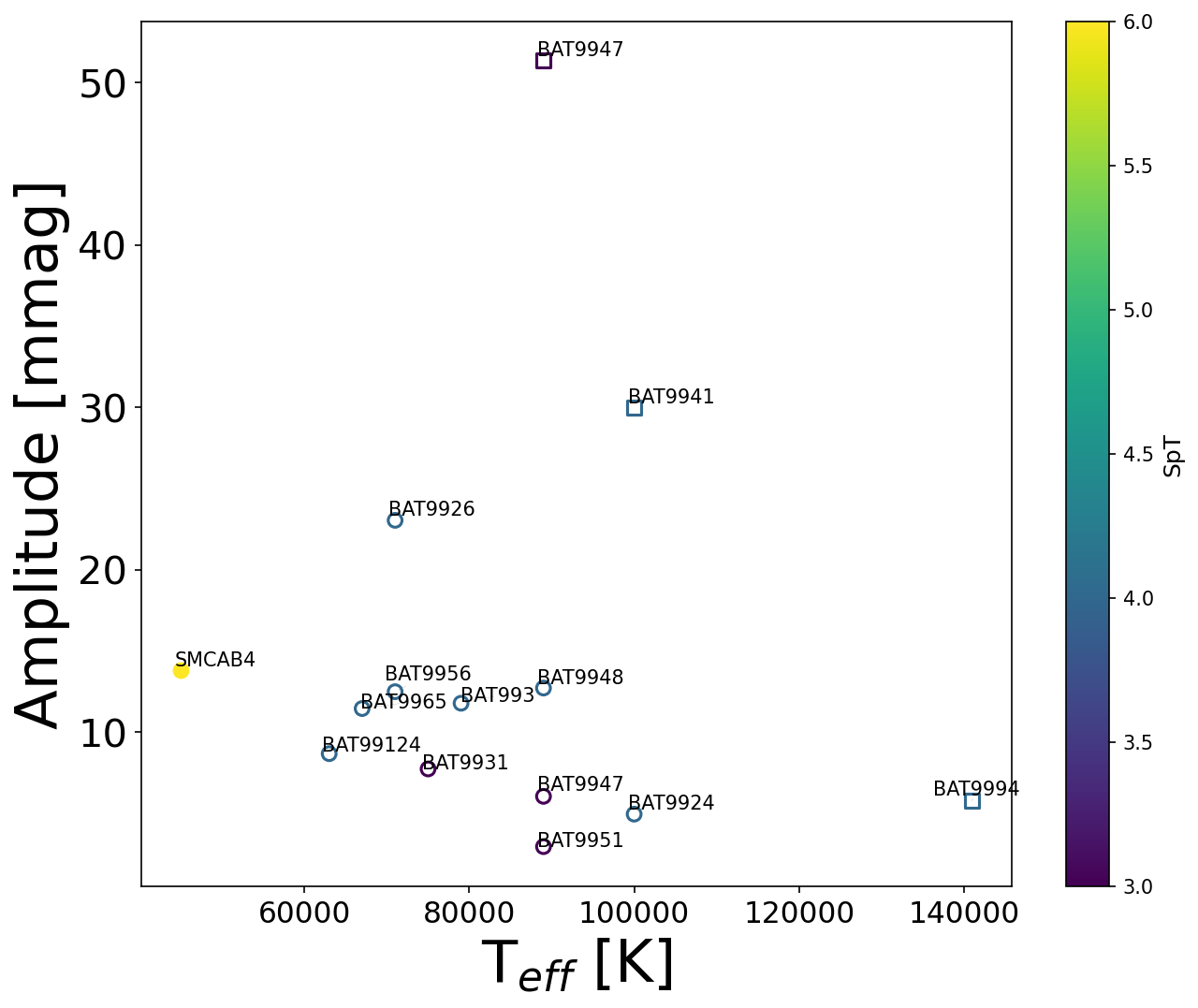}
    \end{subfigure}

    \caption{Variability level as a function of effective temperature and spectral type for the data before (left) and after (middle panel) detrending. 
    The limits between low, moderate, and high variability levels (Sect. \ref{subsec:variabilityCriteria}) are indicated with dashed red lines.
    The right panel displays the amplitude of variability for 13 WN stars (11 showing short-term variability and 3 long-term) from our sample with significant signals as a function of effective temperature and spectral type. Stars with short periods are marked as circles, and ones with long periods as squares. 
    Empty symbols are used to designate stars with no hydrogen in their spectra; otherwise, they are filled. BAT99-47 exhibits variability on both short and long timescales.}
    \label{fig:Amp_Teff}
\end{figure*}

\citealt{Chene2022} claim that the photometric amplitudes of 50 Galactic WRs show a clear decline with increasing surface temperature (T$_{eff}$), and a similar trend is seen in terminal velocity. 
In this work, we compare the photometric variability properties to effective temperatures from \citealt{Hainich_2014} and \citealt{Hainich2015}.
The overall level of variability in the sample does not appear to depend on temperature or spectral type (see left and middle panels in Fig.~\ref{fig:Amp_Teff}), but a slight correlation ($\sim 49\%$) is observed between the amplitude of short-term periodic variability and $T_{eff}$ (right panel in Fig.~\ref{fig:Amp_Teff}). Despite the very limited statistics, this correlation appears more pronounced ($\sim 96\%$) for the stars exhibiting long-term periodicities (represented as squares in the figure).
Interestingly, all of the LMC stars of our sample are hydrogen-depleted, whereas the only two SMC stars show hydrogen in their spectra.

Of the nine stars exhibiting outbursts, six (BAT99-1, BAT99-26, BAT99-31, BAT99-48, BAT99-124, and SMC-AB4) show high variability on long timescales, due to recurrent but non-strictly periodic outburst-like structures observed in their light curves (as can be seen in Appendix \ref{app:outbursts}). 
Seven stars (BAT99-3, BAT99-31, BAT99-37, BAT99-65, BAT99-124, BAT99-128, and BAT99-131) show slow long-term ($>$1 yr) brightness changes, but no significant periods were identified in their GLS periodograms. On the other hand, three stars $-$ BAT99-41, BAT99-47, and BAT99-67 $-$ exhibit a high level of (overall) variability associated with significant peaks in their periodograms. Finally, even though the overall variability of BAT99-94 is of a low level, there is an evident trend in its light curve and its GLS analysis shows a highly significant periodicity. 
To date, there is a lack of long-period binaries (most WN binary systems are short-period, as is suggested by \citealt{Dsilva2023}). However, given the consistency between the OGLE and MACHO data and the good appearance of the folded curves, we suggest that the three stars BAT99-41, BAT99-47, and BAT99-94 are candidates for being long-period binaries (as is stated in Table~\ref{table:nature}), since other physical mechanisms would produce shorter or time-evolving periodic signals. Further spectroscopic studies could confirm the binary origin of these long-term changes.
Focusing on the 11 stars with short-term periodicities detected by OGLE and confirmed by another survey (OGLE + MACHO, OGLE + TESS) \footnote{OGLE+MACHO: BAT99-24, BAT99-26, BAT99-48, BAT99-51, BAT99-65, and BAT99-124.
OGLE+TESS: BAT99-3, BAT99-31, BAT99-47, BAT99-56, and SMC-AB4.}, we classify their variability patterns with the aim of identifying the likely physical origins of their changes (see Table \ref{table:nature}).

\begin{table*}
\centering
\caption{Proposed nature of the stars with confirmed variability.}

\begin{tabular}{ll}
\hline
Proposed nature & Name \\
\hline
Winds & SMC-AB4 \\
Clumps & BAT99-26 \\
Pulsation & BAT99-47 \\
CIRs & BAT99-3, BAT99-24, BAT99-48, BAT99-56, BAT99-65, BAT99-124 \\
Binary candidate & BAT99-31, BAT99-51, SMC-AB4 \\
Long-period binary candidate & BAT99-41, BAT99-47, BAT99-94\\
\hline
\end{tabular}

\label{table:nature}

\end{table*}
Corotating interaction regions appear to be a likely explanation for stars that exhibit a period and its harmonic as found by \citealt{St-Louis2020}, \citealt{Toala2022}, and \mbox{\citealt{David-Uraz2012}}. In our sample, this applies to six stars (BAT99-3, BAT99-24, BAT99-48, BAT99-56, BAT99-65, and BAT99-124). 
Interestingly, the stars potentially exhibiting CIRs belong to the same spectral type: WN4 (Table ~\ref{tab:tableA}). All of these stars exhibit periods of similar values ($\sim 20$\, days), except BAT99-3 (albeit it also has a sub-harmonic at $~$20 days) and BAT99-24.
The discrepancy in the MACHO and OGLE folded light curves of BAT99-48, BAT99-65, and BAT99-124 likely arises from the time-evolving nature of CIRs, which can lead to changes in the amplitude or shape of the variability over time \citep{Toala2022}, making the signals less coherent when folded over extended time baselines. 

Two independent short-period peaks are detected in BAT99-47, which might be indicative of pulsations. A similar behavior has been observed in the Galactic star WR 134 \citep{Naze2021b}, with both stars showing comparable amplitudes of variation and detected frequencies. This resemblance is particularly noteworthy given that WR pulsations are not yet fully understood theoretically. Identifying additional candidates with similar behavior could help shed light on this phenomenon.

In contrast, the presence of several periods may indicate a different phenomenon. The variability of BAT99-26 could be explained by clumps in its stellar wind. The irregular occurrence of these clumps produces multiple variations in the light curves, leading to a ``forest of peaks" in the frequency spectrum. Its signals are not time-coherent, resulting in a time-frequency diagram similar to that observed in WR 40 \citep{Ramiaramanantsoa2019}, where the variability is associated with clumps. 
SMC-AB4 also likely exhibits multiple variability timescales across the different surveys. The 6.5-day period, as suggested by \citealt{Foellmi_2003a} and confirmed here by MACHO data, could correspond to the star's rotation and possibly be linked to a large-scale structure in its wind. The 52-day period detected by \citealt{Schootemeijer2024} and also confirmed here by OGLE and TESS might be associated with binarity (supported by excesses in the Gaia astrometric parameters). \citealt{Pauli_2023} suggested that this binary system could potentially involve a compact object as a companion. The lack of eclipses in the analyzed photometric data does not contradict this hypothesis. BAT99-51 exhibits coherent variability across the OGLE campaigns, with the same period detected consistently in both OGLE and MACHO, and both datasets folding well with this period. Therefore, we suggest that this target is a good binary candidate and a spectroscopic analysis is now necessary to confirm it.

Although the variability of SMC-AB9 has not been confirmed by other surveys, its OGLE light curve shows signs of periodic variability, while its Gaia astrometric parameters suggest a possible binary nature. Along with BAT99-31, SMC-AB9 displays V-shaped dips in its light curve, has periods shorter than 30 days, and has previously been identified as a binary suspect. Therefore, for these two objects, atmospheric eclipses are considered a possible origin of their variability. To this end, we performed a fit of the Lamontagne model \citep{Lamontagne1996} considering, for each star, its specific terminal velocity, mass, and radius, as reported by \citealt{Hainich_2014} and \citealt{Hainich2015}. We employed a grid of intensity ratios (0.1, 0.2, 0.5, 1, and 2) and companion masses (0.5, 2, 5, 10, 20, and 30 M$_{\odot}$), treating the mass-loss rate as a free parameter. Its fit values were then compared to the theoretical predictions provided in the Hainich studies. The best-fit results yielded low inclination angles and excessively large mass-loss rates. When the latter was fixed, the best fit led to intensity ratios that converged toward negative values, which is not physical. Therefore, 
we find it unlikely that the atmospheric eclipse explanation accounts for the observed variability. 
\citealt{Foellmi_2003b} claimed that BAT99-31 may be a binary with a short orbital period and low inclination, but the mass-loss rate required for an atmospheric eclipse scenario with such characteristics would need to be an order of magnitude higher than the value reported by \citealt{Hainich_2014}. 
This supports our hypothesis that an atmospheric eclipse is not the cause of its variability.
SMC-AB9 also fails to fit the Lamontagne model, ruling out an atmospheric eclipse as the origin of its variability.
It is worth mentioning that we classify two objects (BAT99-31 and SMC-AB4) as outbursting and at the same time as binary candidates. Those features are probably independent. Indeed, the mass transfer in a binary system can lead to eruptive outbursts \citep {Smith2011}, but these outbursts are much less frequent (on the order of 100 years) than the outbursts found in our stars.
Notably, two of our binary candidates (SMC-AB4 and SMC-AB9) had already been flagged as binary candidates in Sect. \ref{sec:Gaia}. Furthermore, two others (BAT99-31, BAT99-47) had previously been suggested as binary systems by other authors (see Sect. \ref{sec:spec_surveys}).
The addition of new (candidate) binaries, both long- and short-period systems, is interesting as it helps clarify the multiplicity incidence among WR stars, which may or may not differ from that of their O-star progenitor depending on the evolutionary paths, such as mergers.

In conclusion, our OGLE study shows that variability is ubiquitous amongst WR stars. The changes can be periodic, outburst-like,
or more stochastic. In many cases, the variability itself can evolve with time. To further clarify the nature of the changes, the next step clearly is a dedicated spectroscopic campaign focusing on the identified timescales. Moreover, continuing the photometric monitoring is required to further constrain the temporal evolution of the variability. For example, such data would reveal whether the low-variability targets are truly exhibiting small changes,
or whether they were simply observed during a transient low-variability stage, thereby better constraining the incidence of low versus high variability among WR stars. Finally, it would also be worth extending the same analysis to other sources, such as carbon-rich Wolf–Rayet stars. Some of the causes of variability, such as pulsations,
indeed directly depend on physical parameters (e.g., metallicity, or evolution stage), while others do not. A comparison with other sources with a similarly long-timescale campaign would therefore be extremely useful.

\section{Conclusions}

Long-term OGLE monitoring reveals that photometric variability is prevalent among nitrogen-rich Wolf–Rayet stars in the MCs, with around 80$\%$ displaying moderate to high levels of photometric variability. This work identifies periodic, stochastic, and outburst-like forms of variability behaviors, and shows that these patterns can evolve over time, highlighting the intrinsically dynamic nature of these objects. Periodic variability is found in 11 stars -- linked to CIRs, pulsations, or binary effects -- thereby increasing the number of WN stars exhibiting such variability mechanisms. Additionally, nine stars exhibit quasi-periodic outbursts, a form of variability not previously recognized in these stars. Cross-checks with MACHO and TESS were used to support these detections and enable the identification of new binary candidates, including three likely long-period systems. Overall, these results underscore the complexity and diversity of variability in WR stars and emphasize the need for dedicated spectroscopic campaigns and continued photometric monitoring to better understand the underlying physical processes.

\section*{Data availability}

The figures of non-detrended light curves from our sample, panels containing detrended light curves showing significant signals along with their GLS periodograms, phase-folded and binned phase-folded light curves, as well as comparison plots showing the binned phase-folded OGLE and MACHO light curves overlaid on one another are available on Zenodo (\href{https://zenodo.org/records/17878736}{https://zenodo.org/records/17878736}).

\begin{acknowledgements}
This work has been funded by the Community of Madrid under the framework of the multiyear agreement with the University of Alcalá (Stimulus to Excellence for Permanent University Professors, EPU-INV/2020/008 and EPU-DPTO-INV-UAH/2023/003).
This work has been co-funded by the National Science Centre, Poland, grant No. 2022/45/B/ST9/00243. The OGLE project has received funding from the Polish National Science Centre grant OPUS-28 2024/55/B/ST9/00447 to AU.
M.M.R-D also acknowledges financial support through Spanish grant PID2019-105552RB-C41(MCIU).
This paper utilizes public domain data obtained by the MACHO Project, jointly funded by the US Department of Energy through the University of California, Lawrence Livermore National Laboratory under contract No. W-7405-Eng-48, by the National Science Foundation through the Center for Particle Astrophysics of the University of California under cooperative agreement AST-8809616, and by the Mount Stromlo and Siding Spring Observatory, part of the Australian National University. 
\end{acknowledgements}

\bibliographystyle{aa}

\bibliography{BIBLIOGRAPHY}

\onecolumn
\begin{appendix}
\section{Physical characteristics of the sampled nitrogen Wolf-Rayet stars}
\label{tab:tableA}
\begin{table*}[h!]
\caption{Physical characteristics of the sampled nitrogen WR stars.}
\centering
\begin{tabular}{llllllll}
\hline
Name & Spec. type & T$_\mathrm{eff} (kK) $ & X-ray (erg/s) &  LV &  SV & OGLE variability & Bin? \\
\hline
BAT99-1  & WN3b      & 89 & & H (0.023)& H (0.024)  & F$_s$, O & 
\\
BAT99-2  & WN2b(h)   & 141 &   &M (0.017) & M (0.018) & F$_s$, O & \\
BAT99-3  & WN4b      & 79 & &H (0.042) & H (0.038) & P$_s$+subH, Ch &   \\
BAT99-5  & WN2b      & 141 & &M (0.020) & M (0.022) & F$_s$, O &   
\\
BAT99-7  & WN4b      & 158 & &L (0.005) & L (0.005) &  & \\
BAT99-18 & WN3(h)    & 71 & &L (0.009) &L (0.009) & & \\
BAT99-23 & WN3(h)    & 71 & &L (0.016) & L (0.016) &  & \\
BAT99-24 & WN4b      & 100 & &L (0.009) & L (0.009) & P$_s$+H & \\
BAT99-25 & WN4ha     & 67 & &L (0.009) & L (0.010) &  & \\
BAT99-26 & WN4b      & 71 & &H (0.035) & H (0.034) & F$_s$, O &  
\\
BAT99-30 & WN6h      & 47  & $<1.9 \times10^{33}$ & H (0.020) & H (0.022) & & \\
BAT99-31 & WN3       & 75  & diffuse &H (0.032) & H (0.030) & P$_s$, O, Ch & 1\\
BAT99-35 & WN3(h)    & 71  & &L (0.009) & L (0.010) & & \\
BAT99-37 & WN3o      & 79  & &H (0.033) & H (0.029) & Ch & \\
BAT99-40 & WN4       & 63  & $< 2 \times10^{30}$ &M (0.012) & M (0.013) & & Gaia \\
BAT99-41 & WN4b     & 100  & &H (0.033) & H (0.022) & P$_l$, Ch &  \\
BAT99-44 & WN8ha     & 45  & $<1.6 \times10^{33}$ &M (0.017) & M (0.019) && \\
BAT99-46 & WN4o      & 63  & &M (0.011) & L (0.010) & \\
BAT99-47 & WN3   & 89  & (4.2 -1.6)$\times10^{33}$ &H (0.047) & H (0.028) & P$_l$, P$_s$, Ch &  2 \\
BAT99-48 & WN4b      & 89 & & H (0.029) & H (0.028) & P$_s$+subH, O &  \\
BAT99-50 & WN4h      & 56 & &M (0.014) & M (0.016) & & \\
BAT99-51 & WN3b    &  89 & & M (0.012) & M (0.011) & P$_s$ &   \\
BAT99-54 & WN8ha    & 38 & $<1.5 \times 10^{33}$  &M (0.015) & M (0.014) & P$_l$& \\
BAT99-56 & WN4b      & 71 & &H (0.035) & H (0.035) & P$_s$+subH &  \\
BAT99-57 & WN4b      & 79  & &M (0.012) & M (0.011) & & \\
BAT99-58 & WN7h      & 47 & $<5.5\times10^{33}$ & M (0.018) & M (0.019) & & \\
BAT99-60 & WN4(h)a   & 63 & &L (0.008) & L (0.008) & & 3 \\
BAT99 62 & WN3(h)   & 71 & &M (0.015) & M (0.013) & P$_l$ & \\
BAT99-63 & WN4ha:    & 63 & &M (0.012) & M (0.013) & & \\
BAT99-65 & WN4o      & 67 & & H (0.043) & H (0.041) & P$_s$+subH, Ch &  \\
BAT99-66 & WN3(h)    & 89 & & M (0.011) & M (0.012) & & \\
BAT99-67 & WN5ha    & 47 & 1.7$\times10^{33}$ 
& H (0.030) & M (0.020) & P$_l$, P$_s$, Ch & 2 \\
BAT99-73 & WN4ha     & 60 & &M (0.010) &  M (0.011) & &  \\
BAT99-74 & WN3(h)a   & 79 & & L (0.008) & L (0.009) & & \\
BAT99-75 & WN4o      & 71 & &H (0.030) & H (0.032) & & \\
BAT99-81 & WN5h      & 47 &  &M (0.014) & M (0.015) & & \\
BAT99-82 & WN3b      & 100 & 1.6$\times10^{33}$
&M (0.021) & M (0.021) &  O & 2 \\
BAT99-89 & WN7h      & 50 & $<7.4 \times10^{33}$ &M (0.011) & M (0.012) &  & \\
BAT99-94 & WN4b     & 141  & & L (0.005) & L (0.004) & P$_l$, Ch &  \\
BAT99-124 & WN4     & 63 & & H (0.033) & H (0.026) & P$_s$+H, O, Ch &  \\
BAT99-128 & WN3b     & 112 & & H (0.028) & H (0.027) & Ch & \\
BAT99-131 & WN4b     & 71 & & H (0.022) & H (0.020) & Ch & \\
SMC-AB1 & WN3ha      & 79  & &M (0.020) & M (0.014) & & \\
SMC-AB2 & WN5ha      & 47  & &M (0.015) & M (0.015) & & \\
SMC-AB4 & WN6        & 45  & &H (0.033) & H (0.029) & P$_s$, O & 2, Gaia \\
SMC-AB9 & WN3ha      &  100 & &M (0.013) & M (0.014) & P$_s$ & 3, Gaia \\
SMC-AB10 & WN3ha     & 100 & &M (0.018) & L (0.010) & & \\
\hline
\end{tabular}
\tablefoot{
Spectral types were taken from \citealt{Neugent_2018}, \citealt{Foellmi_2003a}, \citealt{Foellmi_2003b}; $T_{eff}$ has been extracted from \citealt{Hainich_2014} and \citealt{Hainich2015}, and X-ray emission from \citealt{Gosset2005}, \citealt{Guerrero_2008a}, \citealt{Guerrero_2008b} and \citealt{Schnurr2008}. The columns LV and SV mark whether a star is classified as low (L), moderate (M), or highly (H) variable overall or on short timescales, respectively. Numbers in parentheses next to it are the scatter in mag. The seventh column provides information about the variability detected in OGLE data (see Sect.~\ref{subsec:OGLE} for details). The letter "O" indicates outbursting cases, whereas “Ch” denotes cases of changing variability. If a star exhibit one significant peak in its periodogram, letters "P" (for a small number of peaks)  or "F" (for a forest of peaks) are used, with subscripts s or l for short- and long-term signals, respectively (see Table~\ref{tab:ephemeris} for ephemeris). Furthermore, P+(sub)H indicates the presence of (sub-)harmonic. The last column indicates whether the star was previously suspected to be a binary according to the following references: [1] \citealt{Foellmi_2003b}, [2] \citealt{Hainich_2014}, [3] \citealt{Shenar_2019} [4] \citealt{Morgan_1991}, or if its Gaia astrometric parameters suggest a possible binary nature (indicated as "Gaia").}
\end{table*}

\section{Non-detrended light curves} \label{app:LCs}

\begin{figure*}[!h]
\centering
  \includegraphics[width=17cm]{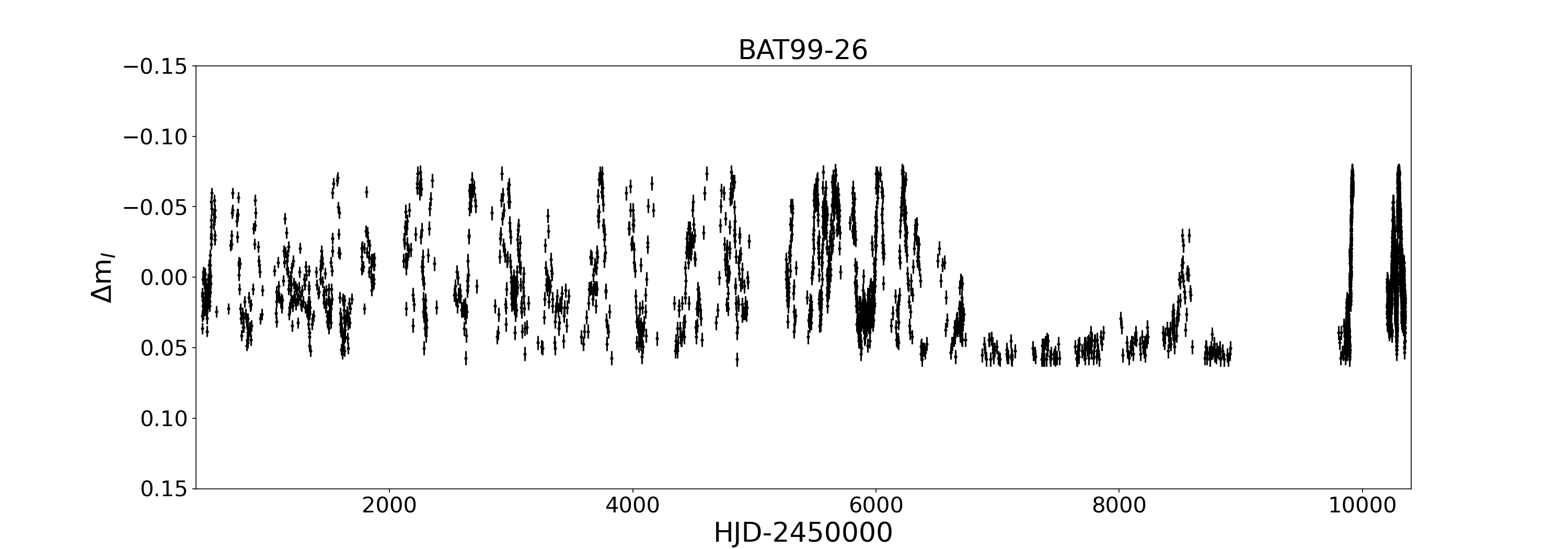}
     \caption{Example of non-detrended light curves: BAT99-26}
     \label{fig:nondetrended-BAT9926}
\end{figure*}

\begin{figure*}[!h]
\centering
  \includegraphics[width=17cm]{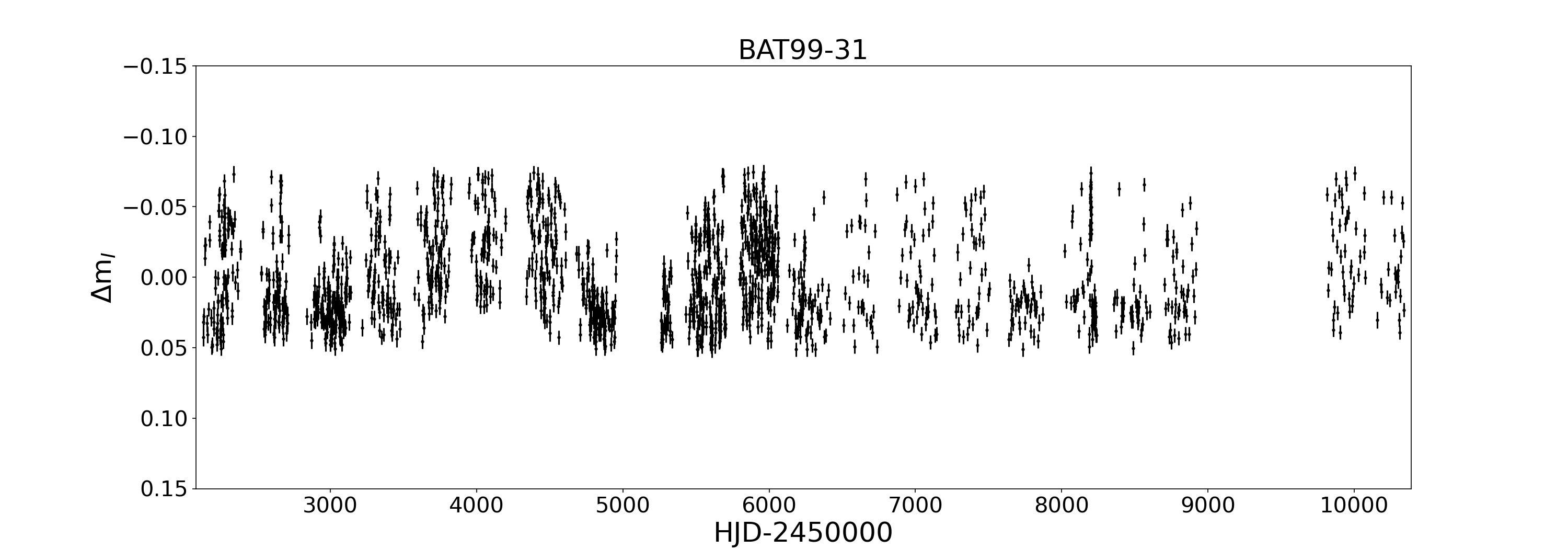}
     \caption{Example of non-detrended light curves: BAT99-31}
     \label{fig:nondetrended-BAT9931}
\end{figure*}

\FloatBarrier

\section{Ephemeris of stars with the OGLE dataset (S/N$\geq$4)} \label{app:ephemeris}

\begin{table}[H]

\scriptsize
\begin{minipage}[t]{0.48\textwidth}
\caption{Ephemeris of stars with a \text{S/N} $\geq 4$ in the OGLE dataset considering long-term variability.}
\centering
\begin{tabular}{cccc}
\hline
Star & Period [d] & T$_{0}$[HJD-2\,450\,000] & Amplitude [mmag] \\
\hline
BAT99-41  & 4789 $\pm$ 300  & 3677 $\pm$ 700  & 30 $\pm$ 2  \\
BAT99-47  &  6800 $\pm$ 600 & 2930 $\pm$ 2000 & 59 $\pm$ 2 \\
BAT99-54  &  303 $\pm$ 4 & 3780 $\pm$ 20 & 10
$\pm$ 2 \\
BAT99-62  & 489 $\pm$ 5  & 7706 $\pm$ 30 & 9 $\pm$ 2  \\
BAT99-67  & 2145 $\pm$ 60 & 5695 $\pm$ 200 & 24 $\pm$ 2 \\
BAT99-94  & 6791 $\pm$ 600 & 5660 $\pm$ 2000 & 6 $\pm$ 1 \\
\hline
\end{tabular}
\tablefoot{T$_0$ corresponds to minimum brightness.}
\label{tab:ephemeris_long}

\end{minipage}
\hfill
\begin{minipage}[t]{0.48\textwidth}

\scriptsize
\centering
\caption{Ephemeris of stars with a \text{S/N} $\geq 4$ in the OGLE dataset for short-term variability. }
\begin{tabular}{cccc}
\hline
Star & Period [d] & T$_{0}$[HJD-2\,450\,000] & Amplitude [mmag] \\
\hline
BAT99-1  &  81.27 $\pm$ 0.09 & 6036 $\pm$ 3 & 10 $\pm$ 2 \\
BAT99-2  &  59.17 $\pm$ 0.05 & 4835 $\pm$ 7 & 5.5 $\pm$ 0.8 \\
BAT99-3  & 10.810 $\pm$ 0.002  & 2184 $\pm$ 2 & 12 $\pm$ 2  \\
BAT99-5  & 42.18 $\pm$ 0.02 & 4784 $\pm$ 4 & 7 $\pm$ 1 \\
BAT99-24 & 6.1826 $\pm$ 0.0004 & 10342.66 $\pm$ 0.09  & 5.1 $\pm$ 0.3  \\
BAT99-26 &  33.7 $\pm$ 0.4 & 8669.0 $\pm$ 0.8 & 24.7 $\pm$ 0.7 \\
BAT99-31 & 29.91 $\pm$ 0.02 & 2132 $\pm$ 9 & 10 $\pm$ 2 \\
BAT99-47 & 1.80463 $\pm$ 0.00004  & 2186.9 $\pm$ 0.7  & 6 $\pm$ 2  \\
BAT99-48 & 23 $\pm$ 2 & 2635.6 $\pm$ 0.6 & 18 $\pm$ 3 \\
BAT99-51 & 6.1053 $\pm$ 0.0005 & 2239.6 $\pm$ 0.3 & 3.5 $\pm$ 0.6 \\
BAT99-56 & 27.44 $\pm$ 0.02 & 5997 $\pm$ 2 & 15 $\pm$ 2 \\
BAT99-65 & 23.118 $\pm$ 0.007 & 5602.7 $\pm$ 0.7 & 19 $\pm$ 3 \\
BAT99-67 & 99.4 $\pm$ 0.2  & 2194 $\pm$ 4  & 6 $\pm$ 1  \\
BAT99-124 & 25.011 $\pm$ 0.008 & 6255 $\pm$ 2 & 10 $\pm$ 2 \\
SMC-AB4  & 51.76 $\pm$ 0.06  & 467 $\pm$ 7  & 16 $\pm$ 2  \\
SMC-AB9  & 13.487 $\pm$ 0.002  & 1763.8 $\pm$ 0.8  & 5.7 $\pm$ 0.4  \\
\hline
\end{tabular}
\tablefoot{T$_0$ corresponds to minimum brightness.}
\label{tab:ephemeris}

\end{minipage}

\end{table}

\FloatBarrier

\clearpage

\section{Detrended panels for stars with S/N$\geq$4} \label{app:detrend_panels}
\vspace{-0.5cm}
\begin{figure*}[!h]
\centering
   \includegraphics[width=16cm, height=0.42\textheight]{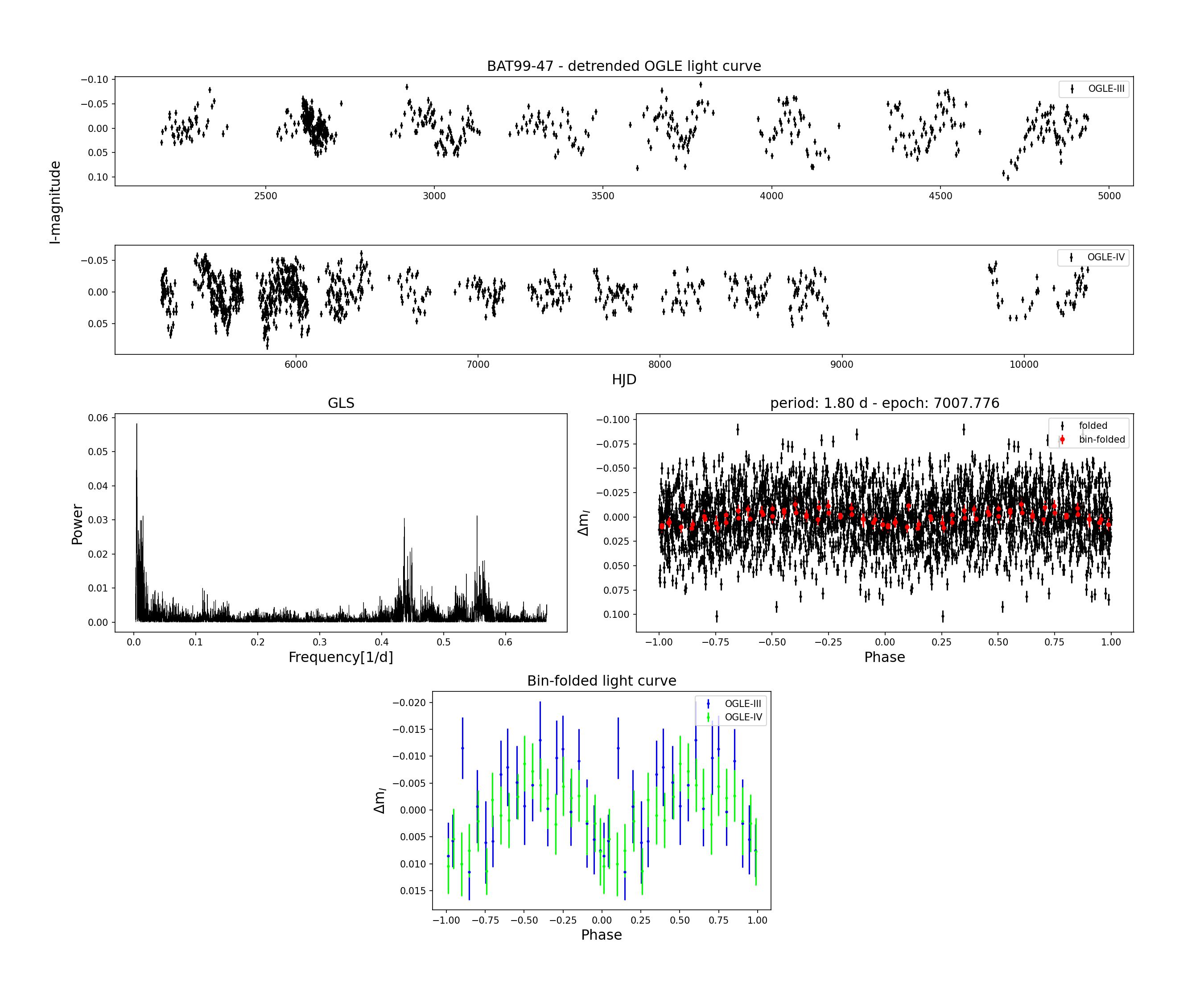}
     \caption{BAT99-47. Top panels: Detrended OGLE light curve. Middle-left panel: GLS frequency spectra. Middle-right panel: Light curve phase-folded to the best period found with GLS presented in black. The binned version is superimposed on it and colored in red. Bottom panel: Binned phase-folded light curve, presented in different colors for different OGLE phases.}
     \label{fig:BAT99-47_panel}
\end{figure*}
\vspace{-0.8cm}
\begin{figure*}[!h]
\centering
   \includegraphics[width=16cm, height=0.42\textheight]{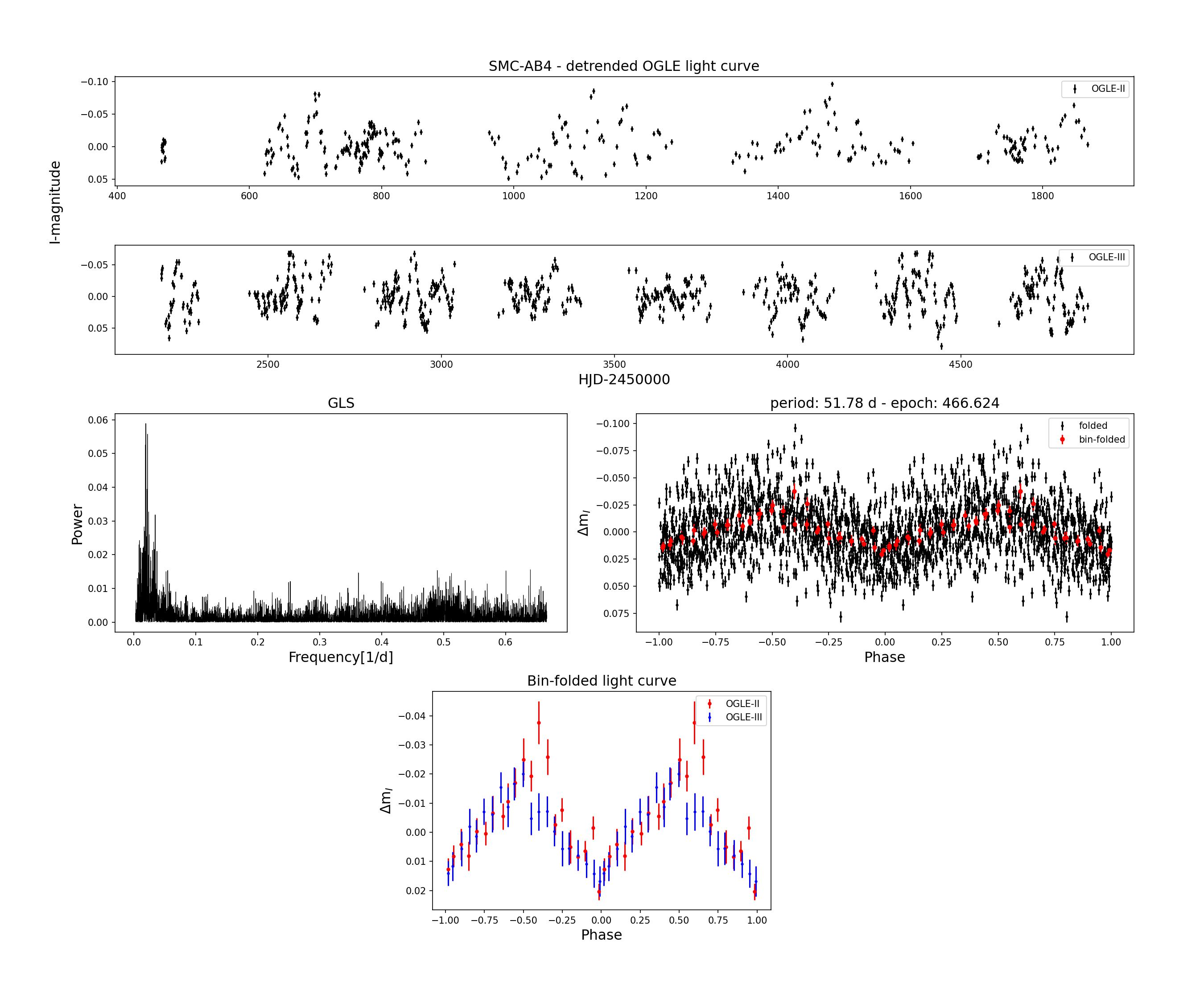}
     \caption{Same as Fig.~\ref{fig:BAT99-47_panel} but for SMC-AB4.}
     \label{fig:SMCAB4-panel}
\end{figure*}

\section{Time-frequency diagrams for detrended data with S/N \texorpdfstring{$\geq$}{>=} 4} 
\label{app:TFD}

\begin{figure*}[!h]
    \centering
    \begin{subfigure}[t]{0.48\textwidth}
        \centering
        \includegraphics[width=\textwidth]{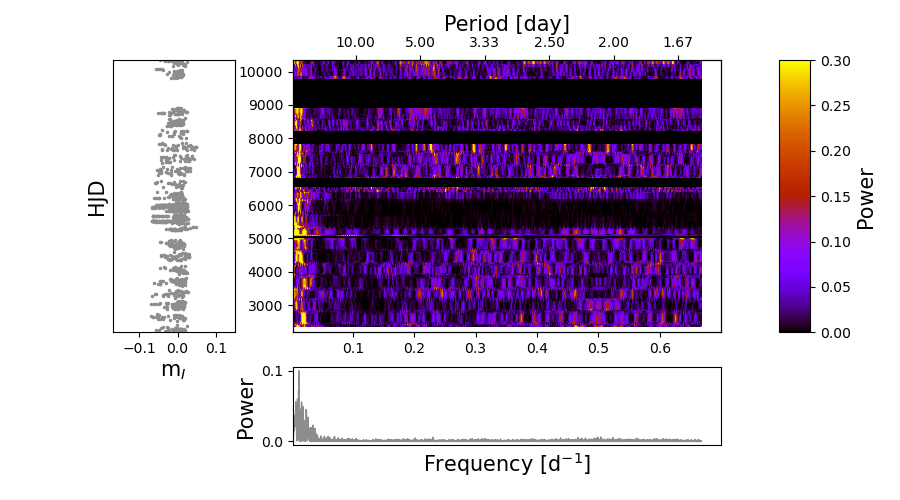}
        \caption{}
        \label{fig:BAT991-TF}
    \end{subfigure}
    \begin{subfigure}[t]{0.48\textwidth}
        \centering
        \includegraphics[width=\textwidth]{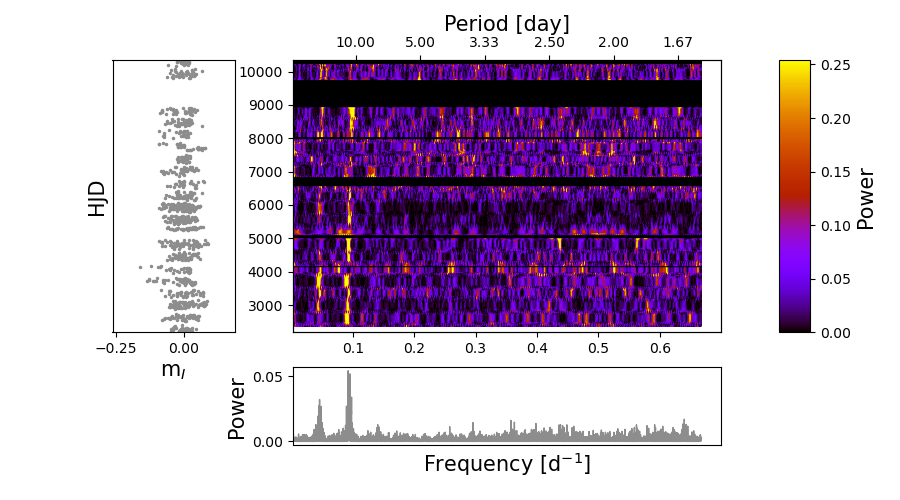}
        \caption{}
        \label{fig:BAT993-TF}
    \end{subfigure}
    \begin{subfigure}[t]{0.48\textwidth}
        \centering
        \includegraphics[width=\textwidth]{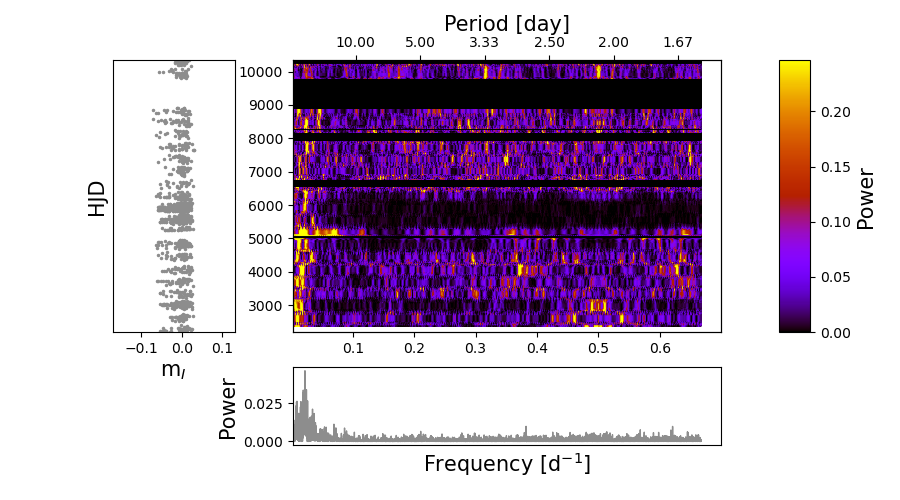}
        \caption{}
        \label{fig:BAT995-TF}
    \end{subfigure}
    \begin{subfigure}[t]{0.48\textwidth}
        \centering
        \includegraphics[width=\textwidth]{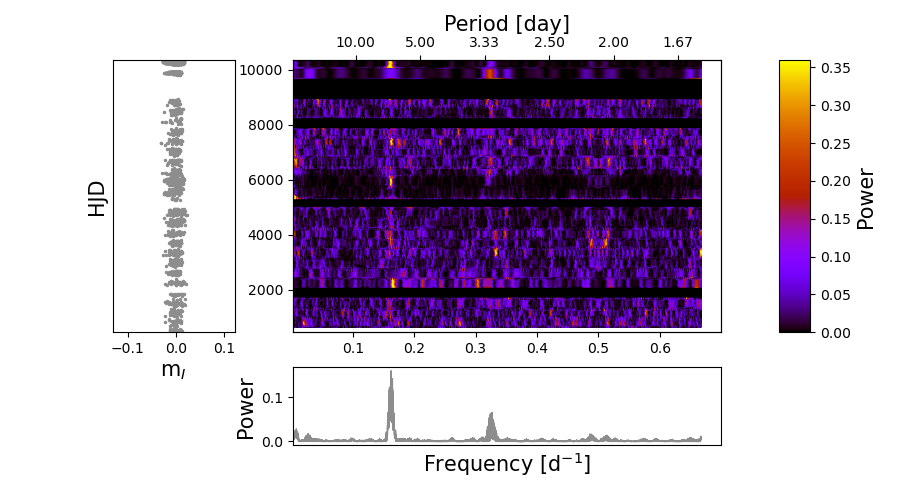}
        \caption{}
        \label{fig:BAT9924-TF}
    \end{subfigure}
    \begin{subfigure}[t]{0.48\textwidth}
        \centering
        \includegraphics[width=\textwidth]{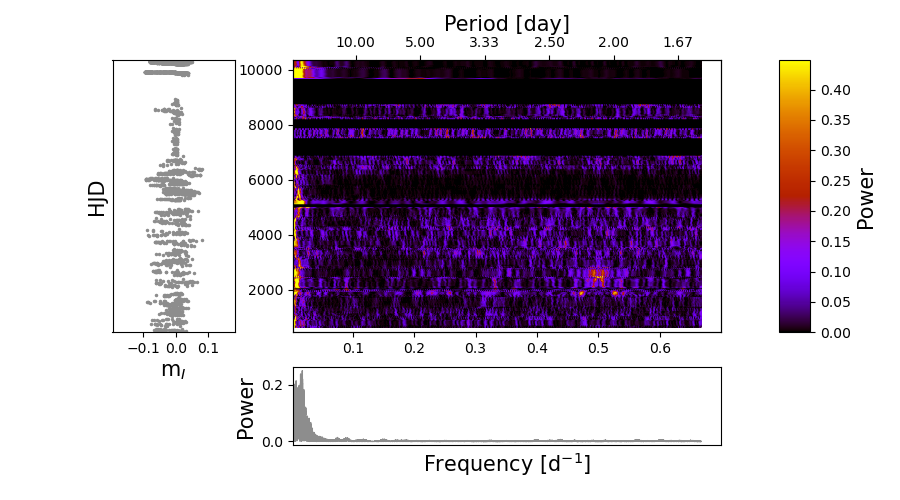}
        \caption{}
        \label{fig:BAT9926-TF}
    \end{subfigure}
    \begin{subfigure}[t]{0.48\textwidth}
        \centering
        \includegraphics[width=\textwidth]{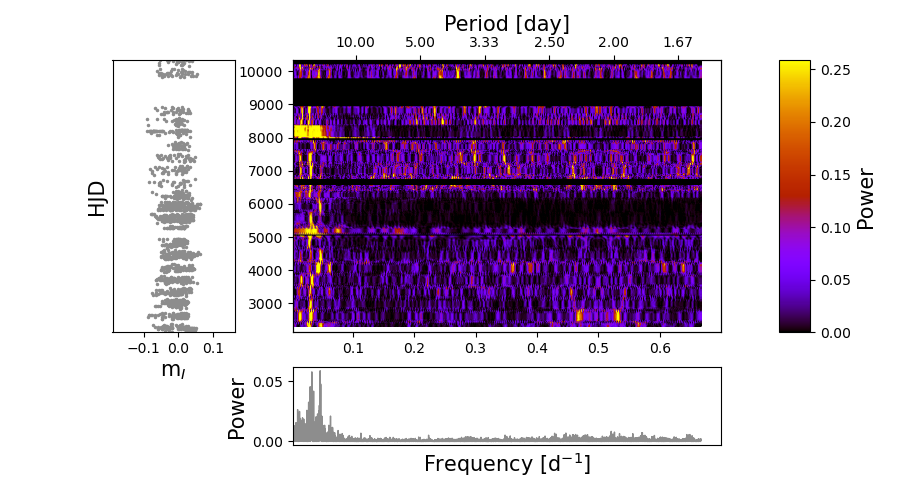}
        \caption{}
        \label{fig:BAT9931-TF}
    \end{subfigure}
    \begin{subfigure}[t]{0.48\textwidth}
        \centering
        \includegraphics[width=\textwidth]{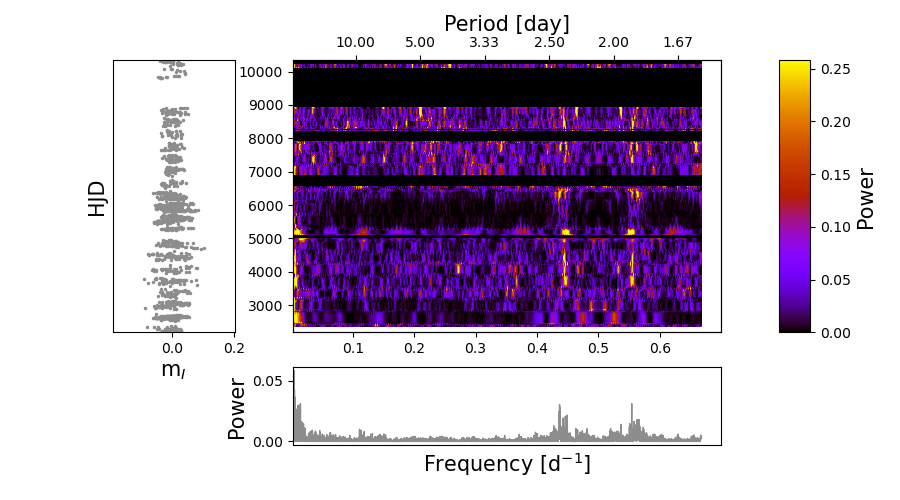}
        \caption{}
        \label{fig:BAT9947-TF}
    \end{subfigure}
    \begin{subfigure}[t]{0.48\textwidth}
        \centering
        \includegraphics[width=\textwidth]{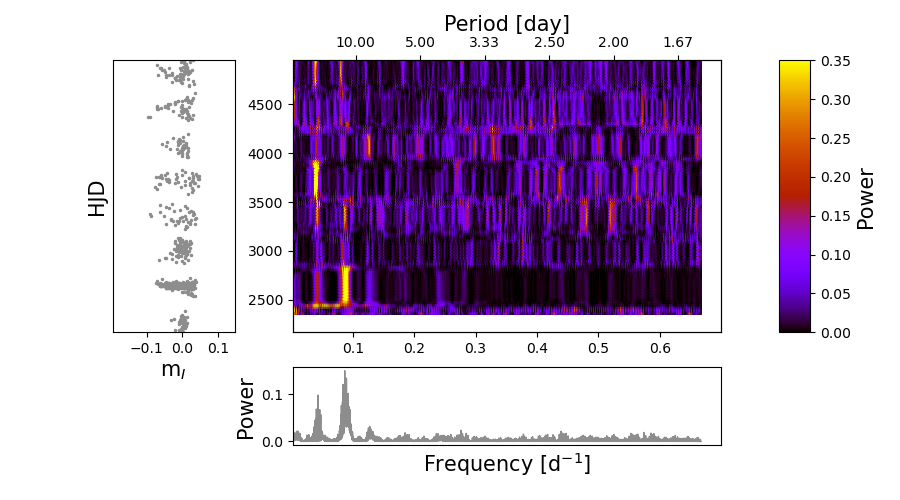}
        \caption{}
        \label{fig:BAT9948-TF}
    \end{subfigure}
    \hfill
    
    \caption{Time-frequency diagrams for the BAT99-1 (a), BAT99-3 (b), BAT99-5 (c), BAT99-24 (d), BAT99-26 (e), BAT99-31 (f), BAT99-47 (g), and BAT99-48 (h). Left panel: Corresponding detrended light curve. Bottom panels: Global GLS periodogram. Middle panel: GLS of the individual light-curve segments.}
    \label{fig:allTF1}
\end{figure*}

\begin{figure*}[ht!]
    \centering
    \begin{subfigure}[t]{0.48\textwidth}
        \centering
        \includegraphics[width=\textwidth]{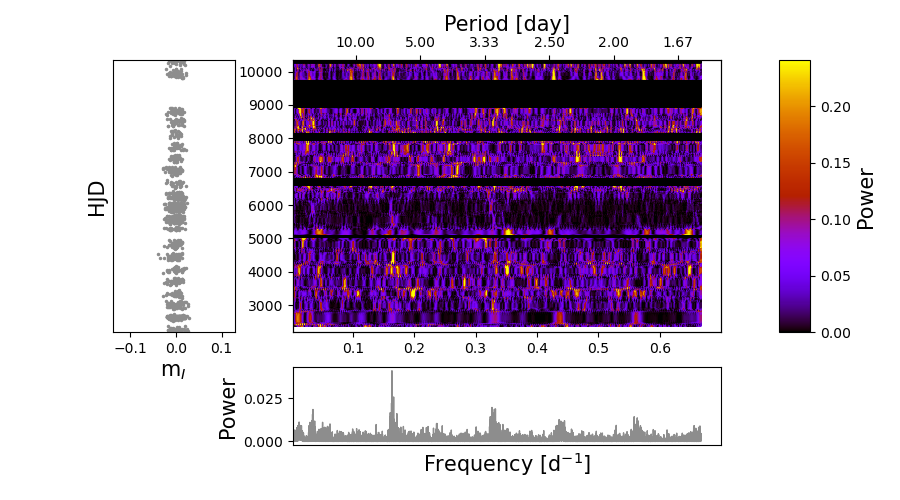}
        \caption{}
        \label{fig:BAT9951-TF}
    \end{subfigure}
    \begin{subfigure}[t]{0.48\textwidth}
        \centering
        \includegraphics[width=\textwidth]{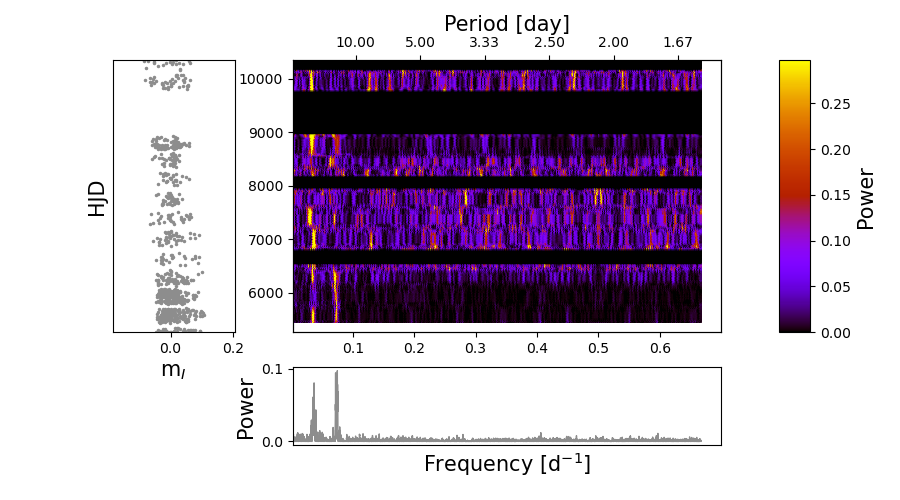}
        \caption{}
        \label{fig:BAT9956-TF}
    \end{subfigure}

    \begin{subfigure}[t]{0.48\textwidth}
        \centering
        \includegraphics[width=\textwidth]{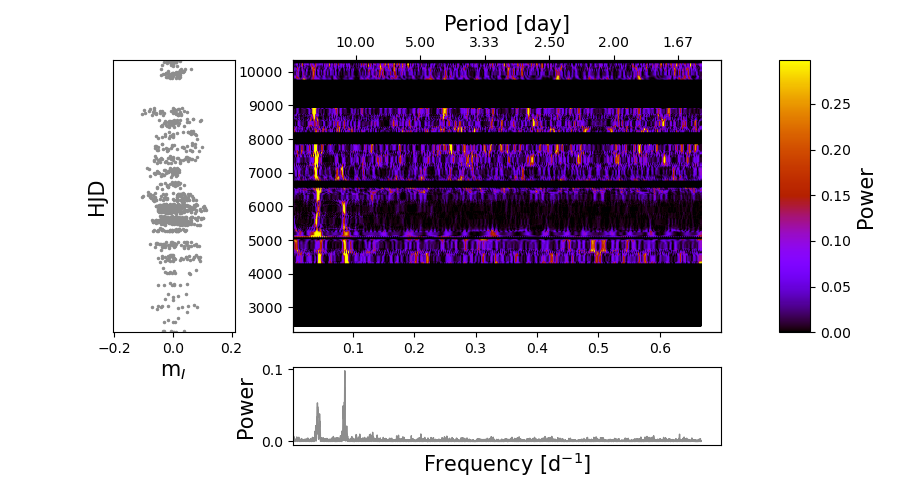}
        \caption{}
        \label{fig:BAT9965-TF}
    \end{subfigure}
    \hfill
    \begin{subfigure}[t]
    {0.48\textwidth}
        \centering
        \includegraphics[width=\textwidth]{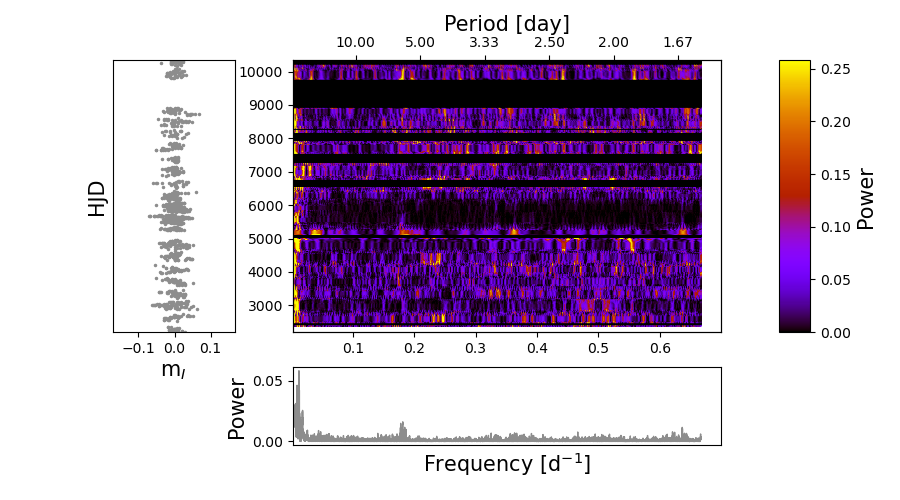}
        \caption{}
        \label{fig:BAT9967-TF}
    \end{subfigure}
    \hfill
    \begin{subfigure}[t]{0.48\textwidth}
        \centering
        \includegraphics[width=\textwidth]{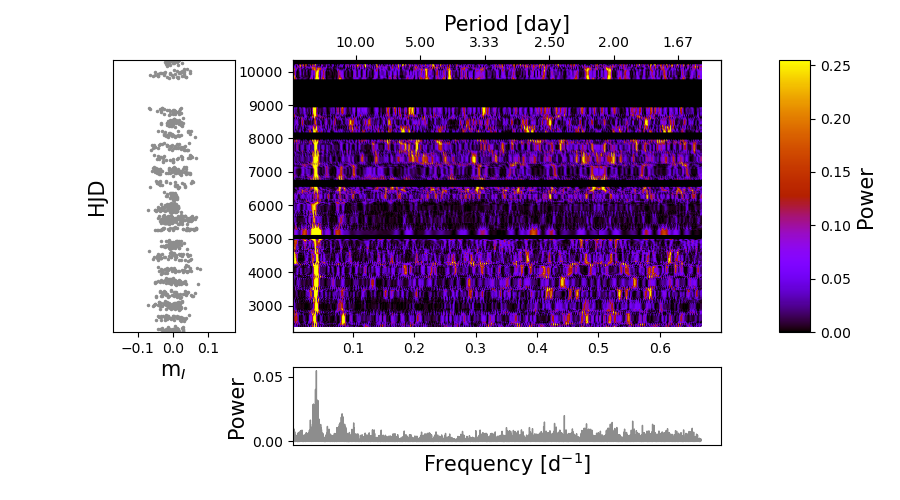}
        \caption{}
        \label{fig:BAT99124-TF}

        \centering
        \includegraphics[width=\textwidth]{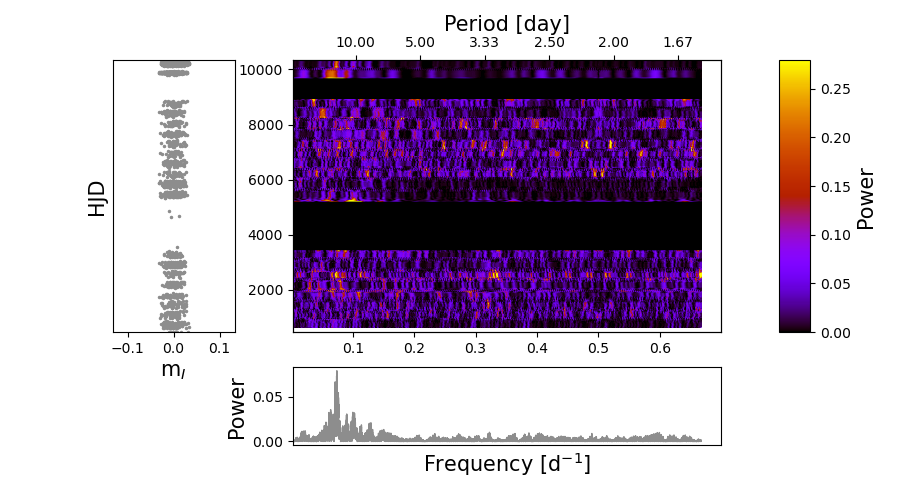}
        \caption{}
        \label{fig:SMCAB9-TF}
    \end{subfigure}
    \hfill
    \begin{subfigure}[t]{0.48\textwidth}
        \centering
        \includegraphics[width=\textwidth]{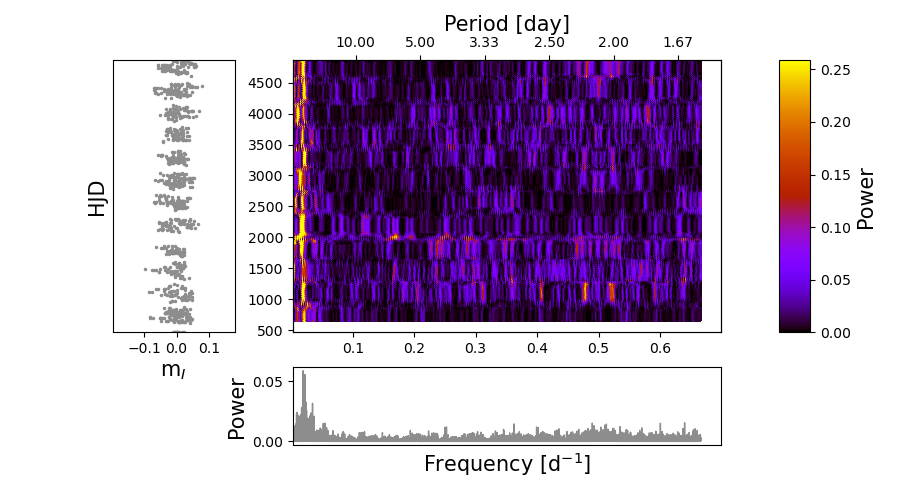}
        \caption{}
        \label{fig:SMCAB4-TF}
    \end{subfigure}

   \caption{Time-frequency diagram for BAT99-51 (a), BAT99-56 (b), BAT99-65 (c), BAT99-67 (d), BAT99-124 (e), SMC-AB4 (f), and SMC-AB9 (g). Left panel: Corresponding detrended light curve. Bottom panels: Global GLS periodogram. Middle panel: GLS of the individual light-curve segments.}
    \label{fig:TF3}
\end{figure*}

\clearpage
\FloatBarrier

\section{Outbursts}
\label{app:outbursts}

\begin{figure*}[!h]
\centering
   \includegraphics[width=17cm, height=0.45\textheight]{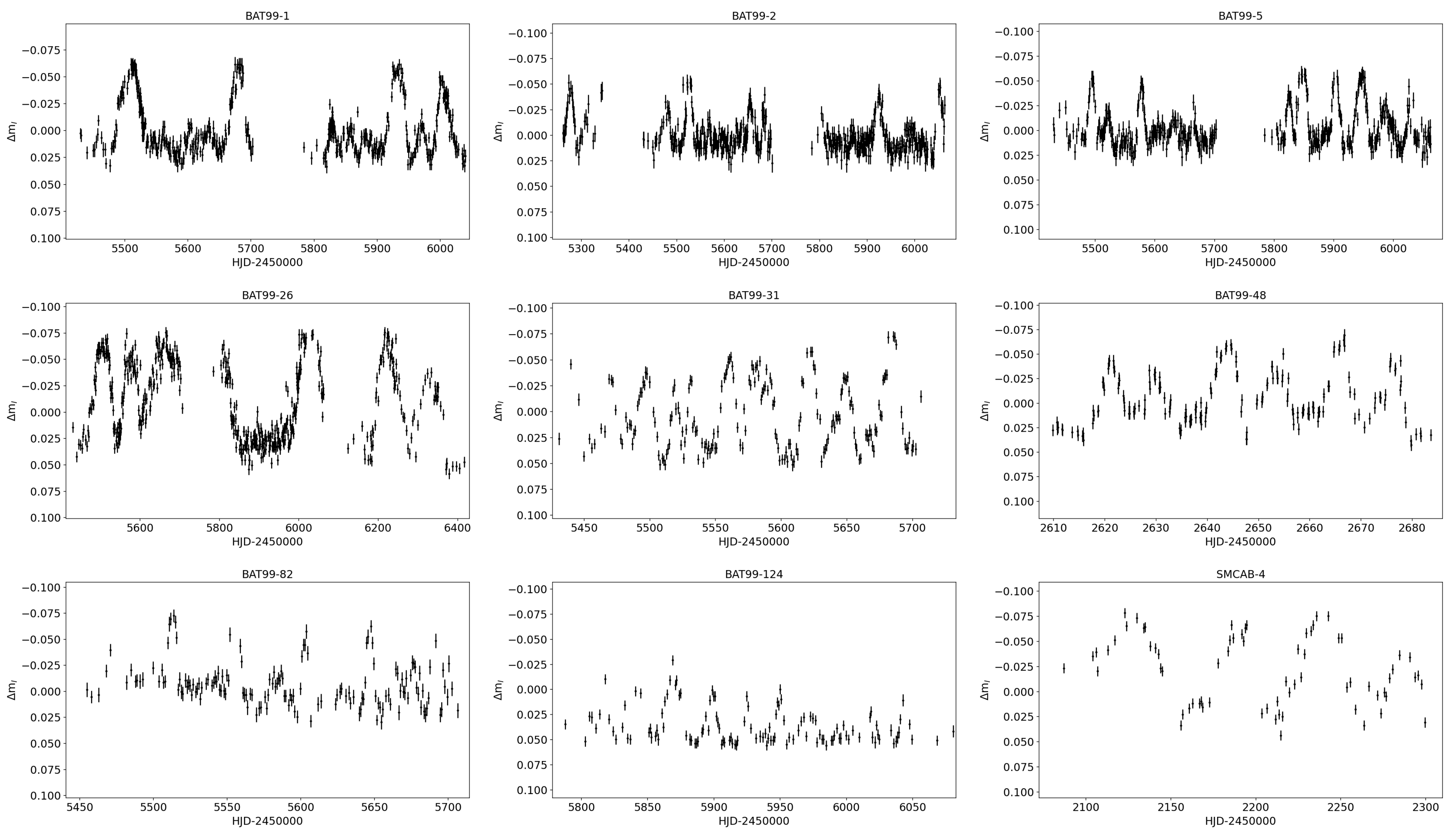}
     \caption{Examples of light curves of the nine stars exhibiting frequent outbursts.}
     \label{fig:outbursts}
\end{figure*}

\FloatBarrier
\twocolumn

\section{Comments on individual stars}
\label{app:comments}

\noindent BAT99-1: 
This star shows no radial velocity (RV) variation but it is classified as a potential runaway by \citealt{Foellmi_2003b} based on the deviation of its mean RV scatter from the one of the LMC. However, \citealt{Neugent_2018} explained that this could be due to the rotation curve of the LMC, as the star is located on the far western side of the Galaxy, where rotational effects lead to lower radial velocities.
The light curve of BAT99-1 exhibits high overall variability and high levels of short-term variability, likely due to recurrent outbursts (Fig.~\ref{fig:outbursts}). The phase-folded plot, using the best-fit period of (81.27$\pm$ 0.09) days from the GLS analysis, shows that the periodic variability is mainly present during OGLE-IV phase. OGLE-III data do not show any significant variability with this period.
\\ 

\noindent \noindent {BAT99-2}:
\citealt{Hainich_2014} suggested that it could be a runaway object. The GLS of the detrended OGLE light curve of this moderate variable detected a forest of peaks with the highest signal at (59.17$\pm$0.05) days. \\

\noindent {BAT99-3}: 
This star is highly variable, both  on overall and short timescales. However, a significant periodicity is detected only on short timescales (10.810 ± 0.002 days). A sub-harmonic at twice this period is also present but is less prominent. 
The $\chi^2$ test applied to the folded and binned light curve does not show consistency between OGLE-III and OGLE-IV.
The signal seems present during the whole observation campaign, as indicated by the time-frequency diagram in Fig.~\ref{fig:BAT993-TF}. An interesting aspect of this star is the possible slight evolution of the signal from approximately 11.36 to 10.5 days, with a similar shift observed in its harmonic.
TESS data from 22 sectors, and folded to the OGLE ephemeris, show a similar variability pattern, with a slightly lower amplitude.
\\

\noindent {BAT99-5}: Recurrent outbursts are visible in the light curve of this star (Fig.~\ref{fig:outbursts}), which is a probable explanation for its moderate variability level both overall and on short timescales. One similarity between BAT99-1 and BAT99-5 is the forest of peaks in the GLS spectrum, which resembles red noise \citep{Naze2021b}, albeit with longer durations (40-100 days centered at (42.18$\pm$0.02) days).  According to the $\chi^2$ test performed on the folded and binned light curve, the same variability is observed in the OGLE-III and OGLE-IV campaigns.
 \\

\noindent {BAT99-24}: 
\citealt{Foellmi_2003b} reported small RV variations, but with no periods found from a Fourier transform. 
This star displays low overall and short-term variability. However, the GLS algorithm identifies a significant short-period peak at (6.1826 $\pm$ 0.0004) days and its less prominent harmonic at half this period value. This signal, although likely present throughout the entire light curve, becomes clearly apparent only in the OGLE-III and OGLE-IV region following the large ``COVID" gap. The light curve, phase-folded to this period and binned, shows that this variability is enhanced in OGLE-III and OGLE-IV, supported by the $\chi^2$ test. The relatively poor cadence of OGLE-II may account for the non-detection of the period in this dataset.
The same significant periodicity is found in MACHO data. 
\\

\noindent {BAT99-26}: This star has small RV variations but no significant peak according to \citealt{Foellmi_2003b}. They reported that the light curve has outbursts visible in MACHO data. 
BAT99-26 exhibits recurrent semi-regular outbursts of larger amplitudes in OGLE data too (Fig.~\ref{fig:outbursts}). Because of them, this star appears among the most variable cases of our sample. BAT99-26 undergoes a relatively quiet phase between HJD $\sim$ 2\,457\,500  and 2\,459\,000 during which no outbursts are observed. The GLS identifies (55.82$\pm$0.03) days as the best period. The prominent variability is evident in the OGLE-IV folded and binned light curve, but is not present in the other two OGLE phases.
According to the $\chi^2$ test performed on the folded and binned data, the variability in OGLE-IV differs from that recorded in OGLE-II and OGLE-III. 

The data collected from 13 TESS sectors do not fold well using OGLE ephemeris (top middle panel in Fig.~\ref{fig:TESSbinned}). However, they were collected during a quiet phase in OGLE data. 
On the other side, they fold well to the period of 33 days, detected in TESS GLS periodogram (top right panel in Fig.~\ref{fig:TESSbinned}).
\\

\noindent {BAT99-31}: 
Some authors classify this star as either a single \citep{Hainich_2014,Shenar_2019} or binary candidate \citep{Foellmi_2003b} with a low inclination angle, though no orbital solution has been established. Its radial velocity scatter is 4.2 times larger than the mean RV scatter in the LMC, supporting the binarity scenario. Additionally, it exhibits a complex Temporal Variance Spectrum (TVS) profile that is not clearly double-peaked. The star also shows diffuse X-ray emission. 

The light curve shows a long-trend sinusoidal feature until HJD 2\,456\,100, when it suddenly becomes quiet; however, there are fewer data points in this region which could prohibit variability detection. Superimposed on this trend, there are a number of outbursts lasting from 10 to 27 days,
making this star highly variable. The GLS algorithm applied to the detrended data identified a (29.91$\pm$0.02) day periodicity. Furthermore, OGLE-III and OGLE-IV light curves are in phase with each other, supported by the $\chi^2$ test performed on the folded and binned data.
The same periodic variability of higher amplitude is detected in TESS data of BAT99-31 as well (left panel in the middle row in Fig.~\ref{fig:TESSbinned}.
We consider this star as a binary candidate.
 \\

\noindent {BAT99-41}: 
The star exhibits high variability over $\sim 13$ years which is clearly visible and identified by the frequency analysis algorithm of the non-detrended light curve (top left panel in Fig.~\ref{fig:longterm}). BAT99-41 has also been observed by MACHO, and their combined folded and binned light curves seem to follow the same trend (left panels in Fig.~\ref{fig:MACHOlongterm}).
No variability pattern similar to that detected in the detrended MACHO data was found in the detrended OGLE data. One possible reason is that MACHO and OGLE data were not taken at the same epoch so the variability detected by MACHO may have evolved (indeed, it is not even visible over the whole MACHO light curve).
\\

\noindent {BAT99-47}: 
\citealt{Foellmi_2003b} did not detect periodic radial velocity variations for this star. Although \citealt{Hainich_2014} suggests binarity due to the presence of X-ray emission, the X-ray detection is described as "tentative" in the analysis by \citealt{Guerrero_2008b}, which is not strong evidence in favor of binarity.
Moreover, \citealt{Shenar_2019} found its radial velocity to be consistent with that of a single star.
BAT99-47 displays high overall variability over $\sim 18.6$ years.
It has also been observed by MACHO, and their combined folded and binned light curves seem to follow the same trend (Fig.~\ref{fig:MACHOlongterm}).
The GLS algorithm performed on detrended light curve detected two signals 
that appear in some parts of time-frequency diagram (Fig.~\ref{fig:BAT9947-TF}): (1.80463 $\pm$ 0.00004) days and (2.34087 $\pm$ 0.00007) days. The variability observed in OGLE-III and OGLE-IV phases is of high level and consistent.
The same 2.3-day periodic variability, but of higher amplitude is detected in the analyzed TESS sector of BAT99-47 (Fig.~\ref{fig:TESSbinned}).
\\

\noindent {BAT99-48}: 
It is not known to have a significant nebula around it. 
This star is found to be highly variable on both long and short timescales. Outbursts are visible in at least some parts of the light curve (Fig.~\ref{fig:outbursts}). The GLS analysis detects a signal at (23$\pm$2) days, along with its harmonic at half this period, both of which persist throughout the entire time-frequency diagram (Fig.~\ref{fig:BAT9948-TF}).
BAT99-48 has also been observed by MACHO and it displays a similar significant periodicity. However, MACHO light curves do not fold well with the OGLE derived periods, suggesting the signals to be epoch dependent.
\\

\noindent {BAT99-51}:  This star that shows medium level variability on both time scales displays a significant peak at (6.1053$\pm$0.0005) days. The $\chi^2$ test reveals that the light curve folded and binned around this period shows a good agreement between OGLE-III and OGLE-IV. Moreover, similar period is observed in the MACHO data, with both datasets folding well. However, the time-frequency diagram of this star in Fig.~\ref{fig:BAT9951-TF} does not show the aforementioned period in OGLE.  \\

\noindent {BAT99-54}: It shows a medium level of variability and a long-term trend with a period of $\sim 303$ days (top right panel in Fig.~\ref{fig:longterm}). 
No significant peak was found on short timescales. 
 \\

\noindent {BAT99-56}: 
\citealt{Foellmi_2003b} marked this star as single since no RV variations were found. BAT99-56 exhibits one of the highest scatter of our sample. The detected period of (27.44 $\pm$ 0.02) days is consistently present throughout the entire time-frequency diagram, along with its first harmonic. \\

\noindent {BAT99-62}: This moderately variable star shows a long-term variability with a period of $\sim 489$ days (left panel in the third row of the Fig.~\ref{fig:longterm}). 
No significant peak was found for short timescales.
\\

\noindent {BAT99-65}: Similar to BAT99-56, this star exhibits one of the highest scatter in our sample, considering both short and overall timescales. Its short-term period is p$_{0}$= (23.118 $\pm$ 0.007) days. It is present throughout the entire time-frequency diagram, along with its harmonic at $1/2$p$_{0}$ (Fig.~\ref{fig:BAT9965-TF}). However, these variations do not appear to be fully coherent, as the OGLE-III light curve is different, both in shape and amplitude from that of OGLE-IV.
With MACHO data we detected a significant periodicity at p$_{0}$.
\\

\noindent {BAT99-67}:  
Radial velocity measurements for this star show small variations, but Fourier transform analysis indicates only marginal evidence of periodicity \citep{Foellmi_2003b}, suggesting it is unlikely to be a short-period binary.
\citealt{Hainich_2014} classify BAT99-67 as a binary suspect due to the high X-ray luminosity reported by \citealt{Guerrero_2008b}.
This star shows a long-term variability with a high scatter at $ \sim 5.9$ years (middle panel in the third row of the Fig.~\ref{fig:longterm}). 
The phase-folded light curves from OGLE-III and OGLE-IV show good agreement between OGLE phases, but not between OGLE and MACHO data (third row panel in Fig.~\ref{fig:MACHOlongterm}). The short-term variability is of moderate level.
\\

\noindent {BAT99-94}: It shows a low-amplitude long-term variability with a significant period detected at $\sim 18.6$ years (right panel in the third row in Fig.~\ref{fig:longterm}). 
BAT99-94 has also been observed by MACHO and its folded light curves seem to follow the same trend, although the OGLE-detected period is longer than the duration of the MACHO dataset (top right panel in Fig.~\ref{fig:MACHOlongterm}).
No significant peak was found for short timescales, where variability is also low. \\

\noindent {BAT99-124}: A prominent peak at p$_{0}$=(25.011 $\pm$ 0.008) days is present throughout the entire detrended highly variable light curve, along with its first harmonic (Fig.~\ref{fig:BAT99124-TF}). A $\chi^2$ test performed on the folded and binned light curve indicates that the variability differs between OGLE-III and OGLE-IV campaigns. \\ A similar significant period is identified in MACHO light curves. However, the MACHO data do not fold well with the OGLE derived ephemeris - meaning that the signals are epoch dependent.\\

\noindent {SMC-AB4}: The periodogram reveals a clear peak at (51.76 $\pm$ 0.06) days, which persists throughout the entire light curve in the time-frequency diagram (Fig.~\ref{fig:SMCAB4-TF}). Additionally, the variability level is among the highest in the sample, and the three OGLE campaigns align well (Fig.~\ref{fig:SMCAB4-panel}). 
\citealt{Schootemeijer2024} also found a 51 day-period from OGLE data although they did not consider it significant. 
\citealt{Moffat_1988}, \citealt{Foellmi_2003b} and \citealt{Crowther_2000} classified this star as single. \citealt{Moffat_1988} based their classification on radial velocity measurements from photographic spectra collected over two 10-day intervals separated by a year, which may have been insufficient to detect a 51-day period. \citealt{Foellmi_2003b} reported a 6.55-day period from MACHO light curve, but this signal is not present in the OGLE data.
Our analysis of MACHO data indeed reveals this short periodicity.
\citealt{Pauli_2023} claimed that empirical position of SMC-AB4 in the HRD  is explained solely by their models of secondaries that have accreted material during mass transfer. They also suggested that the spectroscopic analysis of the oxygen OIV multiplet revealed enhancements consistent with past accretion from a primary companion. According to \citealt{Hainich_2014}, the relatively low hydrogen abundance derived for SMC-AB4, despite its late spectral type, could be explained by past interaction with a close stellar companion, which may have stripped part of the hydrogen-rich envelope from the donor star.
\\

\noindent {SMC-AB9}: This moderately variable star shows a (13.487 $\pm$ 0.002) day periodicity. The phase-folded data show differences in OGLE campaigns, supported by the $\chi^2$ test performed. 
\citealt{Schootemeijer2024} also identified a 13-day period in OGLE data but did not consider it significant.
\citealt{Foellmi_2003b} detected small radial velocity variations in this star; however, its faint magnitude ($v$=15.7) suggests these variations might be instrumental. The star was initially classified as WN2.5+abs by \citealt{Morgan_1991}. The highly blueshifted absorption lines, relative to the SMC's systemic velocity, likely originate from the WR star wind rather than a companion \citep{Foellmi_2003a}.\\ 

For the rest of the stars from the sample, no significant short- or long-term periodicities have been identified in OGLE data. 
Five of them display high short-term and overall variability (BAT99-30, BAT99-37, BAT99-75, BAT99-128, and BAT99-131). Fifteen show moderate variability overall, and thirteen on short timescales. The remaining 7 (9) are low-variable on overall- (short-) time scales.
The following remarks apply to these stars:\\

\noindent {BAT99-7}:
this star has high rotational velocity \citep{Hainich_2014} and is suggested to be a wind-stripped WR \citep{Shenar_2019}. \\

\noindent {BAT99-18}:
According to \citep{Martins2013}, this star is too hot to be reproduced by standard evolutionary tracks. It may be an example of quasi-homologous evolution. \\

\noindent {BAT99-23}:
\citealt{Foellmi_2003b} detected a weak presence of hydrogen. They reported that its He II
line is weaker than in most LMC WN3 stars. \\

\noindent {BAT99-25}:
\citealt{Foellmi_2003b} detected a clear presence of hydrogen in both emission and absorption, originating in the wind of the WR star itself. \\

\noindent {BAT99-30}:
\citealt{Hainich_2014} reported a higher stellar temperature than in previous studies, likely due to line-blanketing effects. They also confirmed the mass-loss rate, although they derived a higher luminosity.\\

\noindent {BAT99-35}:
\citealt{Foellmi_2003b} and \citealt{Hainich_2014} obtained the same classification for this star: WN3(h) \\

\noindent {BAT99-37}:
\citealt{Foellmi_2003b} reported that this star is WN3 and no hydrogen was detected. \\

\noindent {BAT99-40}:
\citealt{Foellmi_2003b} reported X-ray emission in the archival ROSAT data, although \citealt{Guerrero_2008b} listed this object as undetected in ROSAT observations. According to \citealt{Shenar_2019}, the high X-ray luminosity could suggest the presence of colliding winds. \\

\noindent {BAT99-44}:
\citealt{Hainich_2014} analysis agreed with the temperature previously reported for this star, but they did not detect the previously reported infrared excess. The derived luminosity is roughly twice that of previous studies, while the mass-loss rate remains similar.\\

\noindent {BAT99-46}:
\citealt{Hainich_2014} provided the only analysis with model atmosphere. \\

\noindent {BAT99-50}:
\citealt{Hainich_2014} provided the only analysis with model atmosphere. \\

\noindent {BAT99-57}:
According to  \citealt{Hainich_2014}, it is a WN4b star with typical stellar parameters. \\

\noindent {BAT99-58}:
\citealt{Hainich_2014} claimed that their fit does not show a clear infrared excess as previous studies, although a slight mismatch in the infrared photometry could be found. \\

\noindent {BAT99-60}:
\citealt{Shenar_2019} classified this star as a binary candidate. They claimed that absorption lines belonging to He I cannot originate in the hot WR star. This is supported by \citealt{Neugent_2018}. The low RV scatter implies either that the inclination of the system is low or that it is not a spectroscopic binary.  \\

\noindent {BAT99-63}:
it was marked as a binary candidate in the BAT99 catalog, although \citealt{Foellmi_2003b} did not find radial velocity variation. \citealt{Schnurr2008} confirmed that it is a runaway star.  \\

\noindent {BAT99-66}:
\citealt{Hainich_2014} provided the only analysis with model atmosphere.\\

\noindent {BAT99-73}:
\citealt{Foellmi_2003b} found the clear presence of hydrogen in the spectrum of BAT99-73.\\

\noindent {BAT99-74}:
\citealt{Foellmi_2003b} detected the weak presence of hydrogen in their spectrum, which appears to arise in the WR wind. Both absorption and emission lines can be reproduced with \citealt{Hainich_2014} model. \\

\noindent {BAT99-75}:
\citealt{Foellmi_2003b} did not detect hydrogen in the spectrum. \\

\noindent {BAT99-81}: Although no significant peak was found in OGLE data, we detected a significant peak around 6.26 days in MACHO data.  \\

\noindent {BAT99-82}: 
\citealt{Hainich_2014} classified BAT99-82 as a binary suspect due to the X-ray emission detected by \citealt{Guerrero_2008b}.
It is a moderately variable star with outbursts (Fig.~\ref{fig:outbursts}). We note the presence of a ``quiet" region (similar to one in BAT99-26 light curve), during which no outbursts are observed: between HJD $\sim$ 2\,457\,000-2\,458\,100.
The different OGLE epochs exhibit the same variability according to the $\chi^2$ test.
\\

\noindent {BAT99-89}: \citealt{Hainich_2014} found a small difference in temperature compared to previous studies, although they attributed it to the line-blanketing effect. \\

\noindent {BAT99-128}:
\citealt{Foellmi_2003b} suggested that this star might be a runaway.\\

\noindent {BAT99-131}: Although no significant peak was found in OGLE data for this highly variable star, MACHO detects a 12-day peak. This signal is only observed with OGLE data taken between HJD 2\,457\,200 and 2\,458\,900. \\

\noindent {SMC-AB1}:
\citealt{Foellmi_2003a} reported small RV variations, explained by the star being single, short-period binary with a small inclination angle and/or large eccentricity, or long-period binary.\\

\noindent {SMC-AB2}:
\citealt{Foellmi_2003a} reported small RV variations, explained by the star being single, short-period binary with a small inclination angle and/or large eccentricity, or long-period binary. \citealt{Pauli_2023} proposed that this star could have a hidden compact companion. \\

\noindent {SMC-AB10}:
\citealt{Foellmi_2003a} reported small RV variations, explained by the star being single, short-period binary with a small inclination angle and/or large eccentricity, or long-period binary.
\end{appendix}

\end{document}